\documentclass[reprint,superscriptaddress,showpacs,amssymb,amsmath,aps,prd,longbibliography]{revtex4-2}
\usepackage[utf8]{inputenc}
\usepackage{graphicx}
\usepackage{xcolor}
\usepackage{amsmath}
\usepackage{amsthm}
\usepackage{bm}
\usepackage{bbm}
\usepackage{comment}
\usepackage{simpler-wick}
\usepackage{tikz}
\usepackage[compat=1.0.0]{tikz-feynman}

\usepackage{hyperref}

\usepackage{mathdots}
\usepackage{lipsum}
\usepackage{verbatim}
\usepackage{natbib}
\usepackage{nccmath}
\usepackage{amsfonts} 
\usepackage[caption=false]{subfig}

\newcommand{\T}{\text{\footnotesize\ensuremath{T}}}

\newcommand{\swap}{\text{SWAP}}
\DeclareMathOperator{\sgn}{sgn}

\usepackage{amsfonts} 
\DeclareMathSymbol{\shortminus}{\mathbin}{AMSa}{"39}

\newtheorem{theorem}{Theorem}
\newtheorem{lemma}{Lemma}

\newtheorem{corollary}{Corollary}

\newcommand{\smallprod}{\mathop{\raisebox{0.25ex}{\scalebox{0.7}{$\displaystyle \prod$}}}}

\begin{document}

\title{Spacetime quantum mechanics for bosonic and fermionic systems
}

\author{N. L.\  Diaz}

     \affiliation{Center for Nonlinear Studies (CNLS), Los Alamos National Laboratory, Los Alamos, New Mexico 87545, USA}
 \affiliation{Information Sciences, Los Alamos National Laboratory, Los Alamos, NM 87545, USA}
 \affiliation{Departamento de F\'isica-IFLP/CONICET,
		Universidad Nacional de La Plata, C.C. 67, La Plata (1900), Argentina}
	
	\author{R. Rossignoli}
	\affiliation{Departamento de F\'isica-IFLP/CONICET,
		Universidad Nacional de La Plata, C.C. 67, La Plata (1900), Argentina}
	\affiliation{Comisi\'on de Investigaciones Cient\'{\i}ficas (CIC), La Plata (1900), Argentina}
    
\begin{abstract}
  We provide a Hilbert space approach to quantum mechanics where space and time are treated on an equal footing. Our approach replaces the standard dependence on an external classical time parameter with a spacetime-symmetric algebraic structure, thereby unifying the axioms that traditionally distinguish the treatment of spacelike and timelike separations. Standard quantum evolution can be recovered from timelike correlators, defined by means of a quantum action operator, a quantum version of the action of classical mechanics. The corresponding map also provides a novel perspective on the path integral formulation, which, in the case of fermions, yields an alternative to the use of Grassmann variables. In addition, the formalism can be interpreted in terms of generalized quantum states, codifying both the conventional information of a quantum system at a given time and its evolution. 
  We show that these states are solutions to a quantum principle of stationary action grounded in timelike correlations and pseudo-entropies
\end{abstract}

\maketitle

\section{Introduction}\label{sec:intro}
One of the most fundamental assumptions involved in the description of a given system in standard quantum mechanics (QM) is the identification of each possible state of the system with a vector belonging to a properly defined Hilbert space. 
While usually implicit, this very notion of state is linked to a 
definition of ``present'': one assumes that the configuration of the system is specified at a given time such that the physical predictions that one can extract from the state are those corresponding to the same instant. 
Quantum evolution also enters the picture with respect to a pre-established choice of reference frame with the vectors in Hilbert space parameterized by classical time.  By contrast, if two distinguishable quantum systems are separated in space, one defines the joint Hilbert space from a tensor product of the individual Hilbert spaces. General states appear then entangled across space, a characteristic feature unequivocally separating the quantum and classical worlds \cite{bell1964einstein}.

\begin{figure}[t!]
    \centering
    \includegraphics[width=0.95\linewidth]{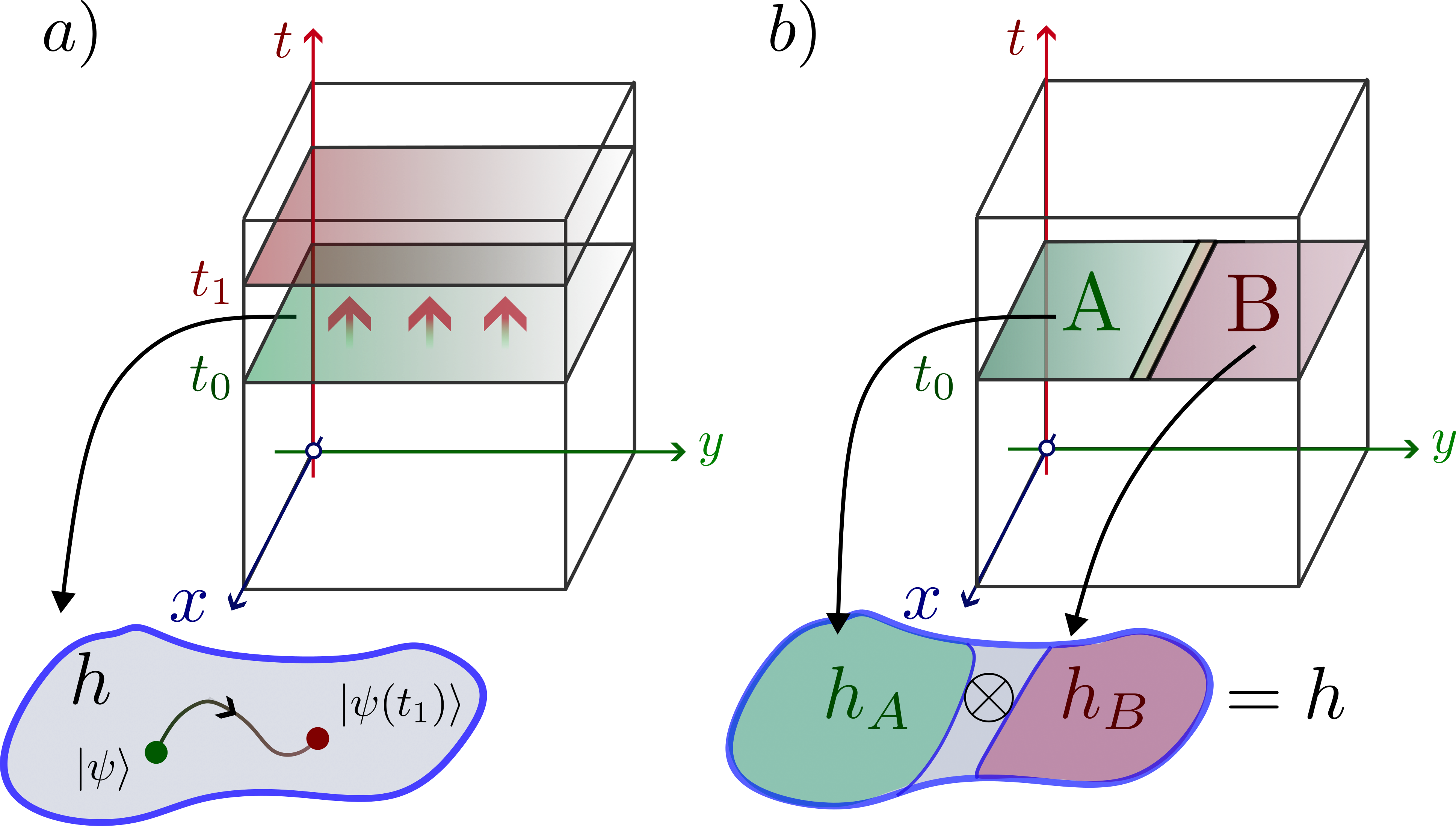}
    \caption{\textbf{Asymmetry in the axioms of standard QM in the treatment of spacelike and timelike separations}. Panel a) depicts how QM assigns a Hilbert space $h$ of possible states of the system at a given time. As time ``flows'', $h$ remains the same while the quantum states change within $h$ according to a classical parameterization. Panel b) represents the scenario of two spacelike separated regions $A$, $B$ (or equivalently systems). QM assigns a different Hilbert space to each region while the joint Hilbert space is constructed from their tensor product, i.e.  $h=h_A\otimes h_B$. }
    \label{fig:intro}
\end{figure}

While for most applications these assumptions are natural, from a foundational perspective this is not entirely satisfactory: firstly,  Einstein's relativity \cite{E.05} suggests that a fundamental description of nature needs to involve a symmetric treatment between space and time. As we depict in Figure \ref{fig:intro}, QM clearly violates this symmetry at the axiomatic level \cite{diaz2024spacetime, DIAZ2025170052}. Secondly, if spacetime is not fundamental, as the discussion about space emerging from quantum correlations suggests \cite{ryu.06,van2010building, evenbly2011tensor, Ca.17}, one should not need to introduce classical information within QM to describe evolution.

In this manuscript, we provide a Hilbert space construction describing general quantum systems in spacetime. Namely, we replace the dependence on an external evolution parameter with the notion of spacetime symmetric algebras which for distinguishable systems unifies the two axioms of Figure \ref{fig:intro}. The formalism builds upon our recent works \cite{diaz2024spacetime}, where we considered a canonical spacetime symmetric approach to classical and quantum field theories (QFTs),  and \cite{DIAZ2025170052}, where we established a deep connection between the formalism and the Path integral (PI) formulation of Feynman \cite{Feynm.1948a} for bosonic systems. Here, we focus on arbitrary quantum systems, including the finite dimensional and non-relativistic cases. In particular, we describe the case of fermions where a notion of tensor product in space and standard PI do not apply, requiring a proper generalization of our previous approach.

The manuscript is organized as follows. In section \ref{sec:prelim} we review the bosonic case developed in \cite{diaz2024spacetime, DIAZ2025170052} but emphasizing finite dimensional systems. In \ref{sec:prelim1} we employ a tensor-product-in-time approach, following our previous work. One of the most important results presented in this section, is a map between quantities in the extended formalism, where time is geometrically defined, and standard QM with classical time. This maps make use of a quantum action (QA) operator \cite{diaz.21}, a quantum version of the classical action codifying evolution through timelike correlators. 
In \ref{sec:prelim2} we discuss an equivalent scheme based on spacetime symmetric algebras. This change of perspective allows us to generalize the formalism to fermions in the next section \ref{sec:fermioniccase}. We develop the fermionic case in detail, as many novelties with respect to bosons arise. In particular, we introduce the fermionic spacetime Hilbert space and fermionic QA. Then, through several Theorems we establish a map to standard QM for discrete time. Finally, we discuss the case of Dirac fermions and continuum spacetime. Several remarks concerning the fermionic PI and Grassmann variables are also included (with a full discussion given in the Appendix \ref{app:PIs}).

In section \ref{sec:implications} we address additional aspects and consequences of the formalism which apply to both bosons and fermions. In \ref{sec:states} we introduce a notion of generalized quantum state applied to spacetime. This notion allows one to interpret timelike correlations as arising from generalized correlations with an environment. The ensuing definition of entropy is then employed to introduce a quantum principle of stationary action in \ref{sec:qaprinciple}. We also discuss in \ref{sec:fbinteractions} the scenario in which both bosons and fermions are present and can interact with each other. In addition, we provide a description on how our formalism can be related to the Page and Wootters (PaW) mechanism \cite{PaW.83}. The latter has recently attracted wide interest \cite{QT.15,b.16,di.19,dia.19,hoh.21,fav.22,diaz2023parallel,giovannetti2023geometric, suleymanov2024nonrelativistic,cafasso2024quantum} as it can be used to replace time evolution with quantum correlations between two quantum systems. While the formalism cannot tackle the asymmetry we described in Figure \ref{fig:intro} \cite{dia.19, diaz.21}, in \ref{sec:paw} we show a notable connection with our proposal: if one applies our scheme to quantum fields, the single particle states are precisely PaW states. We show that this result, previously proposed for bosons \cite{diaz.21, diaz2024spacetime}, hold for both scalar and Dirac fields according with the PaW approach to relativistic particles developed in \cite{dia.19,di.19}. Moreover, in \ref{sec:paw} we show that the QAs, including the Dirac QA, are the second quantized versions of the universe operators of PaW. Finally, we explicitly show that recent proposals \cite{MAP.25,Guo.25} that aim to establish a framework for timelike pseudoentropies \cite{tak.23, har.23, nar.22, chu2023time} can be recovered from our formalism.

In Section~\ref{sec:concl}, we provide a final discussion that highlights both the conceptual implications arising from our formulation and the novel perspectives it opens to explore previously raised foundational questions \cite{fit.15,ho.17,cot.18}. We also discuss how our results lay the foundation for developing novel applications that lie beyond the direct reach of standard QM.

\section{Bosonic spacetime formalism }\label{sec:prelim}

As an introduction to the recently defined spacetime algebras and their relationship with conventional QM and the Path Integral (PI) formulation, we begin by reviewing the bosonic-like scenario previously studied in \cite{diaz.21, DIAZ2025170052, diaz2024spacetime}. Here, we restate and deepen these results with a treatment that makes the fermionic case alike.  
This section also establishes the foundations necessary to develop the novel implications discussed in Section~\ref{sec:implications}, valid for both bosons and fermions.

\subsection{The tensor product-in-time approach}\label{sec:prelim1}

We begin our discussion with the case of qudits (the infinite dimensional case is included). We will refer to these systems as boson-like because they share the composition rule of bosons:
the state space of a global system composed of bosons is the tensor product of the Hilbert spaces of the components. In this section we describe how one can generalize this tensor product structure from spacelike separations to timelike ones.

Consider a $d$ dimensional Hilbert space $h=\text{span}\{|i\rangle\}$ with $i=1,\dots, d$. We introduce a new Hilbert space as the $N$-fold tensor product of $N$ copies of $h$, namely, $\mathcal{H}=\otimes_{t=1}^N h=h^{\otimes N}$, such that 
\begin{equation}
\mathcal{H}=\text{span}\{|i_0i_1\dots i_{N-1}\rangle\}\,,  \end{equation}
having dimension $d^{N}$. Our aim is now to relate this construction to quantum evolution by identifying the Hilbert space label $t$ with time.
In order to do so, it is convenient to 
recall that in quantum computation is quite common to employ tensor copies of a Hilbert space to describe quantities of the original Hilbert space \cite{nie.01,buhrman2001quantum,bultrini2023unifying,holmes2023nonlinear,diaz2023showcasing}. The most common example is the SWAP test \cite{buhrman2001quantum} depicted in Figure 1 and which can be derived from the basic property
\begin{align}\label{eq:swap}
{\rm Tr}\big[\text{SWAP} (A\otimes B)\big]={\rm tr}[BA]\,,\end{align}
which follows from its  definition ${\rm SWAP}|ij\rangle=|ji\rangle$ 
and the linearity of the trace. 
Here $BA$ is the conventional matrix multiplication, i.e., $\langle i|BA|j\rangle=\sum_k \langle i|B|k\rangle\langle k|A|i\rangle$, while ${\rm Tr}$ is the trace in ${\cal H}$ and ${\rm tr}$ that  in $h$.  
Interestingly, equation \eqref{eq:swap} relates traces in $h\otimes h$ with traces in $h$, which can then be related to physical quantities.

\begin{figure}[t!]
    \centering
\includegraphics[width=\columnwidth]{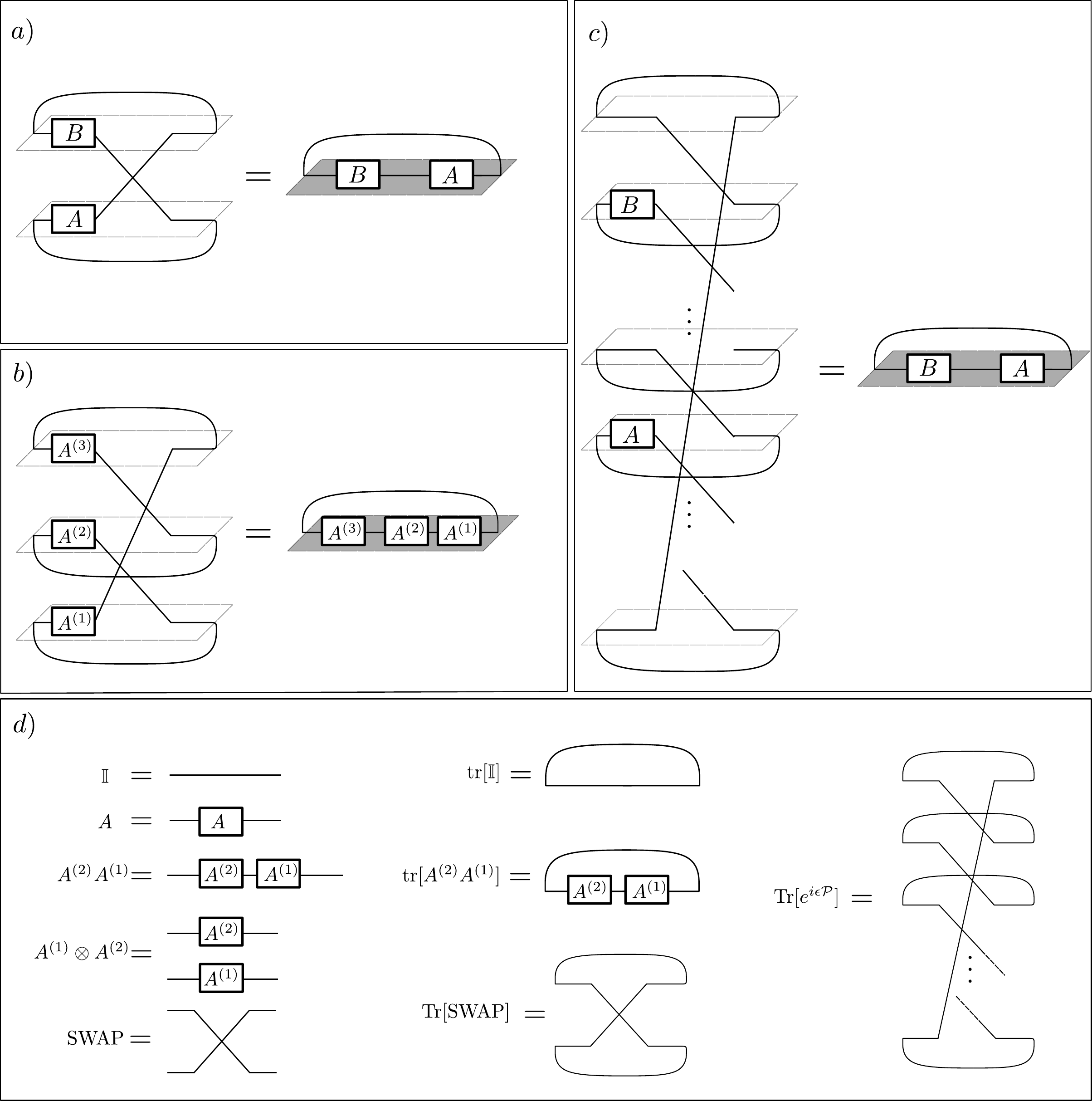}
    \caption{\textbf{Tensor network representation of the map between spacetime traces and conventional traces. } The operator $e^{i\epsilon\mathcal{P}}$ allows one to translate traces in $\mathcal{H}=\otimes_t h_t$ to traces in $h$, as easily seen in tensor network notation. The notation is introduced in d) while the planes in a), b) and c) have been added to emphasize that a Hilbert space is assigned to each time slice.  The panel a)  corresponds to $N=2$ and Eq.\ \eqref{eq:swap} while b) is the case $N=3$ of Lemma \ref{lemma1}. In c) we show the representation of the trace of just two operators with an arbitrarily larger number of time-slices. This corresponds to Lemma \ref{lemma1} with identities places in all but two slices. }
    \label{fig:tn}
\end{figure}

One can easily generalize this expression by defining a time translation operator $e^{i\epsilon\mathcal{P}}$ such that
\begin{equation}\label{eq:eip}
e^{i\epsilon\mathcal{P}}|i_0i_1\dots i_{N-1}\rangle:= |i_{N-1}i_0i_1\dots i_{N-2}\rangle\,.
\end{equation}
Notice that $e^{i\epsilon\mathcal{P}}$ is unitary and that one can recast it as $e^{i\epsilon\mathcal{P}}=\swap_{10}\dots\swap_{N-2,N-3}\swap_{N-1,N-2}$,  with $e^{i\epsilon\mathcal{P}}=\text{SWAP}$ for $N=2$. Here $\epsilon$ is a scale indicating that the generator $\mathcal{P}$ is going to translate a single step (see also below). 
Then, one can prove the following Lemma \cite{DIAZ2025170052} which admits a clear diagrammatic representation shown in Figure \ref{fig:tn}.

\begin{lemma}\label{lemma1}
Consider the time translation operator $e^{i\epsilon\mathcal{P}}$ and general operators $A^{(j)}$. Then the following relation between traces on $\mathcal{H}$ and $h$ holds: 
    \begin{align}\label{eq:treip}
{\rm Tr}\big[ e^{i\epsilon\mathcal{P}} \otimes_{t=0}^{N-1} A^{(t)}\big]={\rm tr}\big[\,\hat{T} \,\smallprod\nolimits_{t=0}^{N-1}A^{(t)}\big]\,.
\end{align}
The product on the r.h.s.\ follows a temporal order, namely,  $\hat{T} \,\Pi_{t=0}^{N-1}A^{(t)}=A^{(N-1)} A^{(N-2)} \dots A^{(1)} A^{(0)}$.
\end{lemma}

Notice that the product of operators on the r.h.s. corresponds to the conventional composition of operators in $h$ (matrix multiplication) and that one might choose $A^{(i)}=\mathbbm{1}$ if one is interested in a product of $n<N$ operators. In other words, the identity holds if one replaces $\otimes_t A^{(t)}\to \otimes_{l} A^{(t_l)}_{t_l}$, for $A^{(t_l)}_{t_l}$ the operator $A^{(t_l)}$ acting on the $t_l$-slice, and $\Pi_t A^{(t)}\to \Pi_lA^{(t_l)}$ with now the (tensor) products running over a subset of the $N$ slices (see also panel c) of Figure \ref{fig:tn}). For clarity, let us remark that in our notation $A_{t}$ acting on $\mathcal{H}$ corresponds to a standard operator acting on the $t+1$ copy of $h$, times identities in the other copies, e.g. for $N=2$ one has $A_0=A \otimes \mathbbm{1}$, $A_1= \mathbbm{1} \otimes A$.

In order to introduce evolution one can define a \emph{quantum action} operator $\mathcal{S}$ as follows: 
\begin{equation}
e^{i\mathcal{S}}:=e^{i\epsilon\mathcal{P}}\otimes_{t=0}^{N-1} e^{-i\epsilon H_t}\,,
\end{equation}
with $H$ a time-independent Hamiltonian and $H_t$ indicating $H$ acting on the copy $t$ of $h$. Here $\epsilon$ is a time spacing so that $T=\epsilon N$ corresponds to the total length of time we are considering. The following important result is thus obtained:
\begin{lemma}\label{th:lemmabosonmap}
Consider the quantum action operator $\mathcal{S}$ for a given Hamiltonian $H$ and general operators $A^{(j)}$. Then
\begin{align}\label{eq:eistr}
{\rm Tr}\big[e^{i\mathcal{S}}
\otimes_{l}A^{(t_l)}_{t_l}\big]={\rm tr}\big[e^{-  iT H}\,\hat{T} \,\smallprod\nolimits_{\,l} A^{(t_l)}(\epsilon t_l)\big]\,,\end{align}
 where $A^{(l)}_{t_l}$ acts on the slice $t_l$, $A(t):=e^{i H t}A e^{-i Ht}$is  the operator $A$ in the Heisenberg picture  and  $\hat{T}$ denotes  the conventional time ordering operator (decreasing order from left to right). 
\end{lemma}

This relation, proven in 
\cite{diaz.21,DIAZ2025170052} by using Lemma 1 and a few properties of the action operator, shows that it is indeed very natural to identify the index $t$ with time. In fact, on the r.h.s. one recognizes the conventional propagators arising e.g. in perturbation theory and in the connections between the PI formulation and canonical QM. For example, for $A^{(0)}=|\psi\rangle\langle \psi|$ and two additional operators $A, B$ one obtains 
\begin{equation}\label{eq:twopointpropagator}
{\rm Tr}\big[|\psi\rangle_0\langle \psi| e^{i\mathcal{S}} A_{t_1}B_{t_2}\big]=\langle \psi,T|\hat{T} A(\epsilon t_1)B(\epsilon t_2) |\psi\rangle 
\end{equation}
with $|\psi,T\rangle\equiv e^{iHT}|\psi\rangle$.
The reason to denote $\mathcal{S}$ quantum action becomes apparent when dealing with systems with a classical analog, in which case the continuum limit of $\mathcal{S}$ takes the form of the classical action in phase-space (see also Eq.\ \eqref{eq:legendre} below). Let us also mention that one can easily define a quantum action for time-dependent Hamiltonians (see  
\cite{diaz.21,DIAZ2025170052} and the fermionic case). We also remark that this result does not use the hermiticity of $H$, implying in particular  that for a given hermitian Hamiltonian one can freely multiply it by complex quantities and then relate the previous to thermal correlators.

Finally, let us notice that if we define ${R_\psi:=|\psi\rangle_0\langle \psi,-T|e^{i\mathcal{S}}}$, and noticing that  $e^{i\mathcal{S}}\to (e^{i\mathcal{S}})^\dag$ leads to inverse time ordering, we can write
\begin{equation}\label{eq:heiscomm}
    {\rm Tr}\big[(R_\psi-R_{\psi}^\dag) A_{t_1}B_{t_2}\big]=\langle \psi|[A(\epsilon t_1), B(\epsilon t_2)] |\psi\rangle 
\end{equation}
where we have assumed $t_1\geq t_2$ (an overall sign needs to be included otherwise). Notably, while on the l.h.s. the operators $A_{t_1}$ and $B_{t_2}$ commute for $t_1\neq t_2$, the r.h.s. of Eq.\ \eqref{eq:heiscomm} contains a standard unequal time commutator between Heisenberg operators.
The role of the operator $R$ will be further discussed in sections \ref{sec:states} and \ref{sec:entanglementintime}.  Notice also that  considering arbitrary matrix elements inserted at the initial time slice leads to an operator relation: ${\rm Tr}_{t\neq 0}\big[(e^{iH_0T}e^{i\mathcal{S}}-e^{-i\mathcal{S}}e^{-iH_0T}) A_{t_1}B_{t_2}\big]=[A(\epsilon t_1), B(\epsilon t_2)]$. Thus, even if operators at different time slices per se are independent, once the quantum action and its adjoint are employed, standard Heisenberg commutators are recovered as ``static'' correlators.

\subsection{The spacetime algebras approach}\label{sec:prelim2}

Let us now comment on how the previous construction defines a spacetime algebra. For simplicity, let us consider first a system composed of $M$ qubits (e.g. a 1D spin chain) so that $$h=\otimes_{x=1}^M h_q=h_q^{\otimes M}=\text{span}\{|i_1 i_2\dots i_M\rangle\}$$ with $i=0,1$. It is interesting to notice the spacelike analogy with our timelike construction in $\mathcal{H}$. Equivalently, we can think of $h$ as the $M$-fold tensor product representation space of the fundamental representation of the group $\mathbb{SU}(2)$ with algebra
\begin{equation}
    \Big[\frac{\sigma_{ix}}{2},\frac{\sigma_{jy}}{2}\Big]=i\epsilon_{ijk}\frac{\sigma_{kx}}{2}\delta_{xy}\,,
\end{equation}
with $\sigma_{ix}=X_{x},Y_{x},Z_{x}$ ($i=1,2,3$) the Pauli matrices acting on the site $x$ (e.g. $\sigma_{32}=Z_{2}\equiv \mathbbm{1}\otimes Z \otimes \mathbbm{1}\dots \otimes \mathbbm{1}$ with $\mathbbm{1}$ the identity matrix in two dimensions).

Consider now the same system and its  description in $\mathcal{H}$. We can write $\mathcal{H}=\otimes_{t,x}h_q=h_q^{\otimes MN}$ showing the symmetry between space and time. Moreover, we now have operators $\sigma_{itx}$, with $t$ the index on the Hilbert space on which the operator acts non-trivially,
satisfying
\begin{equation}\label{eq:qubitsstalg}
    \Big[\frac{\sigma_{itx}}{2},\frac{\sigma_{jt'y}}{2}\Big]=i\epsilon_{ijk}\frac{\sigma_{ktx}}{2}\delta_{xy}\delta_{tt'}\,.
\end{equation}
We refer to this algebra as a \emph{spacetime algebra} for obvious reasons.

The previous idea of extending a given algebra to encompass time as a site index can be applied to any bosonic-like system. As long as commutators are employed (and natural Hilbert space representations), this can be regarded as equivalent to applying a tensor product structure to time. On the other hand, in the treatment for fermionic systems we aim to develop this is no longer the case. For this reason, it is pertinent to consider the case of bosonic particles from the algebraic perspective (as opposed to the equivalent point of view of the previous Lemmas). According to the previous discussion, given a canonical algebra defining creation and annihilation operators $b_i, b_j^\dag$ we extend it as
\begin{equation}\label{eq:bosonicalg}
    [b_i,b_j^\dag]=\delta_{ij}\to [b_{ti},b_{t'j}^\dag]=\delta_{tt'}\delta_{ij}\,,
\end{equation}
with other commutators vanishing. Equivalently,  one can impose $[q_i,p_j]=i\delta_{ij}$ (we take $\hbar\equiv 1$) instead, with $q_i,p_j$ position and momentum operators and define extended position and momentum operators satisfying $[q_{ti},p_{t'j}]=i\delta_{tt'}\delta_{ij}$. Once again, if $i$ is a spatial index, such as in a Quantum Field Theory (QFT) in $d$ dimensions, we have an algebra which is symmetric in spacetime and isomorphic to the algebra of a QFT in $d+1$ dimensions. One can show that the evaluation of traces of the form ${\rm Tr} \big[ e^{i \mathcal{S}}\cdots \big]$ in the eigenbasis of $q_{ti}$ leads to the sum over histories of the PI formulation \cite{DIAZ2025170052}.

It is now interesting to provide an algebraic perspective on the operators involved in the previous Lemmas. Let us first introduce a useful definition: we define Fourier modes in time as 
\begin{equation}\label{eq:ft}
    b_{jn}:=\frac{1}{\sqrt{N}}\sum_{t=0}^{N-1} e^{i \epsilon \omega_n t }b_{jt}\,,
\end{equation}
with $\omega_n=2\pi n/T$,   which is a unitary transformation, namely $[b_{in},b^\dag_{jn'}]=\delta_{ij}\delta_{nn'}$.
Let us remark that this definition is possible only in the extended formalism as the new modes create nonlocal excitations in time. Interestingly, this elementary example shows that we can apply techniques typically associated with ``spacelike'' properties in conventional condensed matter/QFT treatments to time.  
We can now write an explicit representation of the generator of time translations:
\begin{equation}
    \mathcal{P}:=\sum_{n=0}^{N-1} \sum_j \omega_n b^\dag_{jn}b_{jn}\,.
\end{equation}
In fact, one can easily verify that 
\begin{equation}
    e^{i\epsilon\mathcal{P}}b_{jt}e^{-i\epsilon\mathcal{P}}=b_{j,t+1}
\end{equation}
 with periodic boundary conditions which is equivalent to the definition \eqref{eq:eip}. One can also show that
 \begin{equation}\label{eq:legendre}
     \mathcal{P}\approx \epsilon \sum_t p_t (q_{t+1}-q_{t-1})/\epsilon
 \end{equation}
  for small $\epsilon$, which has the form of the Legendre transform of the classical action (in the case of spins, a similar result holds for the mean values of $\mathcal{S}$ evaluated along coherent-spin states). At the same time, for $N=2$, and restricting ourselves to a subspace spanned by Fock states $\{|00\rangle, |01\rangle, |10\rangle , |11\rangle\}$ 
one finds the matrix representation
\begin{equation}e^{i \epsilon \mathcal{P}}= \begin{pmatrix}
    1 &0 &0 &0 \\
    0 & 0 & 1 &0 \\
    0 & 1 &0 &0 \\
    0 &0&0&1
\end{pmatrix}\oplus \dots
\,,\end{equation}
 with $\oplus \dots$ indicating the action of the operator outside of this subspace (on states like $|20\rangle, |12\rangle, |30\rangle,\ldots$ for example), with previous $4\times 4$  block 
 coinciding with the two-qubit SWAP.

As an interesting example on how the previous Lemmas emerge from the bosonic algebra, 
consider a free quadratic Hamiltonian of the form $H=\sum_{i,j}b^\dag_i M_{ij} b_j$. Let us recall first the expression for  conventional two-point spacelike correlators:
\begin{equation}\label{eq:contr}
\begin{split}
    \langle b_k b^\dag_l\rangle:&=\frac{{\rm tr}\big[\exp(-\sum_{i,j} b^\dag_i M_{ij}  b_j)b_k b^\dag_l\big]}{{\rm tr}\big[\exp(-\sum_{i,j} b^\dag_i M_{ij}  b_j)\big]}\\&=\left[\frac{1}{\mathbbm{1}-\exp(-M)}\right]_{kl}\,, 
\end{split}
\end{equation}
which has the form of the Bose-Einstein statistics for $\langle \dots \rangle\equiv {\rm tr}[\rho \dots]$ with $\rho\propto e^{-H}$ the thermal state, and with ${\rm tr}\big[\exp(-\sum_{i,j} b^\dag_i M_{ij}  b_j)\big]=\det\{1/(1-\exp(-M))\}$. Notice that these relations are a consequence of the canonical algebra and hold for $H\to \lambda H$ with $\lambda \in \mathbbm{C}$ by replacing $M\to \lambda M$. We assumed here grand-canonical traces and ${\rm Re}(\lambda M)>0$.

The corresponding quantum action operator acting on $\mathcal{H}$ is
\begin{equation}
    \mathcal{S}:=\epsilon\mathcal{P}-\sum_t \epsilon H_t=\sum_n\sum_{i,j} \epsilon(\omega_n \delta_{ij}-M_{ij})b_{ni}^\dag b_{nj}\,,
\end{equation}
with $\sum_t H_t=\sum_{t,i,j}M_{ij}b^\dag_{ti}b_{tj}$ and where in the last equality we used that $\sum_{t}b^\dag_{ti}b_{tj}=\sum_{n}b^\dag_{ni}b_{nj}$. Interestingly, in $\mathcal{H}$ one can define correlators corresponding to operators with arbitrary time-site positions. At the same time, these can be computed in complete analogy with \eqref{eq:contr} but applied in the extended Hilbert space as there is no algebraic difference between space and time in $\mathcal{H}$. 
To provide a concrete example, let us also consider the action after a Wick rotation. We can write 
\begin{equation}
\begin{split}
    \langle b_{kt_1}b^\dag_{lt_2}\rangle&:=\frac{{\rm Tr}\big[\exp(-\mathcal{S}_E)b_{kt_1} b^\dag_{lt_2}\big]}{{\rm Tr}\big[\!\exp(i\mathcal{S})\big]}
   \\&=\sum_n \left[\frac{e^{-i\omega_n (t_1-t_2)}}{\mathbbm{1}-\exp[i\epsilon(\omega_n \mathbbm{1}+iM)]}\right]_{kl}
   \,,
\end{split}
\end{equation}
where we have defined an ``Euclidean'' quantum action \begin{equation}\mathcal{S}_E:=-i\epsilon\mathcal{P}+\epsilon \sum_t H_t\label{eq:Seuclid}\end{equation} so that $\exp(-\mathcal{S}_E)=\exp(i \mathcal{S})$ when $\mathcal{S}$ has undergone the replacement $H\to -iH$. 
In order to obtain the sum in $n$ we used 
that $\mathcal{S}$ is quadratic so that the contraction can be evaluated via \eqref{eq:contr} with the action replacing the Hamiltonian. The additional sum in the index $n$ comes from inverting \eqref{eq:ft}. Notice that for $\epsilon\ll 1$ the expression $\langle b_{kt_1}b^\dag_{lt_2}\rangle\approx \sum_n ie^{-i\omega_n (t_1-t_2)}\left[1/(\omega_n\mathbbm{1}+iM)\right]_{kl}$ has the form of the thermal propagator obtained via coherent state path integrals (Matsubara expansion). One can actually prove the exact result \cite{DIAZ2025170052}:
\begin{equation}\label{eq:bosontwop}
    \langle b_{kt_1}b^\dag_{lt_2}\rangle=
    \langle \hat{T}_\beta b_k (-i\epsilon t_1)b^\dag_l (-i\epsilon t_2)\rangle_\beta
\end{equation}
for $\langle \dots \rangle_\beta:={\rm tr}[e^{-\beta H}\dots ]/{\rm tr}[e^{-\beta H}]$ denoting a thermal expectation value, 
and where the operators on the r.h.s. are in Heisenberg picture with imaginary time i.e.  $b_k(-it)=e^{tH}b_k e^{-tH}$, $b^\dag_l(-it)=e^{tH}b^\dag_l e^{-tH}$, with $\hat{T}_\beta$ denoting the thermal order operator and where the inverse temperature $\beta=T=\epsilon N$. It is worth emphasizing that while the ladder operators on the l.h.s. are independent for different times, according with Eq.\ \eqref{eq:bosonicalg}, the operators on the r.h.s. are causally connected in the usual sense. In particular, one might recover the standard unequal time commutation relations, as we did in Eq. \eqref{eq:heiscomm} (see also comments below this equation).

Moreover, one can show the following relation between the spacetime trace ${\rm Tr}$ of $e^{-S_E}$ and the conventional partition function,  
\begin{equation}
    {\rm Tr}[e^{-\mathcal{S}_E}]={\rm tr}[e^{-\beta H}]\,,
\end{equation}
where we recall that ${\rm tr}$ is the standard trace. 
Let us make a few comments about these results. 
First of all, notice that while on the r.h.s. of 
\eqref{eq:bosontwop} one needs to evolve the operators, 
the l.h.s. correspond to a single contraction of the form of \eqref{eq:contr} but defined in the extended Hilbert space $\mathcal{H}$.
Secondly, notice that \eqref{eq:bosontwop} is in agreement with Lemma \ref{th:lemmabosonmap} as it corresponds to the case of two operators and a Wick rotated  Hamiltonian. 
Finally, one can in principle extend \eqref{eq:bosontwop} to higher-order contraction by applying Wick's theorem (for gaussian operators) on the extended Hilbert space. One can show that this agrees with the time-dependent Wick's theorem in conventional QM. Thus one can build arbitrary operators from a basic quadratic action, thus recovering Lemma \ref{th:lemmabosonmap} completely via algebraic means, i.e. without explicit reference to the underlying product structure in time. Strictly speaking, this argument is only valid up to all order of  perturbation theory and should be regarded as a consistency check on the previous Lemmas (proven by other means in \cite{diaz2024spacetime,DIAZ2025170052}) and whose rigor might depend on convergence subtleties proper of infinite dimensional systems.

\section{Fermionic spacetime formalism}\label{sec:fermioniccase}

\subsection{Basic definitions and considerations}\label{sec:fermionsdiscrete}

We now begin the development of the formalism for fermions in spacetime. We want to describe, via a spacetime symmetric formalism, fermions conventionally defined by the anticommutation relations  $ \{a_{i},a^\dag_{j}\}=\delta_{ij},
    \{a_i,a_j\}=\{a_i^\dag,a_j^\dag\}=0$, inducing the standard Hilbert space $h=\text{span}\{\prod_i (a_i^\dag)^{n_i}|0\rangle\}$ with $a_i|0\rangle=0$. In particular, if $i,j$ take $L$ values, the dimension of the conventional Hilbert space is $\text{dim}(h)=2^L$. 
For this purpose, we introduce the spacetime fermionic algebra
\begin{subequations}\label{eq:stalg}
\begin{align}
    \{a_{ti},a^\dag_{t'j}\}&=\delta_{tt'}\delta_{ij}\label{eq:stalg1}\\
    \{a_{ti},a_{t'j}\}&=\{a^\dag_{ti},a^\dag_{t'j}\}=0\,.\label{eq:stalg2}
\end{align}
\end{subequations}
It is well-known that a state $|\Omega\rangle$ satisfying $a_{ti}|\Omega\rangle=0$ exists, so that a Hilbert space $\mathcal{H}=\text{span}\{\prod_{t,i}(a_{ti}^\dag)^{n_{ti}} |\Omega\rangle\}$ is obtained. We also assume a finite time window of length $T=\epsilon N$ with $N$ the number of time-slices ($t=0,\dots, N-1$) and $\epsilon$ the time spacing. 
 As a consequence, the dimension of the extended Hilbert space is $\text{dim}(\mathcal{H})=2^{LN}$. Thus, $\mathcal{H}$ as a linear space is isomorphic to $h^{\otimes N}$ just as in the bosonic case. However, the extended operators $a_{ti}$ cannot be obtained from the operators $a_i$ via Kronecker product with identity matrices of $h$ for slices other than $t$. This is a consequence of the intrinsic nonlocal features of a Hilbert space representing fermions (these are evident e.g. when one maps fermions to qubits, such as in the Jordan-Wigner transformation \cite{JW0.28,LSM.61}). Let us also notice that while it is in principle feasible to propose ladder operators that anticommute for equal-times but commute at different times (equivalent to writing e.g. $a'_{1i}:=\mathbbm{1} \otimes a_i$ for $N=2$ with $\{a'_{0i},a'^{\dag}_{1j}\}\neq 0$), this breaks the spacetime symmetry we aim for.

Notably, and despite the fundamental differences with the bosonic case, we can easily obtain a map from spacetime objects to conventional fermionic quantities involving evolution. In order to do so let us first define the Fourier Transform (FT) in time as follows:
\begin{equation}\label{eq:fta}
    a_{ni}:=\frac{1}{\sqrt{N}}\sum_{t=0}^{N-1} e^{i\omega_n \epsilon t}a_{ti}\,,
\end{equation}
with $\omega_n=(2n+1)\pi/T$,  which is a unitary transformation, namely $\{a_{in},a^\dag_{jn'}\}=\delta_{ij}\delta_{nn'}$. 
Now, we can introduce the hermitian operator
\begin{equation}\label{eq:pdef}
    \mathcal{P}:=\sum_{n=0}^{N-1} \sum_j \omega_n a^\dag_{jn}a_{jn}\,,
\end{equation}
which is the \emph{generator of time translations} for fermions. 
In fact, one can easily verify that 
\begin{equation}
    e^{i\epsilon\mathcal{P}}a_{ti}e^{-i\epsilon\mathcal{P}}=a_{t+1,i}
\end{equation}
 with antiperiodic periodic boundary conditions, i.e. $ e^{i\epsilon\mathcal{P}}a_{N-1,i}e^{-i\epsilon\mathcal{P}}=-a_{0,i}\equiv a_{N,i}$. The translations in time are of a geometrical nature, and unrelated to Hamiltonian evolution so far.

It is interesting to compare this definition with the case of qubits. As an elementary but illustrative example consider a single spinless fermion and $N=2$ such that $h$ is isomporphic to the space of a single qubit and $\mathcal{H}$ to $h\otimes h$. For the qubit we know that the time-translation operator is the SWAP. Instead, the natural matrix representation for the fermion in spacetime for which $|\Omega\rangle \equiv (1,0,0,0)^t$ leads to
\begin{equation}\label{eq:fswap}
    e^{i \epsilon \mathcal{P}}=\begin{pmatrix}
    1 &0 &0 &0 \\
    0 & 0 & 1 &0 \\
    0 & -1 &0 &0 \\
    0 &0&0&1
\end{pmatrix}\,,
\end{equation}
and $Pe^{i\epsilon \mathcal{P}}=(e^{i\epsilon \mathcal{P}})^t$. Notice that there is a minus sign which is not present in the matrix representation of the standard SWAP. This subtle difference is important, as it is well-known that gaussian operators together with arbitrary SWAP operations allows for universal (fermionic) unitary transformations (this result can be found in the context of matchgate circuits \cite{jozsa2008matchgates}, i.e. quantum circuits that can be mapped to free fermions). While SWAP can be written as the exponential of a two-body operator (an operator involving the product of up to $4$ creation (annihilation) operators) but not as that of a one-body operator, $e^{i\epsilon \mathcal{P}}$ is clearly gaussian by definition \footnote{It is clear that if we think of $e^{i\epsilon \mathcal{P}}$ as a two-qubit operation, it is a Matchgate with the convention of \cite{jozsa2008matchgates,hebenstreit2019all, diaz2023showcasing}: $e^{i\epsilon \mathcal{P}}=\exp\{i\frac{\pi}{4}(\sigma_2 \otimes \sigma_1-\sigma_1\otimes \sigma_2)\}$ with $\sigma_2 \otimes \sigma_1, \sigma_1\otimes \sigma_2$ both a product of two Majoranas \cite{diaz2023showcasing}. In this sense, $e^{i\epsilon \mathcal{P}}$ is a variation of the fSWAP gate \cite{hebenstreit2019all}. }, an important fact that we exploit below.

Let us pause here to emphasize the differences with the standard single-time picture. To be explicit consider the previous example:
A single fermionic mode has $\dim h=2$, whereas $\dim \mathcal{H}=4$ reflecting the different algebraic structure at play: unequal times (anti) commutation relations in the standard Heisenberg picture are non-trivial because operators at different times are not independent (namely, they are related by the evolution operator), leading to the spacetime asymmetry of Figure \ref{fig:intro}. Instead Eq.\ \eqref{eq:stalg} leads to an enlarged Hilbert space precisely because operators at different times are completely independent at the Hilbert space level. In the following, we develop the necessary results to compatibilize these two different approaches.

 \subsection{Map to conventional fermionic evolution and the  fermionic quantum action}\label{sec:fermionsdiscretemap}
 
Having provided the basic definitions we are now in a position to establish a map between our extended formalism and conventional fermions under unitary evolution.

 Before introducing a Hamiltonian and a fermionic quantum action it is convenient to establish a few fundamental results involving just $\mathcal{P}$. We begin with the case of two-point contractions. 

 \begin{theorem}\label{th:mapeip}
Given ${\rm Tr}$ the trace in ${\cal H}$ 
 and given ${\rm tr}$ that in $h$, the following equalities hold:
     \begin{align}
    {\rm Tr}[Pe^{i\epsilon\mathcal{P} }]&={\rm tr}[\mathbbm{1}]\label{eq:Pmap}\\
    {\rm Tr}[Pe^{i\epsilon\mathcal{P} }a_{t_1i}a^\dag_{t_2j}]&= {\rm tr}[\hat{T}a_i(t_1) a^\dag_j(t_2)]:=\nonumber\\&=\begin{cases}{\rm tr}[a_i a^\dag_j]\,\;\;t_1\geq t_2\\
    -{\rm tr}[a^\dag_ja_i]\,\;\;t_1<t_2 \label{eq:Pmap2}
    \end{cases}\,,    
    \end{align}
 for $P=e^{i\pi\sum_{ti}a^\dag_{ti}a_{ti}}$ the parity operator and where we have introduced the time-order operator $\hat{T}$ defined with respect to $t_1,t_2$ on the l.h.s. 
 The other two-point contractions vanish, i.e.
${\rm Tr}[Pe^{i\epsilon\mathcal{P} }a_{t_1i}a_{t_2j}]= {\rm tr}[\hat{T}a_i(t_1) a(t_2)]=0$ and 
      ${\rm Tr}[Pe^{i\epsilon\mathcal{P} }a^\dag_{t_1i}a^\dag_{t_2j}]= {\rm tr}[\hat{T}a^\dag_i(t_1) a^\dag(t_2)]=0$.
 \end{theorem}
In the Appendix \ref{app:theorems} we provide a proof and we show that $
   {\rm Tr}[Pe^{i\epsilon\mathcal{P} }a_{t_1i}a^\dag_{t_2j}]=\delta_{ij}{\rm tr}[a_i a_j^\dag]\sgn(t_1-t_2)$, which agrees with our definition of $\hat{T}$. 
 Let us notice that our notation is consistent with regarding 
 $a_i(t)$ as a Heisenberg operator in the limit of a null Hamiltonian (see Theorem \ref{th:correlations} below), with $\hat{T}$ the usual time-ordering operator. 
The proof of the Theorem, given in the Appendix \ref{app:theorems}, relies solely on the elementary properties of quadratic fermionic operators, namely on ${\rm tr}[\exp(\sum_k \lambda_k a_k^\dag a_k)]=\prod_k (1+e^{\lambda_k})$ and on the ensuing two-point contraction 
\begin{equation}
    \frac{{\rm tr}[\exp(\sum_{k'} \lambda_{k'} a_{k'}^\dag a_{k'}) a_k a^\dag_l]}{{\rm tr}[\exp(\sum_k \lambda_{k'} a_{k'}^\dag a_{k'})]}=\frac{\delta_{kl}}{1+e^{\lambda_k}}\,,
\end{equation}
 as in the Fermi-Dirac statistics. Here $k,l$ are just arbitrary fermionic indices. To apply this in the current case one can take $k\equiv (t,i)$ or $k\equiv (n,i)$ and use that $Pe^{i\epsilon\mathcal{P}}=\exp\{i\sum_{n,i} [\epsilon \omega_n+\pi]a^\dag_{ni}a_{ni}\}$, with ${\rm tr}\rightarrow {\rm Tr}$. 
 In the Appendix \ref{app:theorems}  we show how this expression leads to the previous Theorem and in Appendix \ref{app:PIs} we discuss its relation with the Matsubara expansions.
 Since our traces only involve quadratic operators, higher-order correlators are automatically obtained through Wick's theorem, leading to the following result.
\begin{theorem}\label{th:mapwick}
Consider operators defined ``on a given time slice'', meaning  $O_{t}\equiv O[a_{ti},a^\dag_{tj}]$. Then
   \begin{equation}
     {\rm Tr}[P e^{i\epsilon \mathcal{P}}\smallprod_{l}O^{(l)}_{t_l}]={\rm tr}[\hat{T}\smallprod_{l}O^{(l)}(t_l)]
   \end{equation}
   where the time ordering operator $\hat{T}$ is defined as usual for fermions (when interchanging two creation (annihilation) operators, an additional minus sign must be added).
\end{theorem}

We remark that the notation $O_{t}\equiv O[a_{ti},a^\dag_{tj}]$ indicates an operator $O$ that is only occupying the mode $t$. A full demonstration is provided in the Appendix \ref{app:theorems}. The essential point is that $P e^{i\epsilon \mathcal{P}}$  is the exponent of a one-body operator and hence gaussian so that Wick's theorem holds, with two-point contractions determined by Theorem \ref{th:mapeip}. Let us provide a simple example that illustrates this: if we introduce the notation $\langle \dots \rangle:={\rm Tr} [P e^{i \mathcal{\epsilon}\mathcal{P}}\dots ]/{\rm Tr} [P e^{i \epsilon\mathcal{P}}]$ and $\langle \dots \rangle_\mathbbm{1}:={\rm tr} [\mathbbm{1} \dots ]/{\rm tr} [\mathbbm{1}]$ we can write
\begin{widetext}\begin{equation}
\begin{split}
\langle a_{t_1i_1}a_{t_2i_2}a^\dag_{t_3i_3}a^\dag_{t_4i_4}\rangle&=\langle a_{t_1 i_1}a^\dag_{t_4 i_4}\rangle \langle a_{t_2 i_2}a^\dag_{ t_3 i_3}\rangle-\langle a_{t_1 i_1}a^\dag_{t_3 i_3}\rangle \langle a_{t_2 i_2}a^\dag_{t_4 i_4}\rangle\\
&=\langle \hat{T}a_{i_1}(t_1)a^\dag_{i_4}(t_4)\rangle_{\mathbbm{1}}\langle \hat{T}a_{i_2}(t_2)a^\dag_{i_3}(t_3)\rangle_{\mathbbm{1}}-\langle \hat{T}a_{i_1}(t_1)a^\dag_{i_3}(t_3)\rangle_{\mathbbm{1}}\langle \hat{T}a_{i_2}(t_2)a^\dag_{i_4}(t_4)\rangle_{\mathbbm{1}}\\
&=\langle \hat{T}a_{i_1}(t_1)a_{i_2}(t_2)a^\dag_{i_3}(t_3)a^\dag_{i_4}(t_4)\rangle_{\mathbbm{1}}\,,\label{eq:4pointcontr}\end{split}\end{equation}
\end{widetext}
where the first identity follows from Wick's theorem (using the nonvanishing contractions),  the second identity from Theorem \ref{th:mapeip}, and the last identity  follows in principle from the standard thermal Wick's theorem \cite{1960NucPh, evans1996wick}, but is also apparent if one time-orders the l.h.s. 
(see Appendix \ref{app:theorems}). The identity \eqref{eq:4pointcontr} agrees with Theorem \ref{th:mapwick} since ${\rm Tr}[P e^{i\epsilon \mathcal{P}}]={\rm{tr}}[\mathbbm{1}]$. 
In each two-point contraction, the temporal order follows the definition in Theorem 1.  
We recall that here one can think of $\hat{T}$ as the usual time-ordering operator in the limit of a vanishing Hamiltonian (as clear from Theorem \ref{th:correlations} below). It is also interesting to remark that in the standard thermal Wick's theorem, the time ordering operator is essentially behaving as a gaussian-like operator. This behaviour can now be understood as a consequence of its explicit representation in the extended Hilbert space $\mathcal{H}$ as  {\it an actual gaussian operator}. These observations also hold for bosons.

We are now in a position to introduce the fermionic quantum actions so that the conventional unitary evolution also appears.

\begin{theorem}\label{th:eisandeip}
Given a time-independent number parity preserving Hamiltonian $H[a_i,a_j^\dag]$  we define the corresponding quantum action operator 
    \begin{equation}
e^{i\mathcal{S}}:=e^{i\epsilon\mathcal{P}}\smallprod_{t=0}^{N-1} e^{-i\epsilon H_t}
    \end{equation}
    with $H_t\equiv H[a_{ti},a^\dag_{tj}]$. Then, given $\mathcal{V}=\prod_{t=0}^{N-1} e^{i \epsilon t H_t}$ the following result holds
    \begin{equation}\label{eq:eisvp}
  e^{i\mathcal{S}}=e^{-i T H_0 }\mathcal{V}^\dag e^{i\epsilon\mathcal{P}}\mathcal{V}\,.
    \end{equation}
\end{theorem}
For clarity, let us recall here that $H_0\equiv H[a_{0i},a_{0j}^\dag]$ indicates that the Hamiltonian acts only on the first slice, not to be confused with a free Hamiltonian.
This important result shows how the time translation operator is related to the quantum action. One can show e.g. that $\mathcal{S}$ has the form of the classical action for Dirac fermions (in the continuum limit; see section \ref{sec:continuumcase}), hence the name. Let us make a few additional remarks on Theorem \ref{th:eisandeip}. First of all, as it is clear from the proof given in the Appendix \ref{app:theorems}, the Hamiltonian $H$ does not need to be hermitian, meaning that one can replace $H\to \lambda H$ for $\lambda\in \mathbbm{C}$. The only difference in the previous expressions is that $\mathcal{V}$ is no longer unitary so that one must replace $\mathcal{V}^\dag\to \mathcal{V}^{-1}$. Moreover, the time dependent case follows by replacing $\mathcal{V}\to \prod_t U^\dag_t(\epsilon t)$ and $e^{i\mathcal{S}}\to e^{i\epsilon \mathcal{P}}\prod_t U[\epsilon(t+1),\epsilon t]$ with $U(\epsilon t)=\hat{T}\exp[i\int_0^{\epsilon t} dt'H(t')]$ the conventional time evolution operator. 
Let us also remark that we only consider parity preserving evolutions, as it is proper for fermionic systems. This implies in particular that evolution operators on different time slices commute, so that the order of the products in time defining $\mathcal{V}$ and $\mathcal{S}$ is not important and one can write e.g. $e^{i\mathcal{S}}=e^{i\epsilon \mathcal{P}}e^{-i\epsilon \sum_t H_t}$.

The last piece of information one needs to establish the fermionic version of the map \ref{th:lemmabosonmap} is the following.
\begin{lemma}\label{lemmaevolv}
Fermionic parity preserving Hamiltonians $H$ satisfy
\begin{equation}\label{eq:commutingH}
    [H_t,H_{t'}]=0\,,
\end{equation}
implying 
    \begin{equation}\label{eq:heisproduct}
\mathcal{V}\smallprod_t O^{(t)}_t\mathcal{V}^\dag=\smallprod_t e^{i\epsilon tH_t }O^{(t)}_te^{-i\epsilon tH_t }\equiv \smallprod_t (O^{(t)}(\epsilon t))_{t}\,.
    \end{equation}
\end{lemma}
The first equation is a direct consequence of the fact that $H_t=H[a_{ti},a^\dag_{tj}]$ must necessarily contain an even number of creation (annihilation) operators for it to be an admissible fermionic Hamiltonian. Since ladder operators of different time-slices anti-commute, an even number commutes. As a consequence, the action of $\mathcal{V}$ on a product in time of operators is the product in time of the same operators in the Heisenberg picture. The amount of evolution of each operators matches its site index, namely if the operator acts on the slice $t$, it is evolved to $\epsilon t$. All of these considerations hold for $H$ non-hermitian and time-dependent evolution.

We are now in a position to state the main theorem of this section. 
\begin{theorem}\label{th:correlations}
Given a time-independent number parity preserving Hamiltonian $H$ and its corresponding quantum action $\mathcal{S}$ the following identity holds:
    \begin{equation}
        {\rm Tr}[P e^{i\mathcal{S}}\smallprod_l O^{(l)}_{t_l}]={\rm tr}[e^{-iT H}\hat{T}\smallprod_l O^{(l)}(\epsilon t_l)]\,.
    \end{equation}
\end{theorem}
Let us make a few remarks on this important result. First of all notice that in the limit of $H\to 0$ we recover Theorem \ref{th:mapwick}. In fact, Theorem \ref{th:correlations} is a direct consequence of the results of Theorem \ref{th:mapwick}, Theorem \ref{th:eisandeip} and of Lemma \ref{lemmaevolv}, with the operator $\mathcal{V}$, whose action relates $\mathcal{S}$ with $\mathcal{P}$, yielding the unitarily evolved operators. Clearly, one can also recover Theorem \ref{th:mapeip} by considering only two operators given by $O^{(1)}_{t_1}= a_{t_1 i}$, $O^{(2)}_{t_2}=a^\dag_{t_2 j}$ and $H=0$. Instead, by adding a density matrix in the initial slice such as $O^{(0)}_{t_0=0}=|\psi\rangle_0 \langle \psi|$ one obtains
\begin{equation}\label{eq:prop}
 \!\!\!{\rm Tr}[Pe^{i\mathcal{S}}   a_{t_1 i}a^\dag_{t_2 j} |\psi\rangle_0 \langle \psi| ]= \langle \psi,T| \hat{T}a_i(\epsilon t_1)a^\dag_j (\epsilon t_2)|\psi\rangle\,,
\end{equation}
namely a two-point contraction for the state $|\psi\rangle$  with $|\psi,T\rangle=e^{i T H }|\psi\rangle$, which is precisely what appears when considering propagators in the PI formulation. One can obtain arbitrary propagators by inserting more operators on the l.h.s. as in Eq.\ \eqref{eq:4pointcontr}.

It is also apparent that by considering operators all acting on the same slice one can obtain standard expectation values of  states (either pure or mixed) at a given time. This is the content of the following Corollary which can be compactly stated by introducing the notation \begin{equation}\label{eq:tildeaction}
    e^{i\tilde{\mathcal{S}}}:= e^{iTH_0}e^{i\mathcal{S}}=\mathcal{V}^\dag e^{i\epsilon\mathcal{P}}\mathcal{V}\,.
\end{equation}
Notice that
$\tilde{\mathcal{S}}$ is unitarily related to $\mathcal{P}$ (as it follows from Theorem \ref{th:eisandeip}) and in particular they share the same spectrum.
\begin{corollary}\label{cor:heis}
    The Heisenberg and Schrodinger pictures are recovered by considering operators acting at a given time and a proper insertion at the initial slice specifying the initial state: 
    \begin{equation}\label{eq:schrodinger}
    {\rm Tr}[Pe^{i\tilde{\mathcal{S}}}   O_{t} |\psi\rangle_0 \langle \psi| ]=\langle \psi | O(\epsilon t) |\psi\rangle=\langle \psi(\epsilon t)|O|\psi(\epsilon t)\rangle
\end{equation}
with $|\psi(\epsilon t)\rangle =e^{-i \epsilon t H}|\psi\rangle$ a pure state in the Schrodinger picture. 
\end{corollary}

Notice that we have replaced $\mathcal{S}\to \tilde{\mathcal{S}}$ so that the bra is not evolved  (otherwise we would be computing a propagator rather than a direct mean value; see \eqref{eq:eisvp} and \eqref{eq:prop}).

Let us  add some conceptual comments on our results. We started by defining a Hilbert space $\mathcal{H}$ through fermionic algebras that treat space and time equally. In particular, one is regarding fermions in different points in time as independent and hence causally disconnected. Then, in order to accommodate evolution, rather than relying on an external ``time'' parameter, we defined $e^{i \mathcal{S}}$ and considered its correlators in spacetime (feasible only through the extended construction). Remarkably, these correlators are precisely equal to time-ordered correlators where the operators involved are evolved according to the  unitary evolution of standard QM.  
In a very precise sense, we are replacing unitary evolution with correlations, a point we discuss further in section \ref{sec:states} where we introduce a notion of spacetime state.

Similarly, we can recover a conventional unequal time anti-commutator as follows
\begin{equation}\label{eq:fermioniccommheis}
\begin{split}
     &{\rm Tr}\,\left[P\, \big(|\psi\rangle_0\langle \psi|e^{i\tilde{\mathcal{S}}}-e^{-i\tilde{\mathcal{S}}}|\psi\rangle_0\langle \psi|\big) a_{t_1i}a^\dag_{t_2j}\right]\\&=\langle \psi|\{a_i(\epsilon t_1),a^\dag_j(\epsilon t_2)\}|\psi\rangle\,,
\end{split}
\end{equation}
as one can easily see by noting that $e^{-i\tilde{\mathcal{S}}}$ satisfies an anti-time ordered version of Theorem \ref{th:correlations}. Notice that a minus sign is involved in the fermionic time-ordering, thus leading to an anticommutator on the r.h.s. as opposed to the commutator we found for bosons in Eq.\ \eqref{eq:heiscomm}.
We have thus recovered the standard unequal time anti-commutation relations in the Heisenberg picture $\{a_i(\epsilon t_1),a^\dag_j(\epsilon t_2)\}$ from a formalism where Eq.\ \eqref{eq:stalg} is imposed instead.

Let us now remark that Theorem \ref{th:correlations} results can be easily applied to time-dependent and/or non hermitian Hamiltonians as explained in the Appendix \ref{app:Atheorem3}, and as stated below Theorem \ref{th:eisandeip}. In particular, under the replacement $H\to -i H$ one obtains
\begin{equation}
    {\rm Tr}[P e^{-\mathcal{S}_E}]={\rm tr}[e^{-\beta H}]
\end{equation}
 with $\beta \equiv T$, where we have introduced the Euclidean fermionic action
\begin{equation}
\mathcal{S}_E:=-i\epsilon\mathcal{P}+\epsilon \sum_t H    
\end{equation}
 so that $e^{-\mathcal{S}_E}=e^{i \mathcal{S}}$ after the replacement (consistent with the standard conventions for the Wick rotation). Similarly, if we introduce the notation $\langle \dots \rangle_{\mathcal{S}_E}:={\rm Tr} [P e^{-\mathcal{S}_E}\dots ]/{\rm Tr} [P e^{- \mathcal{S}_E}]$ and $\langle \dots \rangle_\beta:={\rm tr} [ e^{-\beta H}\dots ]/{\rm tr} [ e^{-\beta H}]$ we can write
\begin{equation}\label{eq:thermaltwop}
    \langle  a_{t_1 i}a^\dag_{t_2 j}\rangle_{\mathcal{S}_E}= \langle \hat{T}_\beta a_i (-i \epsilon t_1)a_j^\dag (-i \epsilon t_2)\rangle_\beta
\end{equation}
which is a two-point ordered thermal correlator. Arbitrary thermal correlators are clearly obtained by inserting more operators. 

It is interesting to explicitly compute the traces on the l.h.s. in the case of a quadratic Hamiltonian. Let us notice first that for a number preserving quadratic fermionic operator one has the general relation 
\begin{equation}\label{eq:fermidirac}
    \frac{{\rm tr}[e^{-\sum_{i,j}a_i^\dag K_{ij}a_j} a_k a^\dag_l]}{{\rm tr}[e^{-\sum_{i,j}a_i^\dag K_{ij}a_j}]}=\left[\frac{1}{\mathbbm{1}+\exp(-K)}\right]_{kl}\,.
\end{equation} 
Hence, given a Hamiltonian $H=\sum_{i,j} M_{ij}a^\dag_i a_j$ and the corresponding Euclidean action
\begin{equation}
    \mathcal{S}_E=\sum_n\sum_{i,j} \epsilon(-i\omega_n \delta_{ij}+ M_{ij})\,a_{ni}^\dag a_{nj}\,,
\end{equation}
 one can set $K_{ni,n'j}\equiv \delta_{nn'}[\epsilon (-i\omega_n \delta_{ij}+M_{ij})+i\pi\delta_{ij}]$ in \eqref{eq:fermidirac}  to obtain
\begin{equation}\label{eq:matsubaratwop}
\begin{split}
     \langle  a_{t_1 i}a^\dag_{t_2 j}\rangle_{\mathcal{S}_E}&=\frac{1}{N}\sum_n e^{-i \epsilon \omega_n (t_1-t_2)} \times \\
     &\times \left[\frac{1}{\mathbbm{1}-\exp\{i\epsilon (\omega_n \mathbbm{1}+i M)\}}\right]_{ij}\,,
\end{split}
\end{equation}
where we used the definition of the FT (yielding the sum over $n$) and the definition of $P$. Remarkably, this is the same expression that one obtains from the time-sliced fermionic PI, with $\omega_n$ precisely the \emph{Matsubara frequencies} (before considering the limit $\epsilon\to 0$). Notably, we recovered it from a canonical-like formalism without the need of introducing Grassmann variables, typically used to define a fermionic version of the \emph{``sum over histories''} of Feynman. This result is more than a coincidence: one can show that the evaluation of traces of the form ${\rm Tr}[P e^{- \mathcal{S}_E}\dots ]$ in product-in-time coherent basis (defined through Grassmann variables in $\mathcal{H}$) are the PIs. We show this explicitly in the Appendix \ref{app:PIs}, which also clarifies the presence of $P$. Nonetheless, let us emphasize that our formalism provides a treatment of PI-like expressions that does not require the use of Grassmann variables, and which, in particular, admits a matrix representation.

As an additional comment, let us notice that if all the operators involved in Theorem \ref{th:correlations} admit a certain symmetry, the traces can be restricted to sum over states satisfying the symmetry. In fact, one can replace $O_t^{(t)}\to O_t^{(t)}Q_t$ (for all $t$), for  $Q_t$ a projector leading to an overall projector $\Pi_t Q_t$ on the left hand side. On the r.h.s. this corresponds to adding $Q^{N-1}=Q$ ($Q^2=Q$) if all the operators, in particular $H$, commute with it. 
For example, one might be interested in working in the ``canonical ensamble'', i.e. evaluations of the trace in the r.h.s. which correspond to a fixed number of particles $m$. This condition corresponds to evaluating the trace on the l.h.s. with a fixed total number of particles $m N$. This holds in any number preserving extended basis, e.g. in the Fourier in time basis, as can be seen by noting that $\prod_t Q_t\equiv \prod_t \int \frac{d\phi}{2\pi}\, e^{i \phi \sum_i (a^\dag_{ti}a_{ti}-m)}=\int \frac{d\phi}{2\pi} e^{i\phi \sum_{t,i} (a^\dag_{ti}a_{ti}-m)}$, with $\sum_{t,i} a^\dag_{ti}a_{ti}=\sum_{n,i} a^\dag_{ni}a_{ni}$ the total number of (extended) particles. We see that there is a one-to-one correspondence between ensembles in $\mathcal{H}$ and $h$, with Theorem \ref{th:correlations}, when stated without restrictions, corresponding to the grand canonical one.

\subsection{The continuum case and Dirac fermions}\label{sec:continuumcase}

All of the previous Theorems assume spacetime algebras for discrete time. While in principle one could apply those results and then consider the formal limit $\epsilon\to 0$ (at fixed $\epsilon N=T$, and eventually $T\to \infty)$ of the ensuing expressions, the Hilbert space formalism suggest a new approach. Namely, one can define a continuum version of the spacetime algebras (equivalent to taking the limit of Eq.\ \eqref{eq:stalg}) and work directly in a continuum setting. Here we develop this approach.

As a concrete and relevant example, we focus our discussion by considering the case of Dirac fermions in a $4$-dimensional spacetime. The case of general fermions and spacetime dimensions can be developed in complete analogy, as stressed at the end of this section. With the aim of describing Dirac fermions in spacetime, we propose the spacetime algebra
\begin{subequations}\label{eq:stalgcont}
\begin{align}
    \{\psi_a(x),\psi^\dag_{b}(y)\}&=\delta^{(4)}(x-y)\delta_{a b}\label{eq:stalg11}\\
    \{\psi_a(x),\psi_{b}(y)\}&=\{\psi^\dag_a(x),\psi^\dag_{b}(y)\}=0\,,\label{eq:stalg22}
\end{align}
\end{subequations}
with $a,b=0,1,2,3$ spinor indices. This is the continuum version of Eq.\ \eqref{eq:stalg} which might be obtained by considering the limit $\epsilon\to 0$. We are also assuming $T\to \infty$. We remark that in this QFT the extended algebra is indeed spacetime symmetric with time and space on equal footing. 
Note the presence of an additional delta in time with respect to the conventional algebra imposed at equal times, namely $\{\psi_a(\textbf{x}),\psi^\dag_{b}(\textbf{y})\}=\delta^{(3)}(x-y)\delta_{ab}$ while $
    \{\psi_a(\textbf{x}),\psi_{b}(\textbf{y})\}=\{\psi^\dag_a(\textbf{x}),\psi^\dag_{b}(\textbf{y})\}=0$.   It is important to remark that in the extended formalism $x^0,y^0$ are clearly unrelated to evolution parameters, and $\psi(x)\equiv \psi(t,\textbf{x})$ is a completely different operator from the conventional evolved field $\psi_H(t,\textbf{x})$ in the Heisenberg picture. Since the extended field operators \emph{at different times anticommute}, $\psi(t',\bf x)$ is not a function of $\psi(t,\bf x)$ for $t'\neq t$, in contrast with  $\psi_H(t',\bf x)$, which is unitarily related to $\psi_H(t,\bf x)$ (in order to distinguish a field in spacetime from a conventionally evolved field, we specify that the latter is in the Heisenberg picture).

Just as in the discrete time case we can introduce a generator of time translations. In fact, we can write
    \begin{equation}
        e^{i\tau \mathcal{P}_\mu x^\mu}\psi_a(z) e^{-i\tau \mathcal{P}_{\mu} x^\mu}=\psi_a(z+\tau x)\,,
    \end{equation}
with 
\begin{equation}\label{eq:pmu}
\mathcal{P}_\mu:=\int d^4x\, \psi^\dag(x) i\partial_\mu \psi(x)=\int \frac{d^4p}{(2\pi)^4}\,  p_\mu \psi^\dag(p) \psi(p)\,,
\end{equation}
where we have introduced the spacetime Fourier Transform (the continuum version of \eqref{eq:fta})
\begin{equation}\label{eq:diracft}
    \psi_a(p):=\int d^4 x\,e^{ipx}\psi_a(x)\,,
\end{equation}
and $\psi^\dag(x) \partial_\mu \psi(x)\equiv  \sum_a \psi_a^\dag(x) \partial_\mu \psi_a(x)$, $px=p_\mu x^\mu$. We use the metric convention $g_{\mu\nu}=\text{diag}(1,-1,-1,-1)$. Note that these derivatives correspond to a ``site derivative'' namely $\lambda^\mu \partial_\mu \psi(x) \approx \{\psi(x+\epsilon' \lambda)-\psi(x)\}/\epsilon'$ with $\lambda$ a $4$-vector, $\epsilon'\ll 1$) and have purely geometrical meaning, including $\partial_0$. This means, in particular, that $\mathcal{P}_0$ is not associated to any particular evolution. Notice also that the Fourier fields satisfy the algebra 
\begin{equation}
    \{\psi_a(p), \psi^\dag_b(k)\}=(2\pi)^4\delta^{(4)}(p-k)\delta_{ab}\,,
\end{equation}
with other anti-commutators vanishing. This algebra is clearly ``off-shell' and thus not accessible to conventional QFT. 
With a slight change in notation $\partial_0 \psi\to \dot{\psi}$ and  $d^4x=dtd^3x $ 
an important first result becomes manifest:
\begin{equation}
\begin{split}
     \mathcal{P}_0
    &=\int dt\int d^3x\, \psi^\dag(t,\textbf{x})i\dot{\psi}(t,\textbf{x})\\&=\int d^4x\, \bar{\psi}(x)\gamma^0 i\partial_0{\psi}(x)
\end{split}
\end{equation}
has the form of the classical \emph{Legendre transform} for a Dirac field. In fact, the conventional momentum conjugate to the Dirac Field $\psi$ is precisely $i\psi^\dag$, while $\bar{\psi}=\psi^\dag\gamma^0$, for $\gamma^\mu$ the Dirac matrices.

We now introduce the 
\emph{Dirac 
 quantum action operator} in analogy with the discrete time case as follows:
\begin{subequations}\label{eq:dqaction}
\begin{align}
    \mathcal{S}_\tau
    &:=\tau \int dt\, \left[ \int d^3x\,\psi^\dag(t,\textbf{x})i\dot{\psi}(t,\textbf{x})- H_D(t)\right]
    \\
    &=\tau\int \!d^4x\, \bar{\psi}(x)(\gamma^\mu i\partial_\mu-m){\psi}(x)\,,  \label{eq:relaction}
\end{align}
\end{subequations}
for $H_D(t)=\int d^3x\, \psi^\dag(t,\textbf{x})(-i\bm{\alpha}\cdot \nabla+\beta m)\psi(t,\textbf{x})$ the Dirac Hamiltonian (as a function of spacetime operators, at a given time $t$) 
and $\tau$ an arbitrary time scale to be discussed below and which makes $\mathcal{S}_\tau$ adimensional. We also employed elementary properties of the Dirac's matrices to write (\ref{eq:relaction}). This is an hermitian \emph{quantum} operator where the index $t$ corresponds to time-sites with no reference to unitary evolution parameterized by an external time. It is precisely the geometrical nature of time underlying (\ref{eq:stalg}) that allows this ``off-shell'' definition. Notice also that $\mathcal{S}_\tau=\tau \mathcal{P}_0-\tau \int dt\, H_D(t)$ in agreement with the discrete formalism.

We can now discuss how to recover conventional unitary evolution. In analogy to the discrete case, we define
\begin{equation}\label{eq:meanv}
    \langle \mathcal{O} \rangle_\tau:=\frac{{\rm Tr}[P e^{i\mathcal{S}_\tau}\mathcal{O}]}{{\rm Tr}[P e^{i\mathcal{S}_\tau}]}\,,
\end{equation}
    which can be regarded as a thermal-like expectation value of $\mathcal{O}$ at inverse temperature $\beta$ and Hamiltonian $H$ with $\beta H\to -i\mathcal{S}_\tau+i\pi\int d^4x\, \psi^\dag(x)\psi(x)$. In addition, it is clear that the FT of the field operators leads to 
    \begin{equation}\label{eq:diracactionmoment}
    \mathcal{S}_\tau=\tau \int \frac{d^4p}{(2\pi)^4}\psi^\dag(p)\gamma^0(\gamma^\mu p_\mu-m)\psi(p)\,.
\end{equation}
Then, the elementary result \eqref{eq:fermidirac} implies
 \fontsize{9.5}{11.5}
\begin{align}
    \langle \psi(p)\bar{\psi}(k)\rangle_\tau&=\frac{\delta^{(4)}(p-k)}{\mathbbm{1}-\exp[i\tau \gamma^0(\gamma^\mu p_\mu-m)/(2\pi)^4]}\gamma^0\label{eq:pprop}\\
    &=\frac{1}{\tau}\frac{i}{\gamma^\mu p_\mu-m}(2\pi)^4\delta^{(4)}(p-k)+\mathcal{O}(\tau)\,,\label{eq:ppropstau}
\end{align}
\normalsize
where we used the basic property $\gamma^0 \gamma^0=1$ leading to $[\gamma^0 (\gamma^\mu p_\mu-m)]^{-1}=(\gamma^\mu p_\mu-m)^{-1}\gamma^0$.
We see that for small $\tau$ the FT of the Dirac propagator is obtained, with \eqref{eq:pprop} an analytic function for any real $\tau$ taking $m^2\equiv m^2-i\tilde{\epsilon}$. As a result,
\begin{equation}\label{eq:diracprop}
     \lim_{\tau\to 0} \langle \sqrt{\tau}\psi(x)\sqrt{\tau}\bar{\psi}(y)\rangle_\tau=\!\int \!\!\frac{d^4p}{(2\pi)^4}\frac{i(\gamma^\mu p_\mu+m)}{p^2-m^2+i\tilde{\epsilon}}\,e^{-ip(x-y)}
\end{equation}
where we used \eqref{eq:diracft} and the standard property ${(\gamma^\mu p_\mu-m)^{-1}=(\gamma^\mu p_\mu+m)/(p^2-m^2)}$. Notice also that $\sqrt{\tau}\psi(x)$ has the same units as the conventional Dirac field $\psi(\textbf{x})$.
One recognizes in the r.h.s. of \eqref{eq:diracprop} the conventional Dirac propagator (with Feynman's prescription) 
allowing one to state the following theorem. 
\begin{theorem}\label{th:diracprop}
Consider the free Dirac quantum action ${\mathcal{S}_\tau=\tau\int d^4x\, \bar{\psi}(x)(\gamma^\mu i\partial_\mu-m){\psi}(x)}$. The two-point correlator becomes the Feynman propagator in the small $\tau$ limit:
  \begin{equation}
  \begin{split}
      \lim_{\tau\to 0} \langle \sqrt{\tau}\psi(x)\sqrt{\tau}\bar{\psi}(y)\rangle_\tau&=\langle 0|\hat{T}\psi_H(x)\bar{\psi}_H(y)|0\rangle\,,
  \end{split}
\end{equation}
with $|0\rangle$ the ground state of the free Dirac Hamiltonian $H_D$.  
\end{theorem}

Theorem \ref{th:diracprop} shows that the two-point correlators in spacetime, computed with respect to the quantum action and for small $\tau$, yield the conventional Feynman propagators which in conventional QM are associated with unitary evolution and time ordering. The definition of the  temporal order operator $\hat{T}$ is the usual, i.e., $
\hat{T}\psi_H(x)\bar{\psi}_H(y):=\psi_H(x)\bar{\psi}_H(y)\theta(x^0-y^0)
      -\bar{\psi}_H(y)\psi_H(x)\theta(y^0-x^0)
$
with $\bar{\psi}\psi\equiv (\bar{\psi}^t \psi^t )^t$ and $\theta$ the Heaviside step function. Notice also that the other two-point contractions vanish.

In addition, since the free action is a quadratic operator, Wick's theorem implies a similar equality for higher order correlations functions such as 
\begin{align*}
    \lim_{\tau\to 0}\tau^2\langle \psi_{a_1}(x_1)\psi_{a_2}(x_2)\bar{\psi}_{b_1}(y_1)\bar{\psi}_{b_2}(y_2)\rangle_\tau&=\\ \langle 0|\hat{T}\psi_{a_1}(x_1)\psi_{a_2}(x_2)\bar{\psi}_{b_1}(y_1)\bar{\psi}_{b_2}(x_2) |0\rangle\,,
\end{align*}
as in the example for discrete time of Eq.\ \eqref{eq:4pointcontr}. The only difference is that one must take the limit $\tau \to 0$ in the continuum time case. Nonetheless, since a $4$-point contraction can be written as a sum of products of $2$-point contractions one can apply the limit to each $2$-point contraction separately. 
We see that all the spacetime correlators of the exponential of the action have a clear meaning: %for each field operators inserted in a spacetime point $x$ one obtains the time-ordered vacuum correlator function with a time-evolved operator  evolved an amount $t_j=x^0_j$. 
For each field insertion at a spacetime point 
$x$, the corresponding time-ordered vacuum correlator function one obtains contains a field operator evolved for a time interval 
$t_j=x_j^0$.
This can be used to introduce interacting terms, as in the discrete case (see the example of section \ref{sec:yukawa}). We can state the following.
\begin{theorem}\label{th:diracprops}
The spacetime correlators of the free Dirac action 
involving fields inserted at different spacetime points are equal to the corresponding Wightman correlation functions of standard QFT in the limit $\tau\to 0$. 
\end{theorem}

Let us make a few comments on these results. First of all, one might wonder why we are obtaining correlation functions corresponding to the ground state of Dirac's Hamiltonian. This can be easily explained by comparing with  Eq.\ \eqref{eq:thermaltwop} and thinking about the continuum case as its limit. In fact, since we are working directly in the unbounded time case, we are essentially working in the limit $\beta \to \infty$ (as evidenced by the continuum values of $p$ defining the FT; see Eq.\ \eqref{eq:ftbetaalg} below). While we are not working with an Euclidean action, we included a factor $\approx -i\tilde{\epsilon}m$ in order to obtain Feynman's prescription, which in the limit $T\to \infty$ is indeed projecting onto the vacuum (from the conventional perspective). Secondly, it is natural to wonder about the interpretation of $\tau$. On one hand, one might regard this construction as a simple mathematical means to consider the limit $\epsilon\to 0$ of the action operator (notice that $\mathcal{P}$ has a well-defined continuum limit, but $e^{i\epsilon \mathcal{P}}$ does not as the notion of ``translating a single step'' is no longer meaningful), and even study how this is related to typical regularization techniques of the PI formulation (see \cite{DIAZ2025170052}). On the other hand, one might want to develop an interpretation in terms of a canonical formalism in a higher dimension, with $\mathcal{S}_\tau$ taking the role of a Hamiltonian and $\tau$ an evolution parameter. Then, our results might be reinterpreted as an holographic correspondence. However, developing this is not straightforward since one would have to deal with the fact that $\mathcal{S}_\tau$ is not a positive definite operator. Moreover, if one regards $\tau$ as a genuine evolution parameter and consider those evolutions which for finite $\tau$ map to the standard theory, one finds that interacting theories are highly non-local, as commented in \cite{DIAZ2025170052, diaz2023parallel}.

Let us also remark that we have derived these results without the need of diagonalizing the quantum action. The diagonalization of the action leads one to define a notion of \emph{extended} Dirac particle. The  interested reader in this can see the discussion of section \ref{sec:paw}. Therein this notion of particle is also related with the Page and Wootters \cite{PaW.83} mechanism and its extensions \cite{di.19,dia.19}. See also Appendix \ref{app:actiondiag}.

Let us now consider the case of a finite time window $T$. In this case, the algebra \eqref{eq:stalgcont} is not modified but the FT is, with 
$
    \psi_n(\textbf{p}):=\int_0^T \frac{dt}{\sqrt{T}}\int d^3x\, e^{it \omega_n } e^{-i\textbf{p}\textbf{x}}\psi(x)\,,
$
 so that 
 \begin{equation}\label{eq:ftbetaalg}
     \{\psi_n(\textbf{p}),\psi^\dag_{n'}(\textbf{p}')\}=\delta_{nn'} (2\pi)^3\delta^{(3)}(\textbf{p}-\textbf{p}')\,.
 \end{equation} We see that the Fourier in time modes have a discrete index. 
 Notice that if one considers a FT only in the spatial components instead $\psi(t,\textbf{p})=\int d^3x\,  e^{-i\textbf{p}\textbf{x}}\psi(t,\textbf{x})$ the ensuing algebra has the form $$\{\psi(t,\textbf{p}),\psi^\dag(t,\textbf{p}')\}=\delta(t-t') (2\pi)^3\delta^{(3)}(\textbf{p}-\textbf{p}')\,.$$ 
Let us  now focus on the Euclidean Dirac action defined by a Wick rotation on the Hamiltonian such that $\mathcal{S}_E=-i{\mathcal{S}_\tau=-i\tau \mathcal{P}_0+ \tau \int dt H_D}$. The FT now leads to
\begin{equation}\label{eq:euclidaction}
    \mathcal{S}_E=\tau \sum_n \!\int \!\frac{d^3p}{(2\pi)^3}\Big\{\psi_n^\dag(\textbf{p})\big[-i\omega_n+(\bm{\alpha}\cdot \textbf{p}+\beta m)\big]\psi_n(\textbf{p})\Big\}.
\end{equation}
In this basis one can easily compute correlators such as 
$
    \langle \psi_n(\textbf{p})\psi^\dag_{n'}(\textbf{k})\rangle_\tau=\delta_{nn'}\frac{\delta^{(3)}(\textbf{p}-\textbf{k})}{1-\exp[i\tau \{\omega_n+i(\bm{\alpha}\cdot \textbf{p}+\beta m)\}/(2\pi)^3]}
    =\frac{1}{\tau}\frac{-1}{i\omega_n-(\bm{\alpha}\cdot \textbf{p}+\beta m)}(2\pi)^3\delta^{(3)}(\textbf{p}-\textbf{k})+\mathcal{O}(\tau)\,.
$
This can be used to show that one is indeed obtaining thermal correlators, in analogy with the discussion of section \ref{sec:fermionsdiscretemap}. 
A particular case corresponds to the expectation number of particles at a given time. In the spacetime formalism this is computed by inserting operators on the same time-slice as follows: 
 \fontsize{9.5}{11.5}
\begin{equation}
\begin{split}
{\displaystyle   \lim_{\tau \to 0}\langle \scalebox{0.9}{$\sqrt{\tau}$}\psi(t,\textbf{p})\scalebox{0.9}{$\sqrt{\tau}$}\psi^\dag(t,\textbf{k})\rangle_\tau}&\!={\frac{V}{T}}{\displaystyle\sum_n \frac{-1}{i\omega_n \mathbbm{1}-(\bm{\alpha}\cdot \textbf{p}+\beta m)}}\\
    &=\!V {\displaystyle  \frac{1}{\mathbbm{1}+\exp\{\scalebox{0.9}{$T$}(\bm{\alpha}\cdot \textbf{p}+\beta m)\}}}\,,
\end{split}
\end{equation}
\normalsize
with $V=(2\pi)^3 \delta^{(3)}(\textbf{0})=(2\pi)^3\int \frac{d^3x}{(2\pi)^3}$ the volume of space and with the first equality a direct consequence of \eqref{eq:euclidaction} as stated above. The second equality holds since the series is precisely the Matsubara expansion \cite{altland.2010} of the Fermi-Dirac statistic at inverse temperature $\beta\equiv T$ with $\omega_n$ the Matsubara frequencies.

Let us notice that all  previous results hold for general fermions with a standard algebra $\{a_i, a_j^\dag\}=\delta_{ij}$, for $i,j$ indices which are either discrete, continuum or a combination of both (here we are using the notation of section \ref{sec:fermionsdiscrete}). The corresponding spacetime algebra is 
\begin{equation}\label{eq:fermioncontinuumalg}
    \{a_i(t), a_j^\dag(t')\}=\delta(t-t')\delta_{ij}\,,
\end{equation}
 with the  Dirac algebra of Eq.\ \eqref{eq:stalgcont} a particular case (with $a_i\to \psi_a(\textbf{x})$).
The spacetime algebra allows one to define a continuum FT in time and the generator of time translations $\mathcal{P}=\sum_i\int dt\, a^\dag_i(t) i\dot{a_i}(t)$. Then, given a Hamiltonian $H[a_i,a_j^\dag]$, the ensuing quantum action is simply 
\begin{equation}
    \mathcal{S}_\tau=\int dt\, \Big[\tau\sum_i a^\dag_i(t) i\dot{a_i}(t)-H(\sqrt{\tau}a_i(t),\sqrt{\tau}a_j^\dag(t))\Big]\,.
\end{equation}
 The map to conventional QM is obtained by considering the small $\tau$ limit of the spacetime correlators of $e^{i\mathcal{S}_\tau}$, just as in the Dirac's case.
 As a final remark, we notice that one can recover the continuum operators from the discrete operators $a_{ti}$ of section \ref{sec:fermionsdiscrete} by defining $a_i(\epsilon t)\equiv a_{ti}/\sqrt{\epsilon}$. It is also feasible to introduce the parameter $\tau$ at the discrete level.

\section{Additional aspects and implications}\label{sec:implications}
In this section we address additional aspects of the formalism which apply to both bosons and fermions. In particular, we show how it allows one to introduce a quantum principle of stationary action related to a generalized notion of state.
We also discuss the scenario in which both bosons and fermions are considered at the same time and can interact with each other. In addition, we unveil a connection between the Page and Wootters mechanism and the quantum action introduced in section \ref{sec:continuumcase}. 
Furthermore, we rederive recent proposals \cite{MAP.25,Guo.25} that aim to ground the notion of timelike pseudoentropies \cite{tak.23, har.23, nar.22, chu2023time} showing that they are particular instances of the formalism we are presenting.

\subsection{Spacetime states}\label{sec:states}

We have successfully defined Hilbert spaces in spacetime and established a map to conventional QM both for bosons and fermions. It is natural to consider the possibility of reinterpreting this map in terms of a notion of state generalized to spacetime. In this section we show how this can be done extending the preliminary results presented in \cite{diaz2024spacetime}.

Let us consider the bosonic-like case first. As in the fermionic  case (see Eq. \eqref{eq:tildeaction}) we introduce the notation \footnote{One can show that while the quantum action is related to Feynman PIs, the operator $\tilde{\mathcal{S}}$ is instead related to Schwinger–Keldysh PIs \cite{schwinger1961brownian}. We employ the second to discuss generalized states since here we are interested in expectation values at a single initial time. See also section \ref{sec:entanglementintime}.}
\begin{equation}
    e^{i\tilde{\mathcal{S}}}:= e^{iTH_0}e^{i\mathcal{S}}=\mathcal{V}^\dag e^{i\epsilon\mathcal{P}}\mathcal{V}\,.
\end{equation}
We can now write an interesting Theorem. 
\begin{theorem}\label{th:states}
Consider an ``environment'' E isomorphic with the system so that $\mathcal{H}_E=\mathcal{H}$, and the states
\begin{align}
|\Psi\rangle&:=\big[(\rho_0e^{i\tilde{\mathcal{S}}/2}\otimes \mathbbm{1}_E)\big]\,|\Phi^+\rangle\\
|\overline{\Psi}\rangle&:=\big[(  e^{i\tilde{\mathcal{S}}/2} \otimes \mathbbm{1}_E)^\dag\big]\,|\Phi^+\rangle\,,
\end{align} 
where $\mathbbm{1}_E$ indicates the identity in $\mathcal{H}_E$ and $|\Phi^+\rangle=\sum_{\textbf{i}}|\textbf{i}\rangle |\textbf{i}\rangle_E$ is a maximally entangled state (unnormalized) in the  system-environment partition. Then, we can define a \emph{generalized state}
\begin{equation}
    R:=\frac{|\Psi\rangle \langle \overline{\Psi}|}{\langle \overline{\Psi}|\Psi\rangle}
\end{equation}
satisfying $R^2=R$, ${\rm Tr}[R]=1$, and 
\begin{align}
    {\rm Tr}_E [R]&=\frac{1 }{{\rm tr}[\rho]}\rho_0 e^{i\tilde{\mathcal{S}}} \label{eq:partialtrtR}\\
        {\rm Tr}_{t'\neq t, E} [R]&=\frac{\rho(t)}{{\rm tr}[\rho]} \label{eq:partialtrtrho}\,,
\end{align}
with $\rho(t)=e^{i\epsilon t H}\rho e^{-i\epsilon t H}$. 
\end{theorem}
Notably, the formalism led us to a \emph{generalized notion of pure states} as those satisfying $R^2=R$, in analogy with $\rho^2=\rho$, but without the hermiticity condition that distinguishes orthogonal projectors from non-orthogonal projectors (here $R^\dag$ describes the correlators with anti-time ordering; equal-time correlators computed with $R $ and $ R^\dag$ coincide). In this sense, the operator $R_\psi\equiv {\rm Tr}_E[R]$ can be interpreted as a mixed generalized state arising from the pure but correlated $R$. In the spacetime approach the correlations between the system and the environment are thus responsible for the causal structure and evolution of physical theories (see also Corollary below). 
Notice that we can think of this construction as a generalized purification of the action. In fact, let us recall that a conventional purification of a density matrix $\rho$ can be obtained as $|\sqrt{\rho}\rangle:=(\sqrt{\rho}\,\otimes\, \mathbbm{1})|\phi^+\rangle$ so that $\rho={\rm tr}_2\big[|\sqrt{\rho}\rangle \langle \sqrt{\rho}|\big]$ for $|\phi^+\rangle=\sum_i |ii\rangle$ a maximally entangled state in $h\otimes h$.

Theorem \ref{th:states} allows us to restate the map of Lemma \ref{th:lemmabosonmap} in a new interesting form that provides spacetime correlators new operational meaning. In fact, by using 
\begin{equation}
    \frac{{\rm Tr}[\rho_0e^{i\tilde{\mathcal{S}}} \otimes_t O^{(t)}_t ]}{{\rm Tr}[\rho_0 e^{i\tilde{\mathcal{S}}}]}={\rm Tr}[R\; (\otimes_t O^{(t)}_t ) \otimes \mathbbm{1}_E]\,,
\end{equation}
which a direct consequence of Theorem \ref{th:states}, one obtains the following Corollary of Theorems \ref{th:states} and \ref{th:correlations}. 
\begin{corollary}\label{cor:weakv}
    Spacetime correlators are weak values defined in $\mathcal{H}\otimes \mathcal{H}\equiv h^{2N}$:
    \begin{equation}\label{eq:meanvaluecor}
    \begin{split}
        {\rm Tr}[R\; (\otimes_t O^{(t)}_t ) \otimes \mathbbm{1}_E]&=
        \frac{ \langle \overline{\Psi}|(\otimes_t O^{(t)}_t ) \otimes \mathbbm{1}_E|\Psi\rangle}{\langle \overline{\Psi}|\Psi\rangle} \\    &=\frac{{\rm tr}[\rho \,\hat{T}\prod_t O^{(t)}(\epsilon t)]}{{\rm tr}[\rho]}\,.
    \end{split}
    \end{equation} 
\end{corollary}

Notably, one can think of spacetime correlators in terms of weak values, a measurable quantity of foundational interest \cite{yak.88, dres.14}. Notice that we require a pair of two different states in order to purify the non-hermitian operators involved in Theorem \ref{th:states}. Thus rather than a conventional mean value, as those associated with space-like properties, a more general setting is needed to accommodate time-like properties. It is also clear that 
if all the operators act on a given time-slice, one can replace the weak value with a traditional pure mean value as  follows from \eqref{eq:partialtrtrho}. 
These results generalize those already presented for free bosonic QFTs in \cite{diaz2024spacetime}.

Let us also mention the recent interest in this kind of generalization of the traditional purification in conventional non-extended QM, where it appears associated to the dS/CFT correspondence and holographic time-like entanglement \cite{tak.23, har.23, nar.22, chu2023time}.

The purification of Theorem \ref{th:states} is clearly not unique and in particular one can make the replacement 
\begin{align}
    |\Psi\rangle&\to |\Psi'\rangle=
    (\rho_0 e^{i\tilde{\mathcal{S}}}\;\otimes \mathbbm{1}_E)\,|\Phi^+\rangle,  \nonumber \\
    |\overline{\Psi}\rangle&\to |\Phi^+\rangle 
\end{align}in the definition of $R$ and all the statements of the Theorem hold. One can then write 
\begin{equation}
    {\rm Tr}[\mathcal{O}\rho_0 e^{i\tilde{\mathcal{S}}} ]=\langle \Phi^+|\mathcal{O}\otimes \mathbbm{1}_E|\Psi'\rangle\equiv \langle \bm {\mathcal O}^\dag|\bm{\rho_0}{e^{\bm i\tilde{\mathcal{S}}}} \rangle
\end{equation}
where we introduced the notation $|\bm A\rangle:=(A\otimes \mathbbm{1}_E)|\Phi^+\rangle$ indicating the Choi–Jamiołkowski isomorphism of the operator $A$. 
We then obtain yet another interpretation of the map. 
\begin{corollary}\label{cor:choi}
Spacetime correlators are equal to the inner product between the Choi state representing $\rho_0 e^{i\tilde{\mathcal{S}}}$ in $\mathcal{H}\otimes \mathcal{H}\equiv h^{2N}$ and a product-in-time Choi state representing the operators:
\begin{equation}
    \langle \otimes_t (\bm{O}^{(t)}_t )^\dag|\bm{\rho_0}{e^{\bm i\tilde{\mathcal{S}}}} \rangle={\rm tr}[\rho \,\hat{T}\smallprod_t O^{(t)}(\epsilon t)]\,.
\end{equation}
\end{corollary}
Notice that $|\otimes_t (\bm{O}^{(t)}_t)^\dag\rangle$ is a product-in-time state  as the extended Choi state is also separable-in-time. In fact, we can write in general $|\Phi^+\rangle=\otimes_t |\phi^+\rangle_t$ so that ${|\otimes_t A^{(t)}_t\rangle=\otimes_t [(A^{(t)}\otimes \mathbbm{1})|\phi^+\rangle_t]}$.  
Instead, $|\bm{\rho_0}{e^{\bm i\tilde{\mathcal{S}}}} \rangle$ is clearly \emph{entangled-in-time} and contains all the time-like correlations which are, in principle, dependent on both the initial state and on the particular physical theory described by the action.

Let us also notice that the previous results hold under Wick rotations. In particular, for $\rho=e^{-TH}$ and $H\to -iH$ we obtain $\rho_0e^{i\tilde{\mathcal{S}}}=e^{-\mathcal{S}_E}$ meaning that one can replace $$|\bm{\rho_0}e^{\bm i\tilde{\mathcal{S}}} \rangle\to |e^{\bm -\mathcal{S}_E}\rangle$$ in Corollary \ref{cor:choi} and rewrite Theorem \ref{th:states} without $\rho$ and with $i\tilde{\mathcal{S}}\to -\mathcal{S}_E$. 
It is now interesting to consider a few examples. For simplicity and ease of notation we focus on the Euclidean action case which describes thermal propagators of the state $e^{-T H}$ (see section \ref{sec:prelim2}).
 \vspace{0.1cm}

\emph{Example 1 (Bosons-Euclidean)}.
Consider bosons in spacetime as those defined in section \ref{sec:prelim2}.
Analogously, we introduce the environment operators $[\tilde{b}_{tj},\tilde{b}^\dag_{t'j'}]=\delta_{tt'}\delta_{jj'}$ and their vacuum $|\Omega\rangle_E$. The Choi state associated with this Fock basis can be written as
\begin{equation}
  |\Phi^+\rangle=\exp\Big{\{}\,\sum_{t,j} b^\dag_{tj} \tilde{b}^\dag_{tj}\,\Big{\}}|\Omega\rangle\rangle
\end{equation}
with $|\Omega\rangle\rangle:=|\Omega\rangle\otimes  |\Omega\rangle_E$. Then, one can immediately show that 
\begin{equation}\label{eq:bosonstates}
\begin{split}
|\Psi\rangle&= |\,\,e^{\bm -\mathcal{S}_E/2}\,\,\rangle\,\,=\exp\Big{\{}\sum_{t,j} 
    b^\dag_{t+1/2,j}(i\epsilon/2) \,\tilde{b}^\dag_{tj}\Big{\}}|\Omega\rangle\rangle\\
     |\overline{\Psi}\rangle&=|(e^{\bm -\mathcal{S}_E/2})^\dag\rangle\!=\exp\Big{\{}\sum_{t,j}  b^\dag_{t-1/2,j}(i\epsilon/2)\, \tilde{b}^\dag_{tj}\Big{\}}|\Omega\rangle\rangle
\end{split}
\end{equation}
where we used that 
\begin{equation}
    e^{-\mathcal{S}_E/2}\,b^\dag_{tj}\, e^{\mathcal{S}_E/2}=b^\dag_{t+1/2,\, j}(i\epsilon/2)\,,
\end{equation}
 with
$b_{t+1/2,j}$ a well-defined (via FT) annihilation operator. We also assumed for simplicity that the action $\mathcal{S}_E$ preserves the vacuum.

It is clear that for quadratic Hamiltonians both $|\Psi\rangle$ and $|\overline{\Psi}\rangle$ are bogoliubov vacua having the form of thermo-field dynamics states.
In this case, one can easily verify that
$
    \langle \overline{\Psi}|\Psi\rangle={\rm tr}[e^{-TH}]
$
by using that the overlap of two gaussian states \cite{ring2004nuclear} or by direct inspection.

\emph{Example 2 (Qubits-Euclidean)}. 
Consider qubits in spacetime as defined by Eq.\ \eqref{eq:qubitsstalg}. We introduce rising (lowering) operators 
\begin{equation}
s^{\pm}_{tx}:=\frac{(\sigma_{1})_{tx}\pm i(\sigma_2)_{tx}}{2}\,.
\end{equation}
The lowest weight state is $|\Omega\rangle\equiv |0\rangle^{\otimes N}$ so that $s^{-}_{tx}|\Omega\rangle=0$ for all spacetime sites. Analogously, we have operators $\tilde{\sigma}_{itx}$, $\tilde{s}^{\pm}_{tx}$ and the environment state $|\Omega\rangle_E$. Then, it is easy to see that 
\begin{equation}
    |\Phi^+\rangle=\exp \Big{\{}\sum_{t,x} s^+_{tx} \tilde{s}^+_{tx}\Big{\}}|\Omega\rangle\rangle\,.
\end{equation}
Considering now 
an Euclidean action we have
\begin{equation}\label{eq:statesspin}
\begin{split}
       |\Psi\rangle&=|e^{\bm -\mathcal{S}_E/2}\rangle=\exp \Big{\{}\sum_{t,x} 
       s^+_{t+1,x}(i\epsilon) \tilde{s}^+_{tx}\Big{\}}|\Omega\rangle\rangle \\
        |\overline{\Psi}\rangle&=| \mathbbm{1}\rangle=\exp \Big{\{}\sum_t s^+_{t,x} \tilde{s}^+_{tx}\Big{\}}|\Omega\rangle\rangle=|\Phi^+\rangle\,.
\end{split}
\end{equation}
Notice that we have slightly modified the puritication of $R$ to avoid translating ``a half-step'' in time (see comments below Theorem \ref{th:states}).

It is interesting to notice the similarities between the two previous examples, namely between Eqs. \eqref{eq:bosonstates} and \eqref{eq:statesspin}. 
Notice also that these expressions are particularly adequate for approximating $R$  
at small $\epsilon$, e.g. by simplifying the evaluation of evolved creation (rising) operators by Trotterization. 
Let us remark in addition that these states are correlated both in space and time. As a matter of fact, the evolution operator is generating spacelike correlations while the time translations are generating timelike correlations.

The bosons and qubits examples suggests an immediate generalization to fermions. Consider an enlarged fermionic system so that $\{\tilde{a}_{ti},\tilde{a}_{t'j}^\dag\}=\delta_{tt'}\delta_{ij}$ with other anticommutators vanishing, including e.g. $\{a_{ti},\tilde{a}^\dag_{t'j}\}=0$. 
We can define the fermionic state
\begin{equation}
  |\Phi^+\rangle:= \exp\Big{\{}\sum_{t,j} a^\dag_{tj} \tilde{a}^\dag_{tj}\Big{\}}|\Omega\rangle\rangle
\end{equation}
for $|\Omega\rangle\rangle$ the global vacuum. Notice that for a single site one has $|\Phi^+\rangle=(1+a^\dag \tilde{a}^\dag)|00\rangle=|00\rangle+|11\rangle$ in analogy with the case of a qubit. Then one can define the fermionic version of a generalized spacetime state.

\begin{theorem}\label{th:fermistates}
Consider the states 
\begin{align}
    |\Psi\rangle&:=(\rho_0 e^{i\tilde{\mathcal{S}}/2})\,|\Phi^+\rangle\\
     |\overline{\Psi}\rangle&:=(  Pe^{i\tilde{\mathcal{S}}/2})^\dag\,|\Phi^+\rangle\,,
\end{align}
defined in the Hilbert space which includes an environment. 
We define a fermionic \emph{generalized state} 
\begin{equation}
    R:=\frac{|\Psi\rangle \langle \overline{\Psi}|}{\langle \overline{\Psi}|\Psi\rangle}
\end{equation}
satisfying $R^2=R$, ${\rm Tr}[R]=1$ and 
\begin{equation}\label{eq:thfermieq}
    \frac{{\rm Tr}[\rho_0e^{i\tilde{\mathcal{S}}} \smallprod_t O^{(t)}_t ]}{{\rm Tr}[\rho_0 e^{i\tilde{\mathcal{S}}}]}={\rm Tr}[R\; \smallprod_t O^{(t)}_t ]\,,
\end{equation}
where all the operators are a function of $a_{ti}$ (i.e. they act trivially on the Environment). 
\end{theorem}
The only difference with the bosonic case is the presence of the additional parity operator in $|\overline{\Psi}\rangle$ which act trivially in the environment. Notice also that the Hilbert space is not of the form $\mathcal{H}_S\otimes \mathcal{H}_E$ as in the bosonic case (that construction is also feasible for fermions). We show in the Appendix \ref{app:choi} that the statements in the Theorem can be interpreted in terms of a fermionic version of the partial trace and purification, with standard partial trace properties holding.
As a consequence, we see that  fermionic spacetime correlators are weak values:
    \begin{equation}
    \begin{split}
        {\rm Tr}[R \smallprod_t O^{(t)}_t]&=\frac{ \langle \overline{\Psi}|\smallprod_t O^{(t)}_t |\Psi\rangle}{\langle \overline{\Psi}|\Psi\rangle}\\
        &=\frac{{\rm tr}[\rho \,\hat{T}\prod_t O^{(t)}(\epsilon t)]}{{\rm tr}[\rho]}\,,
    \end{split}
    \end{equation}
    where all operators act trivially in the environment. Similarly, one can reinterpret spacetime correlators as an overlap between Choi states as in Corollary \ref{cor:choi}. We include a discussion about a fermionic version of the Choi isomorphism in the Appendix \ref{app:choi} where these comments are also expanded. 

As for the bosonic case, we now provide a couple of illustrative examples.

\emph{Example 3 (Fermions-Euclidean)}. Consider an Euclidean fermionic action which, for simplicity, preserves the vacuum. We obtain
 \fontsize{9.5}{11.5}
\begin{equation}\label{eq:fermionstates}
\begin{split}
|\Psi\rangle&= |\,\,e^{\bm -\mathcal{S}_E/2}\,\,\rangle\,\,\,=\;\exp\Big{\{}\,\sum_{t,j} 
    a^\dag_{t+1/2,j}(i\epsilon/2) \tilde{a}^\dag_{tj}\Big{\}}|\Omega\rangle\rangle\\
     |\overline{\Psi}\rangle&=|P(e^{\bm -\mathcal{S}_E/2})^\dag\rangle\!=\exp\Big{\{}\!-\!\sum_{t,j}  a^\dag_{t-1/2, j}(i\epsilon/2) \tilde{a}^\dag_{tj}\Big{\}}|\Omega\rangle\rangle\,.
\end{split}
\end{equation}
\normalsize
Notice that while the expression is almost identical to the bosonic case of \eqref{eq:bosonstates}, the parity operator $P$ add a minus sign in the exponent which defined $|\overline{\Psi}\rangle$.

Notably the previous results can be easily applied to the continuum case. In particular, this was shown for a free Klein-Gordon field in \cite{diaz2024spacetime}. We consider the example of a free Dirac field below.

\emph{Example  (Dirac Fermions-continuum formalism)}.
In analogy with the discrete case, we define the Choi state $|\Phi^+\rangle:=\exp\Big{\{} \int\frac{d^4p}{(2\pi)^4} \sum_\sigma
  \psi_{\sigma}^\dag(p) \tilde{\psi}^\dag_\sigma(p)\Big{\}}|\Omega\rangle\rangle$. Then, by using the expression \eqref{eq:diracactionmoment} of the Dirac quantum action in momentum space we obtain
 \fontsize{9.5}{11.5}
\begin{equation}\label{eq:fermiondstates}
\begin{split}
|\Psi\rangle&\!=\! |e^{i\mathcal{S}_\tau/2}\rangle\\ &\!=\!\exp\Big{\{}\!\int\!\! \frac{d^4p}{(2\pi)^4} \!\sum_{\sigma,\sigma'}\!
    \big[e^{i\tau(\gamma^\mu p_\mu+m)/2}\, \big]_{\sigma \sigma'}\,\psi_{\sigma'}^\dag(p) \tilde{\psi}^\dag_\sigma(p)\Big{\}}|\Omega\rangle\rangle\\
\end{split}
\end{equation}
\normalsize
with a similar expression for $|\overline{\Psi}\rangle=|P(e^{i\mathcal{S}_\tau/2})^\dag\rangle$ with an overall minus sign in the exponent and with $\tau\to -\tau$. One can then directly recover Eq.\ \eqref{eq:ppropstau} from the weak value
\begin{equation}
    \langle \psi(p)\bar{\psi}(k)\rangle_\tau=\frac{\langle \overline{\Psi}|\psi(p)\bar{\psi}(k)|\Psi\rangle}{\langle \overline{\Psi}|\Psi\rangle}\,.
\end{equation}
One can show this explicitly, for example by expanding the exponentials that define the states and contracting the ensuing field operators with the operators $\psi(p)\bar{\psi}(k)$, thus obtaining a geometric series converging to \eqref{eq:ppropstau}.

\subsection{Stationary quantum action principle}\label{sec:qaprinciple}
As is well known, classical mechanics can be formulated from the principle of ``minimal'' action. Notably, the formalism presented in this manuscript allows one to introduce  a \emph{quantum action stationary principle}, in which the generalized notion of state introduced in section \ref{sec:states} plays a central role. In this section, we briefly illustrate this point,  emphasizing the  need for the formalism to introduce this principle.

Let us focus on the thermal case. We have seen that $\frac{e^{-S_E/\hbar}}{Z}$ can be essentially regarded as a non-pure generalized state (we reintroduce the Planck constant $\hbar$ here; in the fermionic case the parity operator is implicitly absorbed in the action). Considering its exponential form we can state the following Theorem. 
\begin{theorem}\label{th:quantumvariationalpr}
    Consider the functional
    \begin{equation}
        F[\Gamma]:=\langle S_E\rangle_\Gamma+\hbar \,{\rm Tr}[ \Gamma\log(\Gamma)]\,,
    \end{equation}
    with $\langle \dots \rangle_\Gamma:={\rm Tr}[\Gamma \dots]$. Under variations of normalized operators ${\rm Tr}[\Gamma]=1$ acting on $\mathcal{H}$, one finds the extremal condition 
    \begin{equation}
        \delta F[\Gamma^\ast]= 0\,,
    \end{equation}
    for $\Gamma^\ast=e^{-S_E/\hbar}/{\rm Tr}[e^{-S_E/\hbar}]$.
\end{theorem}

Remarkably, since all thermal correlators can be obtained from the action, the proper (thermal) dynamics emerge by extremizing the \emph{generalized entropy} (or pseudoentropy \footnote{This type of entropy was  introduced in the context of dS-CFT correspondence \cite{tak.23} (for standard QM).})  $-{\rm Tr}[\Gamma \log(\Gamma)]$ at a fixed quantum action mean value, with $\hbar^{-1}$ the corresponding Lagrange multiplier. One can interpret the entropy as a measure of correlations between the system and an environment, which at the solution $\Gamma=\frac{e^{-S_E/\hbar}}{Z}$ are globally in the state $R$  discussed in the previous section ($\Gamma={\rm Tr}_E R$ for bosons).

\begin{figure}[t!]
    \centering
\includegraphics[width=0.9\linewidth]{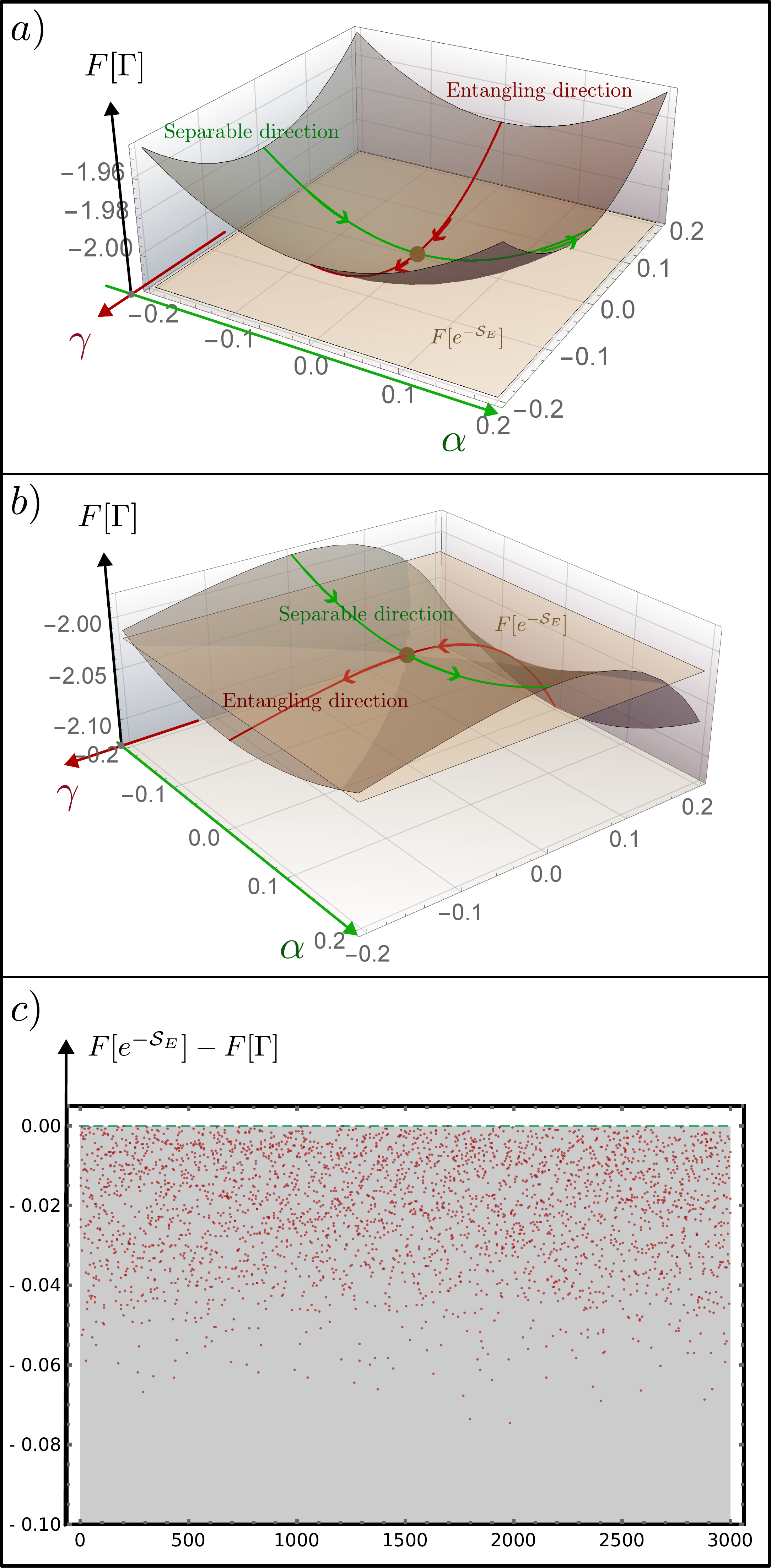}
    \caption{{ \bf Numerical results for the functional $F[\Gamma]$.} In all three panels $H=\lambda Z$ and $N=2$, $\epsilon\equiv 1$ for different test quantum actions. a) $F[\Gamma]$ for a qubit system with $\lambda=1$, $K_1=\alpha (\sigma_1\otimes \mathbbm{1}+\mathbbm{1}\otimes \sigma_1)+\gamma (\sigma_1\otimes \sigma_2+\sigma_2\otimes \sigma_1)$. In both directions $(\alpha,\gamma)$ the functional $F[\Gamma]$ increases when we part from the quantum action (the plane corresponds to $F[e^{-\mathcal{S}_E}]$). 
    b) $F[\Gamma]$ for a qubit system with $\lambda=1$, $K_2=\alpha (\sigma_1\otimes \mathbbm{1}+\mathbbm{1}\otimes \sigma_1)+\gamma (\sigma_1\otimes \sigma_1+\sigma_2\otimes \sigma_2+\sigma_3\otimes \sigma_3)$. In this case, the entangling direction $\gamma$ provides a way to decrease $F[\Gamma]$ showing that $\Gamma=e^{-\mathcal{S}_E}$ is a saddle point of $F$. c) Difference $F[e^{-\mathcal{S}_E}]-F[\Gamma]$ for a spinless fermion. We plot it for $3 \times 10^3$ random values of the parameters defining $\Gamma$, taking possible values $\alpha_{13},\alpha_{14},\alpha_{24}\in (-0.2,0.2)$,  $\lambda\in (-2,2)$ and $\alpha_{12}=0$ (chosen such that $K=K^\dag$ defined in Eq.\ \eqref{eq:Kmajorana}). In this case, all the test quantum actions yield smaller values of $F$ than the solution. }
    \label{fig:functional}
\end{figure}

One of the most notable features of this variational principle is that it allows one to explore many more operators than in standard QM. To explain this point let us consider operators of the form (we set $\hbar=1$ once again as it can be easily restored) $\Gamma=e^{i\epsilon\mathcal{P}}e^{-K}$ with $[\mathcal{P},K]=0$ so that time translation invariance is preserved. Here $\Gamma$ is unnormalized so it must be divided by $Z_\Gamma\equiv {\rm Tr}[\Gamma]$ when considering the associated $F$. One can easily obtain
\begin{equation}\label{eq:Fvariation}
    F[\Gamma]=\frac{1}{Z_\Gamma}{\rm Tr}\Big[\Gamma \,\Big(\sum_t \epsilon H_t - K\Big)\Big]-\log(Z_\Gamma)\,.
\end{equation}
For $K$ hermitian one can immediately prove that $F[\Gamma]\in \mathbb{R}$ (up to a possible constant factor $i\pi$ arising from negative $Z_\Gamma$). 
    Now, in the particular case of a \emph{separable-in-time}  $K=\sum_t H'_t$,  Eq.\ \eqref{eq:Fvariation} leads directly to
    \begin{equation}\label{eq:almostfreenergy}
        F[\Gamma]= T\,{\rm tr}[\rho_{H'}\, (H - H')]-\log(Z_{H'})\,,
    \end{equation} which is $T$ times the standard Helmholtz free energy \cite{balian2005information,balian2006microphysics} computed for the thermal state $\rho_{H'}=e^{-T H'}/Z_{H'}$ with $T=\epsilon N\equiv \beta$. The free-energy, $\propto {\rm tr}[\rho_{H'}\, H]-\beta^{-1}{\rm tr}[\rho_{H'}\log(\rho_{H'})]$ is of course minimized for $H'=H$ in agreement with our statement. However, 
    Theorem \ref{th:quantumvariationalpr} allows one to consider far more general operators, including those for which $K$ is \emph{entangled-in-time}.

    To provide a concrete example, consider the case of a single qubit and $N=2$. Then $\mathcal{H}=h\otimes h$ is isomorphic to the standard Hilbert space of two qubits. This allows one to write a general $K$ as 
    \begin{equation}
        K=\sum_{i=1}^3 \alpha_i \sigma_i\otimes \mathbbm{1}+\sum_{j=1}^3 \beta_j  \mathbbm{1}\otimes \sigma_j+\sum_{i,j=1}^3\gamma_{ij}\sigma_i\otimes \sigma_j\,,
    \end{equation}
    for $\sigma_i$ the Pauli matrices. Those operators with $\gamma_{ij}=0$
    lead to separable-in-time Hamiltonians and thus standard quantum theories. The corresponding variations are those of the free energy  \eqref{eq:almostfreenergy} (for $\alpha_i=\beta_i$ with $H=\sum_i \alpha_i \sigma_i$). Instead, for at least one $\gamma_{ij}\neq 0$, the formalism is exploring novel kind of theories not accessible to canonical QM and associated with 
generalized states being, in principle, more entangled across time. As a matter of fact, in analogy with standard QM,  where these terms correspond to interacting systems, one might interpret these terms as \emph{interactions across time}.
If we also impose $\alpha_i,\beta_j,\gamma_{i,j}\in \mathbb{R}$ and 
$\alpha_i=\beta_i$, $\gamma_{ij}=\gamma_{ji}$, namely $K$ hermitian and homogeneous in time ($[K, \text{SWAP}]=0$), the functional $F$ is a function of $9$ parameters, $3$ correspond to standard QM and $6$ to novel ``directions''. At $\Gamma\propto e^{-\mathcal{S}_E}$ the functional $F$ reaches a stationary point, a property that holds for variation along the novel ``entangling'' directions (see Figure \ref{fig:functional}). One can show that it is possible to move along particular (entangled) directions in operator space that decrease $F$ around the solution. An example is given by $H\propto Z$ and the maximal Schmidt's rank operator \cite{nielsen2003quantum} $K=H+\alpha(\sigma_1\otimes\mathbbm{1}+\mathbbm{1}\otimes \sigma_1)+\gamma \sum_i \sigma_i\otimes \sigma_i$ as shown in panel b) of Fig.\ \ref{fig:functional}. This timelike entanglement is a necessary but not sufficient condition for decreasing $F$ below $F[e^{-\mathcal{S}_E}]$ as shown in panel a) of Fig.\ \ref{fig:functional}.

A similar discussion holds for fermionic systems. As an example  analogous to the qubit one, consider a single spinless fermion and $N=2$. We recall that this is also the example we considered in Eq.\ \eqref{eq:fswap}. For a variation of the form $\Gamma=P e^{i\epsilon \mathcal{P}}e^{-K}$ with $[K,P]=[K,\mathcal{P}]=0$ the expression \eqref{eq:Fvariation} for $F[\Gamma]$ holds.  A simple way to characterize all possible hermitian $K$ operators is by using Majorana operators \cite{diaz2023showcasing} (in spacetime) $c_\mu$ defined to be hermitian operators satisfying $a_t=(c_{2t-1}+ic_{2t})/2$. The corresponding spacetime algebra is $\{c_\mu,c_\nu\}=2\delta_{\mu\nu}$ with $\mu,\nu=1,\dots, 2N$ for $N$ time-slices. Then, we can expand 
\begin{equation}\label{eq:Kmajorana}
K=\sum_{\mu<\nu=1}^4\alpha_{\mu\nu}\,ic_\mu c_\nu+\delta \,c_1c_2c_3c_4\,,
\end{equation}
 for a total of $7$ real parameters (we assume $K=K^\dag$) In addition, we impose time invariance leading to only $5$ independent parameters $\alpha_{12}=\alpha_{34},\alpha_{13}, \alpha_{14}=\alpha_{32}, \alpha_{24}$ and $\delta$ (the difference with the qubit case comes from the parity restriction).  
 Notice that the $\alpha$-terms corresponds to gaussian evolution while the last term is the fermionic parity and does not affect $F$ (non-gaussian quantum actions, that allow for fermionic entanglement across time \cite{gigena2015entanglement}, require a larger $N$). Notice now that given the Hamiltonian $H=Z=-i c_1c_2$ (for a single spinless fermion one has two Majorana operators), the quantum action is $\mathcal{S}_E=-i\epsilon \mathcal{P}-ic_1c_2-ic_3c_4$ so that all the available local-in-time variations correspond to a rescaling of the Hamiltonian ($c_1, c_2$ corresponds to the first time site while $c_3,c_4$ to the second). Instead, all the other terms of $K$ correspond to interactions across time sites. The latter lead to a local maximum of $F$ for fixed Hamiltonian and $\alpha_{12}$ as one can verify by noting that all the eigenvalues but one (the separable $\alpha_{12}$-direction) of the corresponding Hessian are negative for any $H=\lambda Z$. This is illustrated numerically in Fig.\ \ref{fig:functional} c). It is interesting to remark the difference with the qubit case, where it is possible to choose timelike interactions that increase $F$. This different behavior can be associated with the subtle but fundamental differences between bosons and fermions stressed at the end of section \ref{sec:fermionsdiscrete}.

Let us add a few remarks on this novel emerging scheme. We have seen that one can explore more possible ``test'' quantum actions than those that are directly linked to standard QM. Beside being interesting from a foundational perspective, these types of theories naturally appear when developing the operator equivalent to semiclassical approximations \cite{kleinert2006path, puddu1991rpa, rossignoli1998thermal}. As a matter of fact, the latter corresponds to neglecting some of the Fourier modes of Eq.\ \eqref{eq:ft} leading to non-local-in-time features that make use of the full Hilbert space $\mathcal{H}$ \cite{DIAZ2025170052}. Other subtleties appear when considering non-translationally (in time) invariant and/or real-time evolution: on the one hand, one needs to deal with complex entropies (see also \cite{nak.21,har.23}), with a stationary condition holding for both real and imaginary parts of the corresponding functional. On the other hand, imposing an initial (or final) condition, such as those described in section \ref{sec:states} is  subtle, and can in principle be associated with additional constraints on $F$, at least as long as a proper continuum time limit is considered.  These latter subtleties need to be rigorously addressed before one can discuss the relation between the corresponding functional $F$, classical entropies, and the classical action $S_{\text{cl}}$ in the limit $\hbar\to 0$. Such subtle analysis  lies beyond the scope of the present work and is left for future investigation.

\subsection{Fermion-boson interactions}\label{sec:fbinteractions}

Having developed the spacetime formalism for bosonic and fermionic systems separately, it is natural to consider the scenario in which both bosons (or  bosonic-like finite dimensional systems, as in section \ref{sec:prelim}) and fermions are present. This is the subject of this section. It is worth remarking that the approach we follow here corresponds to treating boson-fermion interaction in an all orders of perturbation theory type analysis.

\subsubsection{Discrete time formalism}
 Let us first discuss the discrete time case, following the formalism we presented in Sections \ref{sec:prelim} and \ref{sec:fermionsdiscrete}-\ref{sec:fermionsdiscretemap}.
 Since bosons and fermions are independent physical systems we consider a total Hilbert space $\mathcal{H}=\mathcal{H}_B \otimes \mathcal{H}_F$ with $\mathcal{H}_B$ the spacetime Hilbert space for bosons, defined e.g. by $\mathcal{H}_B=\otimes_t h$ as in section \ref{sec:prelim}, and $\mathcal{H}_F$ defined by the algebra \eqref{eq:stalg}.  It is clear that bosonic and fermionic operators commute with each other, meaning e.g. (for actual bosons) $[b_{t_1i},a^\dag_{t_2 j}]=[b_{t_1i},a_{t_2 j}]=0$ with the ladder operators defined in Eqs.\ \eqref{eq:bosonicalg} and \eqref{eq:stalg}.

It is now very simple to generalize Theorem \ref{th:correlations} to include bosons (using also the results of section \ref{sec:prelim} proved in \cite{DIAZ2025170052}). Consider first a product operator $\mathcal{O}=\mathcal{O}_B \otimes \mathcal{O}_F$. It is clear that 
\fontsize{9.6}{11.5}\selectfont
\begin{equation}\label{eq:bosfersep}
    {\rm Tr}\big[Pe^{i(\mathcal{S}_B+\mathcal{S}_F)}\mathcal{O}_B \otimes \mathcal{O}_F\big]={\rm Tr}\big[e^{i\mathcal{S}_B}\mathcal{O}_B \big]\,  {\rm Tr}\big[Pe^{i\mathcal{S}_F}\mathcal{O}_F\big]\,,
\end{equation}
\normalsize
with $S_B$ ($S_F$) the bosonic (fermionic) quantum action so that $\mathcal{S}_B+\mathcal{S}_F\equiv \mathcal{S}_B\otimes \mathbbm{1}_F+\mathbbm{1}_B \otimes \mathcal{S}_F$. Now one can apply Lemma \ref{th:lemmabosonmap} and Theorem \ref{th:correlations} to each term separately of \eqref{eq:bosfersep} which 
yields 
\begin{equation}\label{eq:bosonfermion}
\begin{split}
     &{\rm Tr}\big[Pe^{i(\mathcal{S}_B+\mathcal{S}_F)}\smallprod_l \,O^{(l)}_{B t_l} \otimes O^{(l)}_{F t_l}\big]=\\
     &={\rm tr}\big[e^{-iT (H_B+H_F)}\hat{T}\,\smallprod_l \,O^{(t_l)}_B(\epsilon t_l)O^{(t_l)}_F(\epsilon t_l)\big]\,,
\end{split}
\end{equation}
where we already rewrote the result as a single trace in $h=h_B \otimes h_F$. Notice that the time ordering is not relevant when comparing bosonic and fermionic operators since they commute with each other.
Now, given a more general operator  
acting on ${\mathcal{H}_B\otimes \mathcal{H}_F}$ one can expand it as a sum over product operators (in the boson-fermion partition). This leads directly to
\fontsize{9.5}{11.5}\selectfont
\begin{equation}\label{eq:bosonfermioncomplete}
    {\rm Tr}\big[Pe^{i(\mathcal{S}_B+\mathcal{S}_F)}\smallprod_l O^{(l)}_{t_l}\big]=
     {\rm tr}\big[e^{-iT (H_B+H_F)}\hat{T}\, \smallprod_l \,\mathcal{O}^{(t_l)}(\epsilon t_l)\big]\,,
\end{equation}
\normalsize
which is the generalization of Theorem \ref{th:correlations} to the case in which bosons are present but they do not interact with fermions.

     We can now discuss the case of interacting bosons-fermions, defined by a Hamiltonian $H=H_B\otimes \mathbbm{1}_F+\mathbbm{1}_B\otimes H_F+H_{\text{int}}$. The corresponding quantum action is
     \begin{equation}\label{eq:bfaction}
         e^{i \mathcal{S}}=e^{i\epsilon (\mathcal{P}_B+\mathcal{P}_F)}\smallprod_t e^{-i\epsilon (H_B+H_F+H_{\text{int}})}\,.
     \end{equation}
     Now, assuming $e^{-i\epsilon (H_B+H_F+H_{\text{int}})}=e^{-i\epsilon (H_B+H_F)}\mathcal{O}_{\text{int}}$, with $\mathcal{O}_{\text{int}}$ having the formal Dyson's expansion $\mathcal{O}_{\text{int}}=\hat{T}e^{-i\int^\epsilon_0 dt' H_I(t')}$, one can write
     \begin{equation}
          e^{i \mathcal{S}}= e^{i (\mathcal{S}_B+\mathcal{S}_F)}\smallprod_t (\mathcal{O}_{\text{int}})_t\,,
     \end{equation}
  allowing us to apply  Eq.\ \eqref{eq:bosonfermioncomplete}. As a result, 
\begin{equation}
\begin{split}
      {\rm Tr}[Pe^{i\mathcal{S}}]&=
     {\rm tr}\big[e^{-iT (H_B+H_F)}\hat{T}e^{-i\int^T_0 dt' H_I(t')}\big]\\
     &= {\rm tr}[e^{-iTH}]\,,
\end{split}
\end{equation}
which is the natural generalization of Theorem \ref{th:correlations} with no operator insertions. If operators are added, Theorem \ref{th:correlations} holds as seen by following the same line of reasoning. In summary we have obtained the following Corollary. 
\begin{corollary}
    Theorem \ref{th:correlations} holds for a bosonic$+$fermionic quantum action as in Eq. \eqref{eq:bfaction} and for general bosonic and fermionic operators. 
\end{corollary}

\subsubsection{Continuum spacetime formalism: Yukawa interaction}\label{sec:yukawa}

The scenario of fermions interacting with bosons is very simply described in the continuum time case. In order to illustrate this scenario, we extend our example of Dirac action of section \ref{sec:continuumcase} to include a Yukawa interaction with Lagrangian density $\mathcal{L}_{\text{int}}=g \phi \bar{\psi}\psi$. The treatment of other interactions are apparent from our discussion of this example.

We consider the total quantum action
\begin{equation}\label{eq:qactiontot}
\mathcal{S}_\tau=\mathcal{S}_{\tau} ^\psi+\mathcal{S}_{\tau}^\phi+\mathcal{S}^{\phi \psi}_{\tau}\,,
\end{equation}
with $\mathcal{S}_{\tau}^\psi$ the free Dirac quantum action of Eq.\ \eqref{eq:dqaction}, 
\fontsize{9.5}{11.5}\selectfont
\begin{equation}\label{eq:kleingordonaction}
    \mathcal{S}_{\tau}^\phi=\tau \!\int d^4x \,\Big \{\pi(x)\dot{\phi}(x)-\frac{\pi^2(x)}{2}-\frac{(\nabla \phi(x))^2}{2}-\frac{m^2\phi^2(x)}{2} \Big \}
\end{equation}
\normalsize
a Klein-Gordon free quantum action for a scalar real field 
 and with
 \begin{equation}
     \begin{split}
         \mathcal{S}^{\phi \psi}_{\tau}=g \tau^{3/2}\int d^4x\,\phi(x) \bar{\psi}(x)\psi(x)\,,
     \end{split}
 \end{equation}
 the Yukawa interaction with the $\tau$ factor defined according to $\mathcal{S}^{\phi \psi}_{\tau}=\int d^4x\, \mathcal{L}_{\text{int}}[\sqrt{\tau}\phi(x),\sqrt{\tau} \psi(x)]$. 
We refer the reader to \cite{diaz2024spacetime} for a complete description of bosonic QFTs in spacetime. Here we simply notice that the essential ideas and small $\tau$ limit are analogous to our results of section \ref{sec:continuumcase}. In particular, if one imposes the commutator spacetime algebra
\begin{equation}\label{eq:stalgbosons}
    [\phi(x),\pi(y)]=i\delta^{(4)}(x-y)\,,
\end{equation} it can be shown that
\begin{equation}
     \lim_{\tau\to 0} \langle \sqrt{\tau}\phi(x)\sqrt{\tau}\bar{\phi}(y)\rangle_\tau=\!\int\! \frac{d^4p}{(2\pi)^4}\frac{i}{p^2-m^2+i\tilde{\epsilon}}e^{-ip(x-y)}\,,
\end{equation}
which is the free bosonic Feynman propagator. Here the brackets indicate the quotient of traces as in \eqref{eq:meanv} but with the Dirac action replaced by $S^\phi_\tau$ and no parity operator (same notation as in \cite{diaz2024spacetime}). In addition, bosonic and fermionic fields commute ($[\phi(x),\psi(y)]=0$ and so on),  meaning that the bosonic and fermionic operators act on different distinguishable Hilbert spaces, which in particular implies  $[S^\phi_\tau,S^\psi_\tau]=0$. Instead, since the interacting part couples both fields, the commutators $[S^\phi_\tau,S^{\phi\psi}_\tau]$ and $[S^\psi_\tau,S^{\phi\psi}_\tau]$ are not vanishing. Nonetheless, they are always proportional to some power of $\tau$, which can be chosen arbitrarily small.

A convenient scheme for treating interactions follows. 
In analogy to the notation of \eqref{eq:meanv} we indicate the following quotient of traces as 
 \begin{align}
   \langle \dots \rangle_\tau&:= \frac{{\rm Tr}[Pe^{i\mathcal{S}_\tau}\dots]}{{\rm Tr}[Pe^{i\mathcal{S}_\tau}]}\\ \langle \dots \rangle_{\tau \text{free}}&:=\frac{{\rm Tr}[Pe^{i(\mathcal{S}^\phi_\tau+\mathcal{S}^\psi_\tau)}\dots]}{{\rm Tr}[Pe^{i(\mathcal{S}^\phi_\tau+\mathcal{S}^\psi_\tau)}]}\,,
 \end{align}
 where the first one involves the complete interacting action \eqref{eq:qactiontot} while the second only the free fermionic+bosonic parts.
Now, for small $\tau$,  we can separate the interaction from the free part of the actions:
 \begin{equation}
     \langle \dots \rangle_\tau=\frac{\langle e^{i\mathcal{S}_{\tau}^{\phi \psi}}\dots \rangle_{\tau \text{free}}}{\langle e^{i\mathcal{S}_{\tau}^{\phi \psi}} \rangle_{\tau \text{free}}}+\mathcal{O}(\tau^2)
 \end{equation}
as it follows from  $e^{i\mathcal{S}_\tau}= e^{i(\mathcal{S}^\psi_\tau+\mathcal{S}_\tau^\phi)}e^{i\mathcal{S}_\tau^{\phi\psi}} +\mathcal{O}(\tau^2)$.
Assuming $g$ small, a perturbative expansion is then readily obtained by expanding $\langle e^{i\mathcal{S}_{\tau}^{\phi \psi}}\dots \rangle_{\tau \text{free}}$ in orders of $g$. For example the first order is given by  $\langle e^{i\mathcal{S}_{\tau}^{\phi \psi}}\dots \rangle_{\tau \text{free}}=\langle \dots \rangle_{\tau \text{free}}+g \tau^{3/2}\langle \int d^4y \phi \bar{\psi}\psi\dots \rangle_{\tau \text{free}}+\mathcal{O}(g^2)$. Now one can apply Wick's theorem inside each $\langle \dots \rangle_{\tau \text{free}}$ which agrees with Feynman diagrammatic expansion. In order to illustrate this agreement, let us consider a single three level contraction arising in a $4$-point correlation function:
\begin{widetext}
     \begin{equation}\label{eq:yukawacontraction}
     \begin{split}
\tau^2\langle\psi(x_1)\psi(x_2)\bar{\psi}(x_3)\bar{\psi}(x_4) \rangle_\tau&\simeq\frac{\langle e^{i\mathcal{S}_{\tau}^{\phi \psi}}\tau^2\psi(x_1)\psi(x_2)\bar{\psi}(x_3)\bar{\psi}(x_4) \rangle_{\tau \text{free}}}{\langle e^{i\mathcal{S}_{\tau}^{\phi \psi}} \rangle_{\tau \text{free}}}=\tau^2\langle\psi(x_1)\psi(x_2)\bar{\psi}(x_3)\bar{\psi}(x_4) \rangle_{\tau \text{free}}\\&+g^2 \tau^5 \int dy dz \,\wick{\c2 \psi(x_1) \c3\psi(x_2) \c1\phi(y)\bar{\c2 \psi}(y)\c4\psi(y)\c1\phi(z)\bar{\c3\psi}(z)\c5\psi(z)\bar{\c4\psi}(x_3)\bar{\c5\psi}(x_4)} \\&+\text{other $\mathcal{O}(g^2)$  contractions}+\mathcal{O}(g^4)
     \end{split}
     \end{equation}
\end{widetext}
with the contractions defined as 
\begin{align}
    \wick{\c1\phi(x) \c1 \phi(y)}&:=\langle \phi(x) \phi(y)\rangle_{\tau \text{free}}\\
    \wick{\c1\psi(x)  \bar{\c1\psi}(y)}&:=\langle \psi(x) \bar{\psi}(y)\rangle_{\tau \text{free}}\,.
\end{align}
Since we know that these contractions yield the bosonic (fermionic) propagators, it is clear that the small $\tau$ limit of the l.h.s. of Eq.\ \eqref{eq:yukawacontraction} is precisely what is obtained by applying the standard Feynman rules. In particular, the contraction of \eqref{eq:yukawacontraction} corresponds to the diagram 
\begin{figure}[h!]
    \centering
\includegraphics[width=0.56\linewidth]{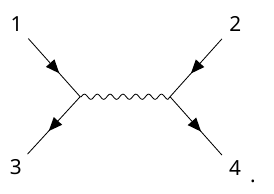}
\end{figure}

\noindent Here solid lines represent fermionic propagators while the wavy line is the bosonic one. 
Notice that there are precisely $5$ contractions, matching the power of $\tau$ accompanying the $g^2$ term. Other contractions are to be treated similarly, including the ``vacuum bubbles'' \cite{P.18}, with the usual factorization between connected and disconnected diagrams holding (the denominator $\langle e^{i\mathcal{S}_\tau^{\phi\psi}}\rangle_{\tau \text{free}}$, which is to be expanded perturbatively as well, cancels all  disconnected diagrams; the factorization holds for finite $\tau$). 

The treatment of other contractions and different interactions can be easily developed along these lines. The essential point is that for small $\tau$ one can separate the exponential of the action in the product of two exponentials, one for the free action part and one for the interacting one (lowest order of the Baker–Campbell–Hausdorff formula; convergence is assumed as usual). Then, assuming a perturbative approach is possible ( small $g$), one can expand the exponential of the interaction in power series and apply Wick's theorem to the ensuing contractions. In this way, one recovers the Feynman rules from the spacetime approach.

\subsection{Second quantization of the Page and Wootters mechanism and quantum time}\label{sec:paw}

It has been recently noticed \cite{dia.19,diaz.21} that a direct \emph{second quantization} of the Page and Wootters (PaW) \cite{PaW.83} formalism leads directly to spacetime algebras for bosons. Moreover, the second quantized universe operators are free bosonic quantum actions \cite{diaz.21,DIAZ2025170052,diaz2024spacetime}. Here we extend these results to fermions and add some considerations on relativistic bosons thus providing a complete connection between the quantum actions characterizing the spacetime formalism and the notion of \emph{quantum time}.

Let us specify first that with ``second quantization'' we indicate the mathematical scheme that allows one to construct a Fock space from a given single particle (sp) space. Such a Fock space describes an arbitrary number of these particles which are also assumed to be indistinguishable. For this reason it can be thought as an (anti)-symmetrization of a direct sum of spaces of different number of particles. This point of view is equivalent to considering the following rules: given an orthonormal single particle basis $|i\rangle$ one identifies $|i\rangle=a_i^\dag |0\rangle$ with $a_i$ annihilation operators. 
In addition, the sum over particles of a single particle operator $O=\sum_{i,j}\langle i|O|j\rangle|i\rangle \langle j|$ (e.g. the total kinetic energy) is identified with the Fock space operator (see e.g. \cite{Sc.97} for details) 
\begin{equation}\label{eq:Fockop}
    O_F:=\sum_{i,j} \,\langle i|O|j\rangle \,a_i^\dag a_j\,.
\end{equation}

On the basis of previous considerations, we can state more clearly what we mean by second quantization of the PaW formalism: We consider the mathematical space, usually identified with the space of the system with a ``clock'',  as the space of single particles. The ensuing Fock space is thus describing an arbitrary number of indistinguishable \emph{PaW-particles}. This was discussed for bosons in \cite{dia.19,diaz.21}. 
An interpretation in terms of ``events'' was later proposed (including fermions) in \cite{giovannetti2023geometric}, although without linking them 
to quantum actions.

To describe fermions and relate the results with the fermionic action of \ref{sec:continuumcase}, we focus on the extension of the PaW mechanism developed in \cite{di.19} for a Dirac particle. Therein, it was shown that the Dirac equation can be recovered by imposing the ``universe equation'' $\mathcal{J}|\Psi\rangle=0$ to the operator \begin{equation}
    \mathcal{J}=P_0 \otimes \mathbbm{1}+\mathbbm{1}\otimes H_D\,.
\end{equation}  This operator acts on the Hilbert space 
defined by four-dimensional spinors and algebra $[X^\mu,P_\nu]=i  \delta^\mu_{\;\; \nu} \mathbbm{1}_4$ with $\mathbbm{1}_4$ the $4\times 4$ identity matrix. A basis of the Hilbert space is provided by states $|x,a\rangle$ satisfying $X^\mu |x,a\rangle=x^\mu|x,a\rangle$ and 
\begin{equation}\label{eq:orthpaw}
\langle y,b|x,a\rangle=\delta^{(4)}(x-y)\delta_{ab}    
\end{equation}
 for  $a,b=0,1,2,3$. Notice that this is not the usual Hilbert space of the Dirac equation \cite{T.92}, since an additional \emph{time operator} $X^0$ has been introduced. In the PaW interpretation, one writes $|x,a\rangle=|t\rangle \otimes |\textbf{x},a\rangle$ with $t\equiv x^0$ a label of states of a ``clock system'' and  $|\textbf{x},a\rangle$ a basis of the standard Hilbert space underlying the Dirac equation.

 Now we build a Fock space from this extended notion of particle. We identify
 \begin{equation}
     |x,a\rangle\equiv \psi^\dag_a(x)|\Omega\rangle\,,
 \end{equation}
where  $\{\psi_a(x),\psi^\dag_b(y)\}=\delta^{(4)}(x-y)\delta_{ab}$, in agreement with \eqref{eq:orthpaw} and the second quantization scheme, and $|\Omega\rangle$ is the vacuum of all the $\psi_a(x)$ operators. Notably, the Dirac field operators and the spacetime algebra \eqref{eq:stalgcont} have {\it naturally emerged}  from this scheme. We see that the single particle space corresponding to the spacetime algebra is of the PaW form, in the sense that it can be understood as ``time $\otimes$ space''. This separation does not hold for higher particle number  states, with the spacetime algebra dictating the proper structure.

Furthermore, considering that the canonical commutators yield ${\langle y,b|P_\mu|x,a\rangle=i\partial_\mu \delta^{(4)}(x-y)\delta_{ab}}$,
the matrix elements of the universe operator are
\begin{equation*}
   \langle y,b|\mathcal{J}|x,a\rangle= (i\delta_{ab}\partial_0+i \boldsymbol{\alpha}_{ba}\cdot \boldsymbol{\nabla}+\beta_{ba} m )\,\delta^{(4)}(x-y)\,, 
\end{equation*}
where all the derivatives act on $x$. This leads one directly to the corresponding Fock operator
\begin{equation}
\begin{split}
    \!\!\mathcal{J}_F&=-\int dtd^3x \,\psi^\dag(t,\textbf{x}) (i\partial_0+\boldsymbol{\alpha}\cdot \boldsymbol{\nabla}-\beta m) \psi(t,\textbf{x})\\
      &=-\int d^4x \,\bar{\psi}(x) (i\gamma^\mu \partial_\mu -m)\psi(x)\,,
\end{split}\label{JF}
\end{equation}
where we used that $\mathcal{J}_F=\int dxdy \,\langle y,b|\mathcal{J}|x,a\rangle \psi^\dag_b(y)\psi_a(x)$. Remarkably, we have proven that the second quantized universe operator \eqref{JF} of the PaW mechanism applied to a Dirac particle {\it is exactly the Dirac quantum action operator} \eqref{eq:relaction} up to an overall (irrelevant) factor.

It is also interesting to relate the diagonalization of the quantum action with the PaW notion of particle. For this purpose we define for a given mass $m$ (implicit in the dispersion relation)
\begin{equation}\label{eq:uvmodes}
 a^s(p):=\tfrac{1}{\sqrt{2E_{\textbf{p}}}} u^{s\dag}_{\textbf{p}}\psi(p),\;\; b^s(p):= \tfrac{1}{\sqrt{2E_{\textbf{p}}}}\psi^\dag(-p)v^s_{\textbf{p}}
\end{equation}
satisfying 
\begin{subequations}\label{eq:abalg}
\begin{align}
    \{a^s(p),a^{r\dag}(p')\}&=(2\pi)^4\delta^{4}(p-p')\delta_{rs}\\ \{b^s(p),\,b^{r\dag}(p')\}&=(2\pi)^4\delta^{4}(p-p')\delta_{rs}
\end{align}
\end{subequations}
for $s,r=1,2$, with all other anticommutators equal to zero and where we recall that $\psi(p)$ is the FT of $\psi(x)$ as discussed in section \ref{sec:continuumcase}.
 The proof relies on basic properties of the Dirac's spinors $u^s_\textbf{p},v^s_\textbf{p}$ and is provided in the Appendix \ref{app:actiondiag}. Therein we also show in compact notation (permitted by the new spacetime FT) that (\ref{eq:uvmodes}) corresponds to a simple bogoliubov transformation whose inverse yields
    $
        \psi(p)=\sum_s \frac{1}{\sqrt{2E_\textbf{p}}}( u^s_{\textbf{p}}a^s(p)+v^s_{\shortminus \textbf{p}}b^{s\dag}(\shortminus p))
      \,.
    $
We can also use this to write the Dirac field in spacetime as
\begin{equation}\label{eq:Diracfieldexp}
    \psi(x)\!=\!\int \!\!\frac{d^4p}{(2\pi)^4} \frac{1}{\sqrt{2E_\textbf{p}}}\sum_s\Big( u^s_{\textbf{p}}a^s(p) e^{-ipx}+v^s_{ \textbf{p}}b^{s\dag}(p)e^{ipx}\Big)
      \,.
\end{equation}
    
    Moreover, standard properties of the Dirac Hamiltonian brings the spacetime quantum action to its diagonal form:
\begin{equation}\label{eq:dqactiondiag}
    \mathcal{S}_\tau\!=\!\!\int\! \frac{d^4p}{(2\pi)^4}\, \sum_s \tau(p^0-E_{\textbf{p}}) \big(a^{s\dag}(p)a^s(p)+b^{s\dag}(p)b^s(p)\big)\,.
\end{equation}
We are actually considering the normal ordered operator with respect to $|\tilde{\Omega}\rangle\neq |\Omega\rangle$, the vacuum of all $a^s(p)$, $b^s(p)$ which, as usual, ``eliminates'' the Dirac's sea of negative energies. Notice in fact that negative $p^0$ states also have positive energy: $\int dt\, :H_D(t):=\int dt\,\int d^3p\, E_{\textbf{p}}\sum_s \{a^{s\dag}(t,\textbf{p})a^s(t,\textbf{p})+b^{s\dag}(t,\textbf{p})b^s(t,\textbf{p})\}$ is \emph{positive semidefinite}.  

In particular, states created by on-shell modes with positive $p^0$ are annihilated by $\mathcal{S}_\tau$ (or equivalently, by $\mathcal{J}_F$). This can be used  to compare this formalism with the physical states of \cite{di.19}:
consider an on-shell single particle state of the form (sums over $s,a$ are implied)
\begin{equation}
    \begin{split}
        |\Psi\rangle_e&=\int d^4p \,\delta(p^2-m^2)\theta(p^0)\sqrt{2E_\textbf{p}}\alpha_\textbf{p}a^{s\dag}(p)|\Omega\rangle\\&= \int d^4p\, \delta(p^2-m^2)\theta(p^0)\alpha_\textbf{p} u^s_{\textbf{p}a}|p,a\rangle\,,
    \end{split}
\end{equation}
 where we used \eqref{eq:uvmodes} and $|p,a\rangle \equiv \psi^\dag_a(p)|\Omega\rangle$. This state has precisely the form of the particle physical state in \cite{di.19}, defined according to the PaW condition ${\mathcal{J}|\Psi\rangle=0}$. Notice that we may replace $|\Omega\rangle\to |\tilde{\Omega}\rangle$ here. Instead, the antiparticle state of \cite{di.19} has the form 
 \begin{equation}
    \begin{split}
        |\Psi\rangle_p&=\int d^4p \,\delta(p^2-m^2)\theta(p^0)\beta_\textbf{p} (v^s_{\textbf{p}a})|-p,a\rangle\\&=\int d^4p \,\delta(p^2-m^2)\theta(p^0)\sqrt{2E_\textbf{p}}\beta_\textbf{p}b^s(p)|\Omega\rangle\,.
    \end{split}
\end{equation}
  This state does not correspond to an antiparticle state of the form $\propto b^{s\dag}(p)|\tilde{\Omega}\rangle$. Instead, it is a hole of $|\Omega\rangle$ (which is full of antiparticles, i.e. $|\Omega\rangle\equiv \prod_{p,s} b^{s\dag}(p)|\tilde{\Omega}\rangle$) thus leading to negative energies. In other words, through the spacetime field theory formalism we see that the ``first quantization antiparticles'' (defined in the PaW formalism) correspond to holes. Instead, the proper on-shell particles and antiparticles have positive $p^0$ and energy if the action and Hamiltonian are properly normally ordered. These are defined by the action of modes $a^{s^\dag}(p)$, $b^{s^\dag}(p)$ on $|\tilde{\Omega}\rangle$. We have thus recovered the usual discussion on the Dirac sea and its field theory solution within these spacetime approaches.
  One can also show that $a^{s^\dag}(E_\textbf{p},\textbf{p})$, $b^{s^\dag}(E_\textbf{p},\textbf{p})$ are the modes to be used if one is to consider ``external lines'' in scattering processes. The corresponding Feynman rules can be obtained in analogy with the discussion of section \ref{sec:yukawa}.

  Let us finally notice that just as the sp operator $\mathcal{J}$ can be generalized to second quantization, the quantum time operator $T_{\text{PaW}}=\int dtd^3x \sum_a\, t \,|t,\textbf{x},a\rangle\langle t,\textbf{x},a|$ \cite{di.19} can also be generalized to the Fock operator 
  \begin{equation}
      T_F=\int dtd^3x \sum_a\, t\, \psi^\dag_a (t,\textbf{x})\psi_a (t,\textbf{x})\,,
  \end{equation} according to the discussion above \eqref{eq:Fockop}. Interestingly, the weight assigned to a single point in spacetime, is proportional to the charge density, reproducing the mean values of PaW for single particles. On the other hand, our previous discussion on the Dirac sea is quite relevant here: if we employ the expansion \eqref{eq:Diracfieldexp}, a direct calculation using standard spinor properties (see Appendix \ref{app:actiondiag}) yields 
  \small
  \begin{equation}
      T_F=\!\int\! dt\, t \!\int \!\!\frac{d^3p}{(2\pi)^3}\sum_s \Big(a^{s \dag}(t,\textbf{p})a^{s}(t,\textbf{p})-b^{s \dag}(t,\textbf{p})b^{s}(t,\textbf{p})\Big)\,.
  \end{equation}
  \normalsize
  Notice that the antiparticles have the ``wrong'' sign. While this might be ``fixed'' by means of a particle-hole transformation for antiparticles, this would correspond to a normaly ordered $T_F$ according to the ``vacuum'' state $|\Omega\rangle$ which leads to negative energies. 
  If the more physical choice of a vacuum $|\tilde{\Omega}\rangle$ is made, the PaW time operator necessarily associates a negative time to antiparticles with respect to particles, providing an explicit realization of the Feynman-St\"ueckelber picture \cite{feynman2018theory} of antiparticles traveling backward in time.  Let us stress, however, that the notion of localization in time should now be regarded as emergent: just as in standard QFT the notion of localization in space is not uniquely defined \cite{cel.16} one might propose different time operators.

Let us now focus on the bosonic case for a Klein-Gordon particle which can be related to the action of \eqref{eq:kleingordonaction} in the Yukawa example of \ref{sec:yukawa}. We will employ a different notation from the related results presented in \cite{diaz.21}, which is more adequate to QFT.

Consider an extended particle Hilbert space \cite{dia.19} defined by the algebra ${[X^\mu,P_\mu]=i\delta^{\mu}_{\;\,\nu}}$. This space has bases $|x\rangle, |p\rangle$ which are eigenstates of the operators $X^\mu$, $P_\nu$. 
Consider now in this scheme a particle with relativistic dispersion relation so that $H=\int d^{3}p\, \sqrt{\textbf{p}^2+m^2}\,|\textbf{p}\rangle \langle \textbf{p}|$. The corresponding universe operator   $\mathcal{J}=P_0\otimes \mathbbm{1}+\mathbbm{1}\otimes H$ 
leads one to \cite{diaz.21}
\begin{equation}\label{eq:Kleingfirstq}
    \mathcal{J}_F=-\int d^4p\, \big(p_0-\sqrt{\textbf{p}^2+m^2}\, \big) a^\dag(p) a(p)\,,
\end{equation}
with
\begin{equation}
    [a(p),a^\dag(p')]=(2\pi)^4\delta^{(4)}(p-p')\,,
\end{equation}
similarly to the Dirac case (without spin). 
Notably, $\mathcal{J}_F$ is the Klein-Gordon free action in disguise. To see this we introduce the field and its conjugated (in spacetime) momentum 
\begin{equation}
    \begin{split}
        \phi(x):&=\int \frac{d^4p}{(2\pi)^4}\frac{1}{\sqrt{2 E_\textbf{p}}}\Big(a(p)e^{-ipx}+\text{h.c}\Big)\\
        \pi(x):&=-i\int \frac{d^4p}{(2\pi)^4}\sqrt{\frac{ E_\textbf{p}}{2}}\Big(a(p)e^{-ipx}-\text{h.c}\Big)
    \end{split}
\end{equation}
  satisfying the spacetime algebra \eqref{eq:stalgbosons}. A direct calculation yields 
 \fontsize{9.5}{11.5}
\begin{equation}
    \mathcal{J}_F=\!-\!\int d^4x \,\Big \{\pi(x)\dot{\phi}(x)-\frac{\pi^2(x)}{2}-\frac{(\nabla \phi(x))^2}{2}-\frac{m^2\phi^2(x)}{2} \Big \}
\end{equation}
which is precisely the quantum action of Eq.\ \eqref{eq:kleingordonaction}. It is also easy to verify that the Legendre transform emerges from the $p^0$ term in \eqref{eq:Kleingfirstq} while the rest from its Hamiltonian part.

In summary, the spacetime algebras lead to single particles with the PaW structure ``time $\otimes$ space'', an insight which holds for both  bosons and fermions. While more general states do not share this structure, there is yet another notable relation: the Fock space version of the universe operators are precisely the (free) quantum actions. While we showed this for the familiar Dirac and Klein-Gordon actions, the same holds for non-relativistic systems \cite{diaz.21} as one can easily see by applying our scheme to other scenarios. Finally, let us observe that in the final spacetime approach we do not define a physical subspace. Instead, we compute quantities according to the map established through the main Theorems of section \ref{sec:fermioniccase} (see the discussion in the Appendix \ref{app:classical}).

\subsection{The entanglement in time approach}\label{sec:entanglementintime}

As we discussed throughout the paper, and mentioned in previous work \cite{diaz.21, diaz2024spacetime, DIAZ2025170052}, the spacetime approach is the natural scenario to accommodate timelike correlations. Interestingly, it was recently shown in  \cite{MAP.25} that imposing a tensor product structure across time  can provide meaning to the recently introduced concept of timelike pseudoentropy \cite{tak.23, har.23, nar.22, chu2023time}, previously defined by analytic continuations from space to time and applied in the context of the holographic dS-CFT correspondence  (and in time-dependent spacetimes in AdS-CFT).

In this section, we explicitly show that the approach developed in \cite{MAP.25} is a particular case of our scheme for bosonic systems. In order to show this, let us briefly recall that the fundamental quantity in the ``entanglement in time approach'' is the operator 
\begin{equation}
    T_{AB}:=J (\rho \otimes \mathbbm{1}_B)
\end{equation}
with $J=\sum_{i,j}^{\dim( h)} U^\dag |i\rangle \langle j| U\otimes |j\rangle \langle i|$ the Jamiolkowski state
associated with the quantum channel $\text{Ad}_U(.)=U^\dag \,.\, U$. 
This operator is defined in such a way to obtain Wightman functions, namely
\begin{equation}
\begin{split}
    {\rm Tr}[T_{AB}O_A\otimes O_B]&={\rm tr}[\rho O_A O_B(t)]\\
       {\rm Tr}[T^\dag_{AB}O_A\otimes O_B]&={\rm tr}[\rho  O_B(t) O_A]
\end{split}    
\end{equation}
with $O_B(t)=U^\dag O_B U$. These relations suggest a clear connection with the quantum action (see Lemma \ref{th:lemmabosonmap} proved in \cite{diaz2024spacetime,DIAZ2025170052}) and the concept of state we discussed in section \ref{sec:states}. As a matter of fact, we can state the following result (proved in the Appendix \ref{app:choi} by direct evaluation).
\begin{theorem}\label{th:entanglement}
For bosons and $N=2$ the operator $T_{AB}$ is given by
    \begin{equation}
        T_{AB}=(\rho_0 e^{i\tilde{\mathcal{S}}})^\dag\,,\quad T^\dag_{AB}=\rho_0 e^{i\tilde{\mathcal{S}}}\,.
    \end{equation}
\end{theorem}
In other words, the operator $T_{AB}$ proposed in \cite{MAP.25} to define timelike entanglement is a particular case of our notion of spacetime state discussed in section \ref{sec:states}. When $A,B$ correspond to subsystems one simply needs to consider partial traces of spacetime states. While the authors of \cite{MAP.25} focus on tensor products in time (which for fermions give rise to non-local operators), one can employ our results to immediately define timelike entanglement for fermions. The analogous of $T_{AB}$ is clearly given by our Theorem \ref{th:fermistates} (with the SWAP operator replaced by Eq.\ \eqref{eq:fswap}, in agreement with the discussion of the section).

The case of an arbitrary number of time slices and even continuum time follow then by our scheme. For many such slices the relation between generalized states and the action of classical mechanics is easily revealed (when the quantum system has a classical analogue).
Let us also notice that since (for arbitrary $N$)
\begin{equation}
\begin{split}
     {\rm Tr}\Big\{\left(\rho_0 e^{i\tilde{\mathcal{S}}}-(\rho_0 e^{i\tilde{\mathcal{S}}})^\dag\right) O^{(A)}_{t_1}O^{(B)}_{t_2}\Big\}\\
     ={\rm tr}\big\{\rho\, [O^{(A)}_{H}(t_1),O^{(B)}_{H}(t_2)]\big\}\,,
\end{split}
\end{equation}
if one is only interested in two subregions $A, B$ of spacetime (e.g. in relativistic QFT as in the examples we discussed in sections \ref{sec:continuumcase}, \ref{sec:fbinteractions} and in \cite{diaz2024spacetime}), it is sufficient to consider partial traces over spacetime of the generalized states. As a consequence, generalizations of the causality bounds proposed in \cite{MAP.25} can be obtained from the imagitivity  $||{\rm Tr}_{\bar{A}\bar{B}}\big[\rho_0 e^{i\tilde{\mathcal{S}}}-(\rho_0 e^{i\tilde{\mathcal{S}}})^\dag\big]||_p$, where $||.||_p$ denotes the Schatten $p$-norm and ${\rm Tr}_{\bar{A}\bar{B}}$ denotes partial trace over regions of spacetime outside $A, B$. We remark that spacelike traces (or traces with respect to entangled subsystems) yield spacetime states with a rich structure rendering general imagitivities interesting and non-trivial quantities when multiple slices in time and in space are considered.  Considering Eq.\ \eqref{eq:fermioniccommheis} one might define a fermionic imagitivity as well.

Let us also mention a recent proposal \cite{Guo.25} to generalize \cite{MAP.25} to an arbitrary number of slices. One can show, by a similar procedure to the proof of Theorem \ref{th:entanglement}, that the spacetime density matrix proposed therein is once again given by $(\rho_0 e^{i\tilde{\mathcal{S}}})^\dag$, now for $N$ arbitrary (and bosons), as it follows from the structure of $e^{i\epsilon\mathcal{P}}$ described in section \ref{sec:prelim} (the concatenation of SWAP operators).
It is also clear that one can replace $\rho_0\to |\psi\rangle_0\langle \phi|$ to describe transitions as the ones in \cite{Guo.25} (the example of a single particle propagator with $|\psi\rangle\langle \phi|\to |q\rangle\langle q'|$ has been considered in detail in \cite{DIAZ2025170052}). Interestingly, the connections with the Schwinger–Keldysh PI \cite{schwinger1961brownian} reported by the author clarify the difference between $e^{i\mathcal{S}}$ and $e^{i\tilde{\mathcal{S}}}$: the first one is associated (upon evaluation of the corresponding trace in proper basis \cite{DIAZ2025170052}) with Feynman PIs while the second to Schwinger–Keldysh PIs. For this reason we employed the second to define states, so that expectation values at a given ``in'' time instead of propagators from ``in'' to ``out'' states are obtained (the difference is explicit if one compares Eqs. \eqref{eq:twopointpropagator} with  \eqref{eq:meanvaluecor}). In particular, one can show that by duplicating the number of time slices ($N\to 2N$) one can recover $e^{i\tilde{\mathcal{S}}}$ as a partial trace over half the times of another operator whose trace gives the Schwinger–Keldysh PIs (we stress that in our approach the PIs, rather than an independent formalism, can be seen as emergent; see the Appendices \ref{app:PIs}, \ref{app:consistenth} and \cite{DIAZ2025170052}).

\section{Conclusions}\label{sec:concl}

We provided a general framework to formulate QM treating time on equal footing with other quantum numbers. In order to do so we identified the algebra defining quantum operators, and ensuing Hilbert space representations, as the key ingredient to be promoted to include time itself. In this way, we generalized our previous work \cite{diaz2024spacetime} on bosonic and relativistic systems to arbitrary finite dimensional systems including fermions. This allowed us to introduce a fermionic QA and to establish a map between spacetime correlators, defined in the extended formalism, and evolution-related quantities of standard QM. Moreover, we have shown that this map admits a PI-like interpretation that does not make use Grassmann variables,  and that generalizes recent results \cite{DIAZ2025170052} for bosonic systems. The case where both bosons and fermions are present and can interact has also been considered.

We also showed that the formalism contains standard QM in a natural way:
it is sufficient to consider partial traces over all time slices except one to recover standard quantum states evolving according with unitary evolution. The parameter of evolution is dictated by the time slice index. In this way, the asymmetry in the axioms of standard QM posed in section \ref{sec:intro} (and depicted in Figure \ref{fig:intro}) has been fully addressed. Moreover, the objects one is tracing from can be interpreted as generalized states. These generalized states codify not only the complete information about the system at a given time but also all of its evolution and temporal correlators. As a matter of fact, one can extract all (space) time correlation functions from these states by simply considering \emph{static} expectation values on the extended Hilbert space. In this mathematically precise sense, the formalism replaces evolutions with correlations.

These correlations also admit a generalized purification leading to an interpretation
of timelike correlators in terms of weak values. Moreover, we have shown that the associated entropy allows one to introduce a stationary quantum action principle for variations along arbitrary operators in the extended Hilbert space. The principle is defined as an extremal condition for the entropy with a fixed quantum action mean value and with $\hbar$ playing the role of the Lagrange multiplier. The principle includes both separable-in-time and entangled-in-time variations, meaning that it's only accessible trough the extended formalism.  This principle suggests a novel angle to understand what separates quantum and classical mechanics, an interesting but subtle perspective that lies  beyond the scope of the manuscript. Considerations related to the classical limit (without the quantum action principle) can be also found in \cite{diaz2024spacetime} where the spacetime approach to classical mechanics has also been presented.

We have also explored connections with the PaW mechanism \cite{PaW.83}, showing that, for quantum fields, our approach can be interpreted as a second quantized version of the PaW scheme. In other words, if a QFT is quantized following our formalism, the single particle excitations correspond to PaW particles \cite{di.19,dia.19}. Moreover, the quantum action for free fields, including the fermionic one is the second quantization of the universe operator defining the dynamics in the PaW mechanism. 

Furthermore, we have rederived recent proposals \cite{MAP.25,Guo.25} that aim to provide a basis for timelike pseudoentropies \cite{tak.23, har.23, nar.22, chu2023time} showing that they are particular cases of the formalism we are presenting. As a consequence, the considerations in \cite{MAP.25, Guo.25}, when combined with our results, strongly support our previous claims on the importance of eliminating the asymmetries between space and time in the postulates of QM in order to gain further insights on the emergent nature of spacetime. Let us in fact recall that for CFTs \cite{MAP.25} provides a rigorous meaning to the entanglement pseudo-entropies proposed in \cite{tak.23} in the context of the dS-CFT correspondence. 
Other interesting remarks on the relation of spacetime density matrices 
and timelike entanglement presented in \cite{MAP.25,Guo.25} (in the sense of pseudoentropies) can be readily applied to our approach. Conversely, the general and explicit structure presented in the current manuscript (and the previous applications to QFT \cite{diaz2024spacetime} and PIs \cite{DIAZ2025170052}), together with the properties explored in section \ref{sec:implications}, can be applied to tackle the recent problems  posed in relation to timelike entanglement and to explore new related avenues. In particular, the explicit form of the spacetime states and their connection with the action of classical mechanics suggest a clear pathway to study the scaling of entanglement between spacetime regions (in both non-relativistic QM and QFTs). Let us also recall that the type of entanglement in time provided by standard QM is very structured and limited, as elucidated by the stationary quantum action principle and the numerical examples we presented in section \ref{sec:qaprinciple}.

It is also worth stressing that the spacetime approach is not simply mapping space to time, but applying the same structure to both: In the spacetime formulation space and time are indistinguishable at the Hilbert space level. What separates spacelike from timelike intervals, and determines the causal structure of a given physical theory, is the structure of spacetime states which in turn is defined by the quantum action operator. This structure can also be introduced by means of the generalized purifications which also appears associated to the dS/CFT correspondence and holographic time-like entanglement \cite{tak.23, har.23, nar.22, chu2023time} and with the concept of weak values \cite{yak.88, dres.14}. 
The same structure defines spacelike and timelike entanglement.

It is also natural to explore possible connections to other recent proposals whose motivation is the same type of space-time asymmetry we have treated here \cite{fit.15,ho.17, cot.18, giovannetti2023geometric, parzygnat2023svd, fullwood2024operator}. There is, however, a key factor setting our scheme apart: we focused on spacetime correlators (or Wightman correlation functions in the context of QFT) rather than in measurements and/or expectation values at different times, a typical subject of these other approaches (see also Appendix \ref{app:consistenth}). Our choice was motivated by the PI formulation, and it led us to a clear unification of space and time correlators, intimately related to PIs. At the same time, as we are working in a canonical approach, it is actually straightforward to introduce general quantum channels in the formalism, including measurements: one can include ancillas within the formalism itself, and since general unitary evolution has been treated, one can accommodate e.g. Kraus operators \cite{nie.01} within the definition of the QA (we also recall that the measurement postulates of QM can, in principle, be recovered from a Hilbert space structure, unitary dynamics, and basic assumptions alone \cite{zurek2003environment, zurek2013wave}). These considerations open very interesting perspectives as they show that one can study fundamental problems of QM, as those posed in the literature above, within a framework that incorporate both the advantages of the PI formulation and quantum information.

Let us conclude our manuscript with the following remark: with a spacetime canonical approach fully developed, we are not only providing (by construction) an answer to previously raised questions about the possibility of removing the space--time asymmetry intrinsic to quantum mechanics, but also establishing a rigorous and intuitive framework in which novel notions and questions become meaningful. In particular, as spacetime states satisfy a variational principle (Sec.~\ref{sec:qaprinciple}) and, moreover, they are operators acting on a tensor product-in-time Hilbert space (or the corresponding fermionic construction), one can explore new numerical, possibly variational schemes based on tensor-network methods, where the computational complexity may be naturally tied to timelike correlations. From a foundational perspective, we note that the space of possible quantum actions and spacetime states is, in principle, larger than that of standard QM (e.g., allowing stronger entanglement across time). This both motivates a full characterization of the subset that reproduces ordinary QM and, more speculatively, suggests possible generalizations. In this setting, one may ask whether physical constraints such as causality, positivity, and probability conservation are tight enough to single out standard QM, or whether more exotic but consistent physical theories are possible.

\section*{Acknowledgements}
The authors would like to thank J. M. Matera, Paolo Braccia, Inés Corte, Dario Cafasso and Marco Cerezo for fruitful discussions. N. L. D. was supported by the Center for Nonlinear Studies at Los Alamos National Laboratory (LANL) and by the Laboratory Directed Research and Development (LDRD) program of LANL under project number 20230049DR.
		We also acknowledge support from CONICET (N.L.D.) and  CIC (R.R.) of Argentina.  Work supported  by CONICET PIP Grant 11220200101877CO.

\appendix

\section{Theorems from section \ref{sec:fermioniccase}: Proofs and additional details}\label{app:theorems}

\subsection{Theorem I}

Before giving a formal proof, let us notice that there is a pictorial way of understanding the Theorem. We will denote by $|\circ\rangle$ and $|\bullet\rangle$  the empty and occupied state respectively of a single fermionic mode. We  examine  first the relation ${\rm Tr}[Pe^{i\epsilon \mathcal{P}}]={\rm tr}[\mathbbm{1}]$. 
It is clear that if we compute the trace in the product basis the only states that contribute are $|\circ \circ \dots \circ\rangle$ and $|\bullet \bullet \dots \bullet\rangle$ since they are the only product-in-time eigenstates of $e^{i\epsilon \mathcal{P}}$ (here the different modes correspond to time). The all-empty state has clearly an eigenvalue $1$. Instead, $e^{i\epsilon \mathcal{P}}|\bullet \bullet \dots \bullet\rangle=P|\bullet \bullet \dots \bullet\rangle$ as the last mode is full, yielding a minus sign, and then the corresponding creation operators has to be moved through all the other ones. Thus, the trace is equal to $\prod_j 2={\rm tr}[\mathbbm{1}]$.

\begin{figure}[t!]
\centering
\includegraphics[width=0.95\linewidth]{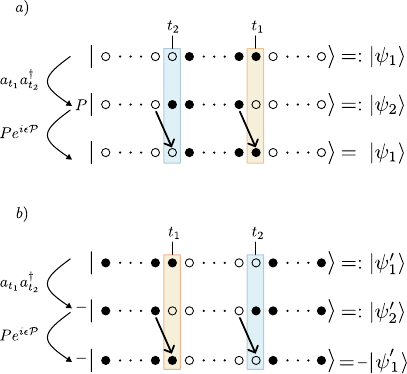}
\caption{\textbf{Schematic proof of the two point contraction relation of Theorem I.}  We show schematically the only states that contribute to the trace ${\rm Tr}[Pe^{i\epsilon\mathcal{P}}a_{t_1}a_{t_2}^\dag]$ in the cases of $t_1>t_2$ (panel a)) and $t_2>t_1$ (panel b)). The black dots indicate a full mode while the white dots indicate an empty mode with all modes correpsonding to a different time.}
\label{fig:theorem1}
\end{figure}

Consider now the two point contraction, depicted in figure \ref{fig:theorem1}. Let us first assume $t_1>t_2$. It is clear that the only states that do not vanish after the action of $a_{t_1}a_{t_2}^\dag$ have the mode $t_2$ empty and $t_1$ full (we ignore the other mode indices). Is is easily seen that this implies that the only contribution comes from the state $|\psi_1\rangle$ which has all modes empty 
except for 
$t_2<t\leq t_1$. 
In fact, after acting with $a_{t_1}a_{t_2}^\dag$ upon the state the mode in $t_2$ is full and $t_1$ is empty. When we translate in time we need an empty mode coming from $t_2-1$ and a full one coming $t_1-1$. But if $t_2-1$ is empty we also need $t_2-2$ empty and so on. A similar argument shows the previous claim for $t_2<t<t_1$ and $t>t_1$. In addition, one can see that $|\psi_2\rangle:=a_{t_1}a_{t_2}^\dag|\psi_1\rangle=P |\circ\dots \circ \bullet_{t_2} \bullet \dots \bullet \circ_{t_1} \circ  \dots \circ\rangle$ 
as one can see by noting that $a_{t_1}$ must pass through the same number of creation operators as contained in $|\psi_1\rangle$. Finally,  $e^{i\epsilon \mathcal{P}}|\psi_2\rangle=P|\psi_1\rangle$
since the last mode is empty in $|\psi_2\rangle$. This means that $\langle \psi_1|Pe^{i\epsilon \mathcal{P}}a_{t_1}a^\dag_{t_2}|\psi_1\rangle=\langle \psi_1|\psi_1\rangle=1={\rm tr}[a a^\dag]$. 

A similar argument holds for the case $t_1<t_2$  with just the state $|\psi'_1\rangle= |\bullet\dots \bullet \bullet_{t_1} \circ \dots \circ \circ_{t_2} \bullet  \dots \bullet\rangle$ (see figure \ref{fig:theorem1}) contributing. This time $|\psi'_2\rangle:=a_{t_1}a^\dag_{t_2}|\psi'_1\rangle, $ with $|\psi_2'\rangle=-|\bullet\dots \bullet \circ_{t_1} \circ \dots \circ \bullet_{t_2} \bullet  \dots \bullet\rangle$ since $a^\dag_{t_2}$ is to be shifted through all creation operators with $t\leq t_1$ while $a_{t_1}$ through all $t<t_1$. Instead, $e^{i\epsilon \mathcal{P}}|\psi'_2\rangle=-P|\psi'_1\rangle$  since the last mode is full yielding a minus sign and then $a_1^\dag$ is shifted through all the other full modes.

Let us now provide the formal demonstration.  
Consider the equality  ${\rm Tr}[Pe^{i\epsilon \mathcal{P}}]={\rm tr}[\mathbbm{1}]$ first.

\emph{Proof.}
 By noting that $$Pe^{i\epsilon\mathcal{P}}=\exp\{i\sum_{n,j}(\pi+\epsilon\omega_n)a^\dag_{nj}a_{nj}\}\,,$$
 where $
\epsilon\omega_n=\frac{2\pi}{N}(n+1/2)$, one can immediately show, using elementary fermionic properties, that
\begin{equation}
    {\rm Tr}[Pe^{i\epsilon \mathcal{P}}]=\smallprod_{n,j} [1+e^{i(\pi+\epsilon\omega_n)}]=\smallprod_{n,j} [1-e^{i(2n+1)\pi/N}]\,.
\end{equation} 
In order to evaluate this product consider the polynomial $p(z)=z^N+1$ whose roots are $z_n=e^{i(2n+1)\pi/N}$ as one can readily verify. This implies $p(z)=\prod_n [z-z_n]$. We see that ${\rm Tr}[Pe^{i\epsilon \mathcal{P}}]=\prod_j p(1)=\prod_j 2={\rm tr}[\mathbbm{1}]$ (the dimension of the conventional Hilbert space).  

\hfill$\square$

\vspace{0.35cm}

Consider now the claim ${\rm Tr}[Pe^{i\epsilon\mathcal{P} }a_{t_1i}a^\dag_{t_2j}]= {\rm tr}[\hat{T}a_i a^\dag_j]$.
\begin{proof}
 By using the fact that the operator $\mathcal{P}$ and the parity operator $P$ are both quadratic we have
 \fontsize{9.5}{11.5}
    \begin{equation}
 \begin{split}
    \!\frac{{\rm Tr}[Pe^{i\epsilon\mathcal{P} }a_{t_1i}a^\dag_{t_2j}]}{{\rm Tr}[Pe^{i\epsilon\mathcal{P} }]}&\!=\!
     \frac{{\rm Tr}\Big[\!\exp\{i\sum\limits_{n,i} (\epsilon \omega_n+\pi)a^\dag_{ni}a_{ni}\}a_{t_1i}a^\dag_{t_2j}\Big]}{{\rm Tr}\Big[\!\exp\{i\sum\limits_{n,i} (\epsilon \omega_n+\pi)a^\dag_{ni}a_{ni}\}\Big]}\\&\!=\!\frac{\delta_{ij}}{N}\sum_{n=0}^{N-1} \frac{e^{-i\epsilon\omega_n(t_1-t_2)}}{1-e^{i\epsilon \omega_n}}\,,
 \end{split}
 \end{equation}
 \normalsize
  where in the last equality we used the Fourier expansion of the creation(annihilation) operators. 
  We can now evaluate this sum exactly by noting that
  \fontsize{9.5}{11.5}
  \begin{equation}
      \begin{split}
          \sum_{n=0}^{N-1} \frac{e^{-i\epsilon\omega_n(t_1-t_2)}}{1-e^{i\epsilon \omega_n}}&=\frac{1}{2}\sum_{n=0}^{N-1} \sum_{t=0}^{N-1}e^{-i\epsilon \omega_n (t_1-t_2-t)}\\&=\sum_{t=0}^{N-1}\frac{e^{-i\frac{\pi}{N}(t_1-t_2-t)}}{2}\frac{1-e^{-i2\pi (t_1-t_2-t)}}{1-e^{-i\frac{2\pi}{N}(t_1-t_2-t)}}
      \end{split}
  \end{equation}
  \normalsize
where the last equality holds for the cases $t=t_1-t_2$ and $t=t_1-t_2+N$ as a limit (with the argument of the sum being strictly equal to $1/N)$. Now, for $t_1, t_2, t$ integers we obtain 
  \begin{equation}
      \sum_{n=0}^{N-1} \frac{e^{-i\epsilon\omega_n(t_1-t_2)}}{1-e^{i\epsilon \omega_n}}=\frac{N}{2}\sum_{t=0}^{N-1}e^{-i\frac{\pi}{N}\Delta_t}\delta_{\Delta_t,0\,{\rm mod}\, N}\,,
  \end{equation}
  with $\Delta_t:=t_1-t_2-t$.
For $t_1-t_2\geq 0$, and   since $0\leq t\leq N-1$, only $\delta_{\Delta_t,0}$ contributes so that $$ \sum_{n=0}^{N-1} \frac{e^{-i\epsilon\omega_n(t_1-t_2)}}{1-e^{i\epsilon \omega_n}}=\frac{N}{2}\,.$$ Instead, for $t_1-t_2<0$ the 
$\delta_{\Delta_t,-N}$ is the only non-vanishing term yielding $ \sum_{n=0}^{N-1} \frac{e^{-i\epsilon\omega_n(t_1-t_2)}}{1-e^{i\epsilon \omega_n}}=-N/2$. We conclude that 
\begin{equation}
    \frac{{\rm Tr}[Pe^{i\epsilon\mathcal{P} }a_{t_1i}a^\dag_{t_2j}]}{{\rm Tr}[Pe^{i\epsilon\mathcal{P} }]}=\frac{\delta_{ij}}{2}\sgn(t_1-t_2)\,,
\end{equation}
for $t_1,t_2=0,\ldots,N-1$, where here  $\sgn(0)=1$.

Let us also notice that   $\theta_t=\frac{1}{N}\sum_{n=0}^{N-1} \frac{e^{-i\epsilon\omega_n(t_1-t_2)}}{1-e^{i\epsilon \omega_n}}$ is the discrete Fourier expansion of the step function 
  $\theta_t:=\left\{^{\;\,\;1/2\;\;t\geq 0}_{-1/2\,\;t<0}\right.$ 
  for $-N\leq t\leq N-1$, thus taking $N$ positive and $N$ negative values. This is easily seen by noting that $\epsilon\omega_n
  =\frac{2\pi}{2N}(2n+1)$, with the Fourier coefficients of $\theta_t$ corresponding to even indices vanishing.

  Now, since ${\rm Tr}[Pe^{i\epsilon \mathcal{P}}]={\rm tr}[\mathbbm{1}]$, while ${\rm tr}[a_i a^\dag_j]={\rm tr}[ a^\dag_j a_i]=\delta_{ij}{\rm tr}[\mathbbm{1}]/2$ we obtain the desired relation.

\end{proof}

\subsection{Theorem \ref{th:mapwick}}
Here we prove the claim of Theorem \ref{th:mapwick}. Notice first that any operator of $\mathcal{H}$ ($h$) is a polynomial in $a_{ti},a^\dag_{tj}$ ($a_i,a_j$). This means that if we assume that
\begin{equation} \label{eq:appth2}
    {\rm Tr}[Pe^{i\epsilon \mathcal{P}}\smallprod_l \psi^{(l)}_{t_l}]={\rm tr}[\hat{T}\smallprod_l \psi^{(l)}(t_l)]\,,
\end{equation}
where $\psi^{(l)}_{t_l}=a_{t_l i_l},a^\dag_{t_l,i_l}$ (an annihilation or creation operator) and $\psi^{(l)}=a_{i_l}, a^\dag_{il}$, i.e. we consider only the product of creation (annihilation) operators, the general Theorem follows (the ``constant'' case is already covered by Theorem \ref{th:mapeip}). We now prove \eqref{eq:appth2}.

\begin{proof}
    It is well-known that Wick's theorem applied to gaussian operators yields a compact expression in terms of Pfaffians for the mean values of a product of annihilation (creation) operators (or equivalently to a product of Majorana operators \cite{jozsa2008matchgates}). We can use it to write
    $
        {\rm Tr}[Pe^{i\epsilon \mathcal{P}}\prod_l \psi_{t_l}^{(l)}]=\text{Pf}(C)$, where $\text{Pf}$ denotes the Pfaffian of the matrix $C$ defined as $C_{ij}:={\rm Tr}[P e^{i\epsilon \mathcal{P}}\psi^{(i)}_{t_i}\psi^{(j)}_{t_j}]$ for $i<j$ and $C_{ji}=-C_{ij}$.
Let us assume first that the product of operators is already time-ordered. We can now write
\begin{equation}\label{eq:apppfaff}
     {\rm Tr}\Big[Pe^{i\epsilon \mathcal{P}}\smallprod_l \psi_{t_l}^{(l)}\Big]=\text{Pf}(C)={\rm tr}\big[\,\smallprod_l \psi^{(l)}(t_l)\big]
\end{equation}
where we used that the two-point contractions defining $C$ are related with two-point contraction in $h$ by Theorem  \ref{th:mapeip}. The last equality is again a consequence of Wick's theorem, now applied in $h$ to the identity operator. 
On the other hand, if the product is not time-ordered we can simply write ${\rm Tr}[Pe^{i\epsilon \mathcal{P}}\prod_l \psi_{t_l}^{(l)}]=(-1)^\gamma{\rm Tr}[Pe^{i\epsilon \mathcal{P}}\prod'_l \psi_{t_l}^{(l)}]$ for $\prod_l'$ the time-ordered product and $(-1)^\gamma$ a sign which takes into account how many permutations we needed. Now, our previous reasoning of \eqref{eq:apppfaff} yields 
\fontsize{9.5}{11.5}
$$(-1)^\gamma{\rm Tr}[Pe^{i\epsilon \mathcal{P}}\sideset{}{'}\smallprod_l \psi_{t_l}^{(l)}]=(-1)^\gamma{\rm tr}[\sideset{}{'}\smallprod_l \psi_{t_l}^{(l)}]={\rm tr}[ \hat{T}\smallprod_l \psi_{t_l}^{(l)}]\,,$$
\normalsize
with the last equality holding because $(-1)^\gamma$ is precisely the extra sign appearing when relating the two products, namely $\prod'_l \psi_{t_l}^{(l)}=(-1)^\gamma\hat{T}\prod_l \psi_{t_l}^{(l)}$. Notice that this holds even for repeated time indices, since in that case no rearrangement is needed in on the l.h.s. in agreement with the definition of $\hat{T}$.

\end{proof}

It is interesting to consider an additional remark regarding our proof: if one assumes the thermal Wick's theorem (applied in the limit of infinite temperature), then one can already conclude the proof by claiming that $\text{Pf}(C)={\rm tr}[\hat{T}\prod_l \psi^{(l)}(t_l)]$, regardless of the order in time used to defined the contractions. In fact, the two-point contractions carry $\hat{T}$ as well, according to Theorem \ref{th:mapeip}. However, our proof does not assume this. On the contrary, the thermal Wick's theorem arises now as a corollary by running our ``map'' in reverse: now that we have proven Theorem \ref{th:mapwick}, one can write ${\rm tr}[\hat{T}\prod_l \psi^{(l)}(t_l)]={\rm Tr}[P e^{i\epsilon \mathcal{P}}\prod_l \psi^{(l)}_{t_l}]=\text{Pf}(C)$, which is a Pfaffian of time-ordered-two-point contractions in $h$ (Theorem \ref{th:mapeip}). If one is interested in thermal states at finite temperature, one can use $\mathcal{S}$ instead (or include a thermal gaussian state $\rho$ in the initial slice) and use this same line of reasoning (see Theorem \ref{th:correlations} and comments below). The important insight is that the gaussian-like properties of $\hat{T}$ on $h$ are now obvious since
in $\mathcal{H}$ its action is indeed associated with actual gaussian operators.

\subsection{Theorem \ref{th:eisandeip}}\label{app:Atheorem3}
In this section we will prove that 
\begin{equation}\label{eq:eiseipsappendix}
    e^{i\mathcal{S}}=U_0(T)\mathcal{V}^{-1} e^{i\epsilon\mathcal{P}}\mathcal{V}
\end{equation} with 
\begin{equation}\label{eqapp:V}
    \mathcal{V}^{-1}=\prod_t U(\epsilon t)
\end{equation} and 
\begin{equation}
e^{i\mathcal{S}}=e^{i\epsilon \mathcal{P}}\prod_t U[\epsilon(t+1),\epsilon t]\,.
\end{equation} Theorem \ref{th:eisandeip} corresponds to the case of a time-independent and hermitian Hamiltonian, i.e. $U(\epsilon t)=e^{-i\epsilon t H}$.  
We will first establish the equivalence between this result and the following expression:
\begin{equation}\label{c1}
    e^{-i\epsilon\mathcal{P}}\mathcal{V}^{-1} e^{i\epsilon\mathcal{P}}=U^{-1}_{N-1}(T) \smallprod_t U_t[\epsilon(t+1),\epsilon t]\mathcal{V}^{-1}\,. 
\end{equation}
\begin{proof}
The proof of the equivalence follows immediately by rewriting (\ref{c1}) as $$\mathcal{V}^{-1} e^{i\epsilon\mathcal{P}}\mathcal{V}=e^{i\epsilon\mathcal{P}}U_{N-1}^{-1}(T)\smallprod_t U_t[\epsilon(t+1),\epsilon t]\,.$$ 
Now, considering that $e^{i\epsilon\mathcal{P}}U_{N-1}^{-1}(T)e^{-i\epsilon\mathcal{P}}=U^{-1}_0(T)$ (the anti-periodic boundary conditions do not yield minus signs since $U$ is parity preserving) we can write $$U_0(T)\mathcal{V}^{-1} e^{i\epsilon\mathcal{P}}\mathcal{V}=e^{i\epsilon\mathcal{P}}\smallprod_t U_t[\epsilon(t+1),\epsilon t]=e^{i\mathcal{S}}\,,$$ which is precisely Eq.\ \eqref{eq:eiseipsappendix} of the main body. 
\end{proof}

The proof of Eq.\ \eqref{eq:eiseipsappendix} in the main body now reduces to proving (\ref{c1}).

\begin{proof}
The action of the translation operator on $\mathcal{V}^{-1}=\prod_{t=0}^{N-1}U_t(\epsilon t)$ in the left-hand side of (\ref{c1}) yields
\begin{align}\label{c2}
   e^{-i\epsilon\mathcal{P}}\smallprod_{t=0}^{N-1}U_t(\epsilon t) e^{i\epsilon\mathcal{P}}&=\smallprod_{t=0}^{N-1}U_{t-1}(\epsilon t)\nonumber=\smallprod_{t=0}^{N-2}U_t[\epsilon(t+1)]\nonumber\\&=U_{N-1}^{-1}(\T)\smallprod_{t=0}^{N-1}U_t[\epsilon(t+1)] 
\end{align}
where we used $U(0)=\mathbbm{1}$, $T=N\epsilon$. Notice also that $[U_t(\epsilon t_1),U_{t'}(\epsilon t_2)]=0$ for parity preserving evolution. On the other hand,
\begin{align}\label{c3}
    \smallprod_{t=0}^{N-1}U_t[\epsilon(t+1)]\mathcal{V}&=\smallprod_{t=0}^{N-1}U_t[\epsilon(t+1)]\smallprod_{t=0}^{N-1}U^{-1}_t(\epsilon t)\nonumber\\&=\smallprod_{t=0}^{N-1}U_t[\epsilon(t+1)]U^{-1}_t(\epsilon t)\nonumber\\&=\smallprod_{t=0}^{N-1}U_t[\epsilon(t+1),\epsilon t]\,.
\end{align}
By multiplying (\ref{c2}) on the right by $\mathcal{V}\mathcal{V}^{-1}=\mathbbm{1}$ and using (\ref{c3}) we recover (\ref{c1}).
\end{proof}

\subsection{Remarks on the remaining theorems}

Having completed the proofs of the main theorems in Section \ref{sec:fermioniccase}, we now provide comments on the remaining results. Unlike the previous theorems, which required full proofs, these results and theorems have already been established in the main text or follow by simple computations of main text results. For completeness, in this section we recall the main text discussions and add additional comments enabled by the details provided by the proofs of the previous subsection.

\vspace{0.1cm}

\emph{Lemma \ref{lemmaevolv}}. As described in the main, Eq.\ \eqref{eq:commutingH} is a direct consequence of $H$ containing an even
number of creation (annihilation) operators for it to be
an admissible fermionic Hamiltonian. Since ladder operators of different time-slices anti-commute, an even number commutes. This holds for a general ``evolution'' operator $U_t$ preserving parity, even if the evolution is imaginary, thus also implying \eqref{eq:heisproduct} for $\mathcal{V}$ defined in \eqref{eqapp:V}. 

\vspace{0.1cm}

\emph{Theorem \ref{th:correlations}}. Theorem \ref{th:correlations} is a direct consequence of the results of Theorem \ref{th:mapwick}, Theorem \ref{th:eisandeip} and of Lemma \ref{lemmaevolv}, with the operator $\mathcal{V}$, whose action relates $\mathcal{S}$ with $\mathcal{P}$, yielding the evolved operators. Notice that this holds for the more general definition of $\mathcal{V}$ provided in Eq.\ \eqref{eqapp:V} (see also statements below this equation) and allowing for imaginary evolution.

\vspace{0.1cm}

\emph{Corollary \ref{cor:heis}}. This corollary is just an application of Theorem 4 where $\smallprod_l O^{(l)}_{t_l}=O_t \,|\psi\rangle_0 \langle \psi|$ with the initial state $|\psi\rangle_0 \langle \psi|=O^{(0)}_0$. We recall that the notation $\tilde{\mathcal{S}}$ has been introduced in Eq.\ \eqref{eq:tildeaction}. For more general $\mathcal{V}$ (Eq.\ \eqref{eqapp:V}) it can be defined as  $e^{i\tilde{\mathcal{S}}}:= U_0(T)e^{i\mathcal{S}}=\mathcal{V}^\dag e^{i\epsilon\mathcal{P}}\mathcal{V}$ leading, in particular, to the time-dependent version of the Corollary.

\vspace{0.1cm}

\emph{Theorem \ref{th:diracprop}}. The Theorem is proven in the main body by direct evaluation.

\vspace{0.1cm}

\emph{Theorem \ref{th:diracprops}}. The Theorem is a direct consequence of Theorem \ref{th:diracprop} and Wick's theorem as explained in the main text.
\vspace{0.1cm}

\section{Spacetime states and quantum principle of stationary action}\label{app:choi}
In this section, we develop the necessary tools to prove the results of section \ref{sec:states}. While the use of the Choi-isomorphism is standard for bosons, we develop here a generalization for fermions. We finally prove the variational quantum action principle.  

\subsection{Bosonic spacetime states}
Let us begin our discussion by stating a few useful facts related to the Choi's isomorphism.

Consider a generic system and an environment with an isomorphic Hilbert space. The maximally entangled state $|\phi^+\rangle\rangle=\sum_{i}|i\rangle_S \otimes |i\rangle_E$ allows one to introduce a set of useful relations: given an operator $A\equiv A_S$ acting on the system, we can write $|A\rangle\rangle=A_S\otimes \mathbbm{1}_E |\phi^+\rangle\rangle$. In particular $|\phi^+\rangle\rangle\equiv |\mathbbm{1}\rangle\rangle$. Then one can straightforwardly prove that
\begin{align}\label{eqapp:choirelations1}
    \langle \langle B|A\rangle\rangle&={\rm tr}[B^\dag A] \\
     \langle \langle B|O|A\rangle\rangle&={\rm tr}[B^\dag O A]\label{eqapp:choirelations2}\\
    AB^\dag&={\rm tr}_E[|A\rangle\rangle\langle\langle B|]\label{eqapp:choitrace}
\end{align}
where all the operators $A,B,O$ act on the system.

Using these tools one can write the standard purification of a density matrix as $|\sqrt{\rho}\rangle\rangle$, namely 
$
    \rho={\rm tr}_E[|\sqrt{\rho}\rangle\rangle\langle\langle \sqrt{\rho}|]
$
which is a direct consequence of \eqref{eqapp:choitrace}. In particular, $\rho=e^{-K}/Z_K$ with $Z_k={\rm tr}[e^{-K}]$ leads to 
\begin{equation}
    \rho= {\rm tr}_E \left(\frac{|e^{-K/2}\rangle\rangle \langle \langle e^{-K/2}|}{\langle \langle e^{-K/2}|e^{-K/2}\rangle\rangle} \right) 
\end{equation}
with the operator inside the trace, an orthogonal projector. It is interesting to remark the similarity between this relation and Eq.\ \eqref{eq:partialtrtR} in Theorem \ref{th:states}, with the role of the projector embodied by $R$ in the extended formalism.

The application of Eqs.\ \eqref{eqapp:choirelations1} to the spacetime formalism lead directly to the Theorems in section \ref{th:states}, as we prove below. 
\vspace{0.1cm}

\emph{Theorem \ref{th:states}}. The property \eqref{eqapp:choitrace} leads directly to ${\rm Tr}_{E}R=\rho_0 e^{i\tilde{S}/2}e^{i\tilde{S}/2}=\rho_0 e^{i\tilde{S}}$ which is the relation \eqref{eq:partialtrtR}, the main result of the Theorem. Notice that, by construction, the dagger in the definition of $|\overline{\Psi}\rangle$ cancels the dagger arising from the bra.

In order to prove \eqref{eq:partialtrtrho} let us notice that we can first perform the partial trace over the environment, so that we have to consider only the operator ${\rm Tr}_{t'\neq t}[\rho_0 e^{i\tilde{S}}]$ (consequence of Eq.\ \eqref{eq:partialtrtR}, which we just proved).
Then, we make use of Lemma \eqref{th:lemmabosonmap}:
by considering only operators acting on the same slice $t$, including $\rho_t$, the l.h.s. just becomes the expectation value of arbitrary operators with respect to ${\rm Tr}_{t'\neq t}[\rho_0 e^{i\tilde{S}}]$. At the same time, the Lemma establish an equality with the expectation value of the same operators with respect to $\rho(t)$. 
Since they also act on the same Hilbert space, it holds that ${\rm Tr}_{t'\neq t}[\rho_0 e^{i\tilde{S}}]=\rho(t)$. We also recall that $\mathcal{H}=h^{\otimes N}$, with each $h$ the standard Hilbert space.

\qed

\emph{Corollary \ref{cor:weakv}}. Since the partial trace over the environment of $R$ leads to the exponential of the quantum actiion operator (previous Theorem \ref{th:states}), the mean values of operators acting trivially on the environment, are equal to the mean values computed with $\rho_0 e^{i\tilde{S}}$. Then, by applying Lemma \eqref{th:lemmabosonmap} one obtains Corollary \ref{cor:weakv}.

\qed

\emph{Corollary \ref{cor:choi}}. As noticed in the main text, the purification of Theorem \ref{th:states} is not unique. As a matter of fact, we have seen that the purification is a direct consequence of Eq.\ \eqref{eqapp:choitrace} with $AB^\dag=\rho_0 e^{i\tilde{S}}$. While the product $AB^\dag$ is fixed, the definition of $A$ and $B$ is clearly not. In particular one can choose $A=\rho_0 e^{i\tilde{S}}$, $B=\mathbbm{1}$, leading to
\begin{equation}
    {\rm Tr}[\mathcal{O}\rho_0 e^{i\tilde{\mathcal{S}}} ]=\langle \Phi^+|\mathcal{O}\otimes \mathbbm{1}_E|\Psi'\rangle\equiv \langle \bm {\mathcal O}^\dag|\bm{\rho_0}{e^{\bm i\tilde{\mathcal{S}}}} \rangle\,.
\end{equation}

\qed

Before discussing Theorems \ref{th:fermistates} and \ref{th:quantumvariationalpr} we include here the proof the main Theorem of section \ref{sec:entanglementintime} about the entanglement in time approach  and its relation to bosonic spacetime states.

\emph{Theorem \ref{th:entanglement}}. The proof is obtained by straightforward calculation.
\begin{proof}
By definition we have
    \begin{align*}
        T_{AB}&=\sum_{i,j} U^\dag |i\rangle \langle j| U \otimes |j\rangle \langle i| \, (\rho\otimes \mathbbm{1})\\
        &=(U^\dag \otimes \mathbbm{1})\,\sum_{i,j}|ij\rangle \langle ji| \,(U\otimes \mathbbm{1})\, (\rho \otimes \mathbbm{1})\\
        &=(U^\dag \otimes \mathbbm{1})\,\sum_{i,j}|ij\rangle \langle ij| \text{SWAP}\, (U\otimes \mathbbm{1})\, (\rho \otimes \mathbbm{1})\\
        &=(U^\dag \otimes \mathbbm{1})\, \text{SWAP} \, (U\otimes \mathbbm{1})\, (\rho \otimes \mathbbm{1})\\
        &=\text{SWAP} \, (U \otimes U^\dag)\, (\rho \otimes \mathbbm{1})\,,
    \end{align*}
    where we used the completeness relation $\sum_{i,j}|ij\rangle \langle ij|=\mathbbm{1}$ in $h\otimes h$.  On the other hand, for $N=2$ we have
\begin{equation*}
    \begin{split}
        \rho_0 e^{i\tilde{S}}&=(\rho \otimes \mathbbm{1}) ((U^\dag )^2\otimes \mathbbm{1})\,\text{SWAP}\, (U \otimes U)\\&=(\rho \otimes \mathbbm{1}) \,\text{SWAP}\, (U^\dag \otimes U)
    \end{split}
\end{equation*}
    from which the Theorem follows. 
\end{proof}

\subsection{Fermionic spacetime states, Choi isomorphism, and Fermionic partial traces}

In this Appendix we discuss several concepts related to the notion of fermionic spacetime state introduced in section \ref{sec:states}. Many of the ideas we develop might also have a more general scope as they cover topics usually not discussed in depth for fermions. In particular, we establish the fermionic analogue to equations \eqref{eqapp:choirelations1}-\eqref{eqapp:choitrace}.

Consider a generic fermionic system defined by $\{a_i,a^\dag_{j}\}=\delta_{ij}$ $\{\tilde{a}_k,\tilde{a}^\dag_{l}\}=\delta_{kl}$ with other anticommutators vanishing, including $\{a_i,\tilde{a}^\dag_{k}\}=0$. The separation between operators $a_i$ and $\tilde{a}_k$ defines a bipartition $S\oplus \tilde S$ of the full sp space.  Any fermionic system can be separated among modes in this way without loss of generality. Consider now a state of the form
\begin{equation}    |\psi\rangle=\sum_{\mu,\nu} \Gamma_{\mu\nu} A^\dag_{\mu} \tilde{A}^\dag_\nu |0\rangle 
\end{equation}
where we have adopted the notation
\begin{equation}
    A_\mu:=\prod_i (a^\dag_i)^{n_i}\,,\;\;\; \tilde{A}_\nu:=\prod_k (\tilde{a}^\dag_k)^{\tilde n_k}
\end{equation}
for $\mu:=(n_1,n_2,\dots)$, $\nu:=(\tilde n_1,\tilde n_2,\dots)$. A ``local'' expectation value, defined by an operator $O_S=O[a_i,a_j^\dag]$ (independent of the tilde-modes), assumed to be even (i.e.\ a product of an even number of operators) can be computed as
\begin{equation}
\begin{split}
    \langle \psi|O_S|\psi\rangle&=\sum_{\mu,\mu'}(\Gamma \Gamma^\dag)_{\mu\mu'}\langle 0|A_{\mu'} O_S A_{\mu}^\dag|0\rangle\\&={\rm tr}[\rho_S O_S]\,,
    \label{eq:red}
    \end{split}
\end{equation}
where we used that 
\begin{equation}
\langle 0|\tilde{A}_\nu \tilde{A}^\dag_{\nu'} |0\rangle=\delta_{\nu\nu'}\label{eq:orth}    
\end{equation}
and we have defined a fermionic reduced density operator  
\begin{align}
     \rho_S:&=\sum_{\mu,\mu'}(\Gamma \Gamma^\dag)_{\mu\mu'}  A_{\mu}^\dag|0\rangle\langle 0|A_{\mu'}\,\label{eq:red2}
     \end{align}
which is a well-defined mixed state for the modes in ${\cal S}$ ($\rho\geq 0$, ${\rm tr}\,\rho=1$). Here the vacuum is to be reinterpreted as the vacuum in $S$.  
It is interesting to notice the similarities with the standard distinguishable case for which a bipartite state $|\psi\rangle=\sum_{i,j}\Gamma_{ij}|ij\rangle$ leads to the local state ${\rm tr}_2[|\psi\rangle \langle \psi|]=\sum_{i,i'}(\Gamma \Gamma^\dag)_{ii'}|i\rangle \langle i'|$.
We also remark that for any fermionic state with definite total number parity 
$e^{i\pi\hat N}|\psi\rangle=\pm|\psi\rangle$, $\langle\psi|O|\psi\rangle=0$ for any odd operator (a product of an odd number of operators), so that \eqref{eq:red} is sufficient for evaluating any local average. On the other hand, 
Eqs.\ \eqref{eq:red}--\eqref{eq:red2} show that any fermionic mixed state $\rho$ can be purified by considering a complementary sp subspace having at least the same dimension as the original one (see also \cite{gigena2016one}).

Let us now assume that the partitions $S$ and $\tilde{S}$ are isomorphic. 
Given the common vacuum state $|0\rangle$ we can then introduce a family of Choi states
\begin{equation}
    |\phi^+\rangle:=\sum_\mu e^{i\gamma_\mu} A^\dag_{\mu} \tilde{A}^\dag_\mu |0\rangle\,. 
\end{equation} 
Here $\gamma_\mu$ is an arbitrary phase introduced for convenience. In particular, a reordering of the sp basis might add some signs. For example, the state $|\phi^+\rangle=\exp(\sum_i a^\dag_i \tilde{a}^\dag_i)|0\rangle$ corresponds to a particular choice of signs. Notice that one can identify the $\tilde{S}$ modes with a fermionic environment, namely the system and the environment are fully fermionic (no tensor product structure separating the two subsystems).

We now introduce a fermionic notion of Choi-isomorphism. We restrict our discussion to the case of operators commuting with parity. Given a fermionic operator $B_S$ (depending solely on the $A_\mu$) we define  
\begin{equation}
    |\bm{B}\rangle:=B_S |\phi^+\rangle\,.
\end{equation}
The condition $[P_S,B_S]_\pm=0$, with $P_S$ the parity operator on the modes $S$, holds iff 
\begin{equation}
    P|\bm{B}\rangle=\pm|\bm{B}\rangle\,,
\end{equation} as a direct consequence of the relation $[P_S,B_S]_\pm|\phi^+\rangle=P|\bm{B}\rangle\mp |\bm{B}\rangle$. We see that fixing the parity of operators corresponds to the standard parity superselection rule in the space of fermionic Choi states.
By using \eqref{eq:orth} one can now prove the relation
\begin{equation}\label{eqapp:fermichoi}
    \langle \bm{C}|O_S|\bm{B}\rangle={\rm tr}[B_SC_S^\dag O_S]\,.
\end{equation}

Having presented these tools we are now in a position to prove the main fermionic Theorem of \ref{sec:states}. 
\vspace{0.1cm}

\emph{Theorem \ref{th:fermistates}}. By making use of Eq.\ \eqref{eqapp:fermichoi} with $B=(\rho_0 e^{i\tilde{\mathcal{S}}/2})$, $C=(  Pe^{i\tilde{\mathcal{S}}/2})^\dag$ and $O_S\equiv \smallprod_t O^{(t)}_t$ we immediately recover Eq.\ \eqref{eq:thfermieq}, the main statement of the Theorem. 
By construction, it is also clear that parity is preserved by all the operators involved. 

\qed

Let us also notice that the discussion below Theorem \ref{th:fermistates} in the main text, stressing the analogies with the bosonic Theorem and Corollaries, is rigorously grounded in the concepts that we have established throughout this section.

\subsection{Variational principle of stationary action}

Here we prove Theorem \ref{th:quantumvariationalpr}. Let us recall first that in standard QM, maximizing the entropy under some constrained expectation values, leads to states having an exponential form \cite{balian2005information, balian2006microphysics}. In particular, fixing the energy leads to the Helmholtz-free energy and ensuing thermal-states (Boltzmann-Gibbs distribution).
The mathematical idea behind Theorem \ref{th:quantumvariationalpr} is essentially the same but with two fundamental novelties: in first place, the space of operators considered is much larger. Secondly, the constraint introduced by the Lagrange multiplier $\hbar^{-1}$ fixes the expectation value of the quantum action operator, an object that can only be defined in our formalism. Moreover, according to the formalism developed in section \ref{sec:states} for spacetime states, we don't impose any hermiticity nor positivity conditions.

Our strategy is to consider the following  statement:
\begin{equation}\label{eqapp:var}
    F\left[\frac{e^{-S_E/\hbar}}{Z}+\delta A\right]-F\left[\frac{e^{-S_E/\hbar}}{Z}\,\right]=\mathcal{O}(\delta A^2)\,.
\end{equation}
where $Z={\rm Tr}[e^{-\mathcal{S}_E/\hbar}]$ and with $\delta A$ an arbitrary traceless operator (${\rm Tr}[\delta A]=0$). As this is a general variation preserving the trace, Theorem \ref{th:quantumvariationalpr} follows from Eq.\  \eqref{eqapp:var}. We now prove the result \eqref{eqapp:var}. 

\begin{proof}
By definition
\begin{align*}
    \delta F&={\rm Tr}\left[\delta A \mathcal{S}_E\right]+\hbar\,{\rm Tr}\left[\left(\frac{e^{-S_E/\hbar}}{Z}+\delta A \right)\log\left(\frac{e^{-S_E/\hbar}}{Z}+\delta A \right)\right]
    \\&-\hbar\,{\rm Tr}\left[\left(\frac{e^{-S_E/\hbar}}{Z}\right)\log\left(\frac{e^{-S_E/\hbar}}{Z}\right)\right]\\
    &={\rm Tr}\left[\left(\frac{e^{-S_E/\hbar}}{Z}+\delta A \right)\mathcal{S}_E\right]+\hbar \log(Z)\\&+\hbar\,{\rm Tr}\left[\left(\frac{e^{-S_E/\hbar}}{Z}+\delta A \right)\log\left(\frac{e^{-S_E/\hbar}}{Z}+\delta A \right)\right]
\end{align*}
where $\delta F$ is the l.h.s.\ of Eq.\ \eqref{eqapp:var} and where we used elementary properties of the logarithm. Now notice that $\log\left(\frac{e^{-S_E/\hbar}}{Z}+\delta A \right)=\log(1+Ze^{\mathcal{S}_E/\hbar}\delta A)-\hbar^{-1}\mathcal{S}_E-\log(Z)$. Inserting this relation in the previous expression yields
\begin{align*}
    \delta F&=\hbar\,{\rm Tr}\left[\left(\frac{e^{-S_E/\hbar}}{Z}+\delta A \right)\log(1+Ze^{\mathcal{S}_E/\hbar}\delta A)\right]\,.
\end{align*}
Now we expand $\log(1+Ze^{\mathcal{S}_E/\hbar}\delta A)=Ze^{\mathcal{S}_E/\hbar}\delta A+\mathcal{O}(\delta A^2)$, which corresponds to $\delta A \ll e^{-\mathcal{S}_E/\hbar}/Z$ (in some operator norm sense). We finally obtain
\begin{align*}
    \delta F&=\hbar\,{\rm Tr}\left[\delta A\right]+\mathcal{O}(\delta A ^2)=\mathcal{O}(\delta A ^2)\,,
\end{align*}
where one can also easily extract the quadratic order from the expansion of the logarithm. 
\end{proof}

Let us also now notice that one might consider instead variations of the form $e^{-(S_E+\delta S)/\hbar}/{\rm Tr}[e^{-(S_E+\delta S)\hbar}]$, namely a variation of the quantum action operator itself as $\mathcal{S}_E\to \mathcal{S}_E+\delta S$. Then, one can easily show that Eq.\ \eqref{eqapp:var} is equivalent to 
\begin{equation}
    F\left[\frac{e^{-(S_E+\delta S)\hbar}}{Z_{\mathcal{S}_E+\delta S}}\right]-F\left[\frac{e^{-S_E\hbar}}{Z_{\mathcal{S}_E}}\,\right]=\mathcal{O}(\delta S^2)\,.
\end{equation}

As a final remark, let us also notice that since the quantum action is general in all of our expressions, everything holds under the replacement $\mathcal{S}_E\to \log(P)+ \mathcal{S}_E$ assuming parity preserving actions. For this reason, the fermionic case is already included in our proof.

\section{Spacetime formalism and the Fermionic Path Integral} \label{app:PIs}
In this section we describe how the formalism developed in the main text provides a Hilbert space embedding of the PI formulation. Since the bosonic case was developed in detail in \cite{DIAZ2025170052}, here we focus on fermions.

Let us recall a few basic facts of PIs for fermionic systems. It is well-known that the standard notion of PIs for fermions requires the introduction of Grassmann variables \cite{altland.2010}, i.e. ``numbers'' $\psi_i$ satisfying $\{\psi_i, \psi_j\}=0$ and anticommuting with operators $a_i$ ($a_i^\dag)$ as well. One can then define fermionic coherent states of the form $|\bm{\psi}\rangle:=e^{-\sum_i \psi_i a^\dag_i}|0\rangle$ such that $a_i |\bm{\psi}\rangle=\psi_i |\bm{\psi}\rangle$. The trace of an operator $O$ (made of an even number of fermionic operators) can then be computed as   \cite{shankar2012principles}
\begin{equation}\label{eq:appgrassmantr}
\begin{split}
    {\rm tr}[O]&=\int \smallprod_i [d\bar{\psi}_i d\psi_i e^{-\bar{\psi}_i\psi_i}] \langle -\bm{\bar{\psi}}|O|\bm{\psi}\rangle\\ &\equiv \int d\bar{\bm{\psi}} d\bm{\psi} e^{-\sum_i\bar{\psi}_i\psi_i} \langle -\bm{\bar{\psi}}|O|\bm{\psi}\rangle\,.
\end{split}
\end{equation}
This is related to the completeness relation $\int d\bar{\bm{\psi}} d\bm{\psi} e^{-\sum_i\bar{\psi}_i\psi_i} |\bm{\psi}\rangle \langle \bm{\bar{\psi}}|=\mathbbm{1}$ (notice that the sign in the bra is only present when computing the trace). 

With the previous properties at hand, one can build time-sliced PIs as usual. To give a concrete example consider the quantity ${\rm tr}[e^{-\beta H}]=\int d\bar{\bm{\psi}} d\bm{\psi} e^{-\sum_i\bar{\psi}_i\psi_i} \langle -\bm{\bar{\psi}}|e^{-\beta H}|\bm{\psi}\rangle$. Then. we can write $e^{-\beta H}=\prod_{t=0}^{N-1}e^{-\epsilon H}$, for $\epsilon=\beta/N$, and insert the completeness relation in between to obtain
\begin{equation}\label{eq:standardpi}
\begin{split}
      {\rm tr}[e^{-\beta H}]&=\int \smallprod_{t=0}^{N-1} d\bar{\bm{\psi}}_t d\bm{\psi}_t e^{-\sum_{i}\bar{\psi}_{ti}\psi_{ti}}\langle \bm{\bar{\psi}}_{t+1}|e^{-\epsilon H}|\bm{\psi}_t\rangle\\
      &=\int \smallprod_{t=0}^{N-1} d\bar{\bm{\psi}}_t d\bm{\psi}_t e^{i\epsilon\sum_{i}i\dot{\bar{\psi}}_{ti}\psi_{ti} }\frac{\langle \bm{\bar{\psi}}_{t+1}|e^{-\epsilon H}|\bm{\psi}_t\rangle}{\langle \bm{\bar{\psi}}_{t+1}|\bm{\psi}_t\rangle}
\end{split}
\end{equation}
with $\langle \bm{\bar{\psi}}_{N}|\equiv \langle -\bm{\bar{\psi}}_0|$ and $\dot{\bar{\psi}}_{ti}\equiv (\bar{\psi}_{t+1,i}-\bar{\psi}_{ti})/ \epsilon$. One can then write the Hamiltonian term up to order $\epsilon$ (not without subtleties) to write the integrand as the exponential of an action having the classical form (but in Grassmann variables).

Consider now the extended formalism. If one allows for Grassmann variables in $\mathcal{H}$ one can construct an extended basis of coherent states, which one might call \emph{trajectory states} as follows: 
\begin{equation}\label{eq:coherentbasisextended}
    |\bm{\psi}_0,\bm{\psi}_1,\dots,  \bm{\psi}_{N-1}\rangle:=e^{-\sum_{t,i}\psi_{ti}a^\dag_{ti}}|\Omega\rangle\,.
\end{equation} 
Then 
\begin{equation}\label{eq:grassmaneigenvalues}
    a_{ti}|\bm{\psi}_0,\bm{\psi}_1,\dots,  \bm{\psi}_{N-1}\rangle=\psi_{ti}|\bm{\psi}_0,\bm{\psi}_1,\dots,  \bm{\psi}_{N-1}\rangle\,,
\end{equation}
as one might easily verify using the extended algebra. 
We can now compute traces in $\mathcal{H}$ according to (the extended version of) \eqref{eq:appgrassmantr}. In particular,
\\

\begin{widetext}
    \begin{equation}\label{eq:extendedpi}
    \begin{split}
        {\rm Tr}[P e^{-\mathcal{S}_E}]&=\int \smallprod_{t=0}^{N-1} [d\bar{\bm{\psi}}_t d\bm{\psi}_t e^{-\sum_{i}\bar{\psi}_{ti}\psi_{ti}}]\langle -\bm{\bar{\psi}}_0,-\bm{\bar{\psi}}_1,\dots,  -\bm{\bar{\psi}}_{N-1}|Pe^{-\mathcal{S}_E}|\bm{\psi}_0,\bm{\psi}_1,\dots,  \bm{\psi}_{N-1}\rangle \\
        &=\int \smallprod_{t=0}^{N-1} [d\bar{\bm{\psi}}_t d\bm{\psi}_t e^{-\sum_{i}\bar{\psi}_{ti}\psi_{ti}}]\langle \bm{\bar{\psi}}_0,\bm{\bar{\psi}}_1,\dots,  \bm{\bar{\psi}}_{N-1}|e^{-\mathcal{S}_E}|\bm{\psi}_0,\bm{\psi}_1,\dots,  \bm{\psi}_{N-1}\rangle 
    \end{split}
\end{equation}

\end{widetext}

where we used that
\begin{equation}
    P|\bm{\psi}_0,\bm{\psi}_1,\dots,  \bm{\psi}_{N-1}\rangle=|-\bm{\psi}_0,-\bm{\psi}_1,\dots,  -\bm{\psi}_{N-1}\rangle
\end{equation}
as one can easily verify by using that $Pa_{ti}P=-a_{ti}$ and $[P,\psi_{ti}]=0$. In addition, by using the definition of $\mathcal{P}$ (Eq.\ \eqref{eq:pdef}) one can easily find the action of time translation the basis of states \eqref{eq:coherentbasisextended}. Taking into account the antiperiodic boundary conditions, one obtains $\langle \bm{\bar{\psi}}_0,\bm{\bar{\psi}}_1,\dots,  \bm{\bar{\psi}}_{N-1}|e^{i\epsilon \mathcal{P}}=\langle \bm{\bar{\psi}}_1,\bm{\bar{\psi}}_2,\dots,  \bm{\bar{\psi}}_{N-1},-\bm{\bar{\psi}}_0|$. As a direct consequence, 
\begin{equation}
\begin{split}
     &\langle \bm{\bar{\psi}}_0,\bm{\bar{\psi}}_1,\dots,  \bm{\bar{\psi}}_{N-1}|e^{-\mathcal{S}_E}|\bm{\psi}_0,\bm{\psi}_1,\dots,  \bm{\psi}_{N-1}\rangle\\ &=\langle \bm{\bar{\psi}}_1,\dots,  \bm{\bar{\psi}}_{N-2},-\bm{\bar{\psi}}_0|\smallprod_t e^{-\epsilon H_t}|\bm{\psi}_0,\bm{\psi}_1,\dots,  \bm{\psi}_{N-1}\rangle  \\&=
     \smallprod_{t=0}^{N-1} \langle \bar{\bm{\psi}}_{t+1}|e^{-\epsilon H}|\bm{\psi}_t\rangle\,,
\end{split}
\end{equation}
which is precisely the term in \eqref{eq:standardpi}, in agreement with ${\rm Tr}[P e^{\mathcal{S}_E}]={\rm tr}[e^{-\beta H}]$ and showing that ${\rm Tr}[P e^{-\mathcal{S}_E}]$, when evaluated in the trajectory basis is precisely the fermionic PI. Moreover, if one assumes as usual $\epsilon\ll 1$ it is now straightforward to show that 
    \begin{equation}\label{eq:actiongrassman}
    \begin{split}
       e^{-\sum_{t,i}\bar{\psi}_{ti}\psi_{ti}}\langle \bm{\bar{\psi}}_0,\bm{\bar{\psi}}_1,\dots,  \bm{\bar{\psi}}_{N-1}|e^{-\mathcal{S}_E}|\bm{\psi}_0,\bm{\psi}_1,\dots,  \bm{\psi}_{N-1}\rangle\\=\exp\Big[-\Big(\epsilon\sum_{t,i}\{\dot{\bar{\psi}}_{ti}\psi_{ti} +H(\psi_{ti},\bar{\psi}_{tj})\}\Big) \Big]+\mathcal{O}(\epsilon)\,,  
    \end{split}
    \end{equation}
with $H(\psi_{ti},\bar{\psi}_{tj})\equiv \langle \bm{\bar{\psi}}_t|H|\bm{\psi}_t\rangle$. If one now replaces \eqref{eq:actiongrassman} in \eqref{eq:extendedpi} the time-sliced PI is apparent. Interestingly, the classical action $S_{\text{cl}}$ emerges as the expectation value of the quantum action. 
While the example we considered corresponds to partition functions, the case of thermal correlators is now readily obtained: consider for example ${\rm Tr}[P e^{-\mathcal{S}_E}a_{t_1 i}a^\dag_{t_2 j}]={\rm Tr}[ a^\dag_{t_2 j}P e^{i\mathcal{S}}a_{t_1 i}]=- {\rm Tr}[P a^\dag_{t_2 j} e^{-\mathcal{S}_E}a_{t_1 i}]$. In this form, the completeness relation leads directly to an integral with the form of \eqref{eq:appPI} with $e^{- \mathcal{S}_E}\to a^\dag_{t_2 j} e^{-\mathcal{S}_E}a_{t_1 i}$. At the same time, 
\begin{equation}\label{eq:appPI}
   \langle \bm{\bar{\psi}}_0,\bm{\bar{\psi}}_1,\dots|a^\dag_{t_2 j} e^{-\mathcal{S}_E}a_{t_1 i}|\bm{\psi}_0,\bm{\psi}_1,\dots\rangle\simeq -e^{-S_\text{cl}}\psi_{t_1 i} \bar{\psi}_{t_2 j}\,,
\end{equation} 
with $e^{- S_\text{cl}}$ the r.h.s. of \eqref{eq:actiongrassman}, where we used Eq.\ \eqref{eq:grassmaneigenvalues}, $\bar{\psi}_{t_2 j} \psi_{t_1 i} =-\psi_{t_1 i} \bar{\psi}_{t_2 j}$ and the equality holds up to $\mathcal{O}(\epsilon)$. The important point is that the usual ``insertion'' of Grassmann variables in the PI, leading to correlation functions, correspond now to the insertion of spacetime operators, in agreement with Theorem \ref{th:correlations}. Similar results are obtained for real time (unitary evolution).

Let us now notice that it is a common technique to define $\psi_{nj}:=\frac{1}{\sqrt{N}}\sum_{t=0}^{N-1} e^{i\omega_n \epsilon t}\psi_{tj}$ with $\omega_n$ the Matsubara frequencies. This classical transformation among Grassmann variables allows one to easily compute PIs associated with free Hamiltonians, and serves as a starting point to develop semiclassical approximations. In particular, for $H=\sum_{k}\lambda_k a_k^\dag a_k$ it leads directly to  
\begin{equation}\label{eqapp:matsu}
    \frac{1}{Z_\beta}{\rm tr}[e^{-\beta H}\hat{T} a_{i}(-it_1)a^\dag_{ j}(-it_2)]=\frac{1}{T}\sum_n  \frac{e^{-\epsilon \omega_n (t_1-t_2)}}{ i\omega_n- \lambda_k}+\mathcal{O}(\epsilon)
\end{equation}
where we have already evaluated the corresponding PI expression. Notably, all of these considerations, based on classical Grassmann variables, have an operator correspondence in our formalism: the Matsubara frequencies are the normal frequencies of $\mathcal{P}$, the FT is a unitary transformation (Eq.\ \eqref{eq:fta}), and the thermal propagator on the l.h.s. of \eqref{eqapp:matsu} has a trace representation (Theorem \ref{th:correlations}) leading directly to the expression on the r.h.s. of \eqref{eqapp:matsu} (by using Eq.\ \eqref{eq:matsubaratwop}). We see that by means of the spacetime Hilbert space one can essentially compute any given PI without using Grassmann variables. Let us also remark that in our approach there is no need to take into account possible transformations of the measure of integration as we are simply computing a trace, an invariant quantity under changes of bases.

Let us finally mention that now that the connection with the time-sliced PIs has been provided, most of the considerations on the continuum time case and the subtleties of the PI formulation developed in \cite{DIAZ2025170052} for bosons can be directly applied to fermions as well.

\section{Spacetime classical formalism and the Dirac equation}\label{app:classical}
Having described how to formulate QM in spacetime it is natural to discuss a spacetime classical approach, and its relation to the quantum formalism. As shown in \cite{diaz2024spacetime} it is quite simple to develop a formulation of classical mechanics in spacetime phase-spaces, meaning that spacetime classical algebras are defined by extended Poisson brackets (PBs). Moreover, one can think of the quantum spacetime algebras and quantum action as a canonical-like quantization of this spacetime classical mechanics. 
Here we apply the formalism developed in
\cite{diaz2024spacetime} to the Dirac action in $3+1$ dimensions as an example, and discuss its relation it to the quantum formalism of \ref{sec:continuumcase}.

Following \cite{diaz2024spacetime} we define \fontsize{9.5}{11.5}\selectfont
\begin{equation}
    \{f,g\}_{\text{PB}}=\sum_a\!\int \!d^{4}x \left[\frac{\delta f}{\delta \psi_a(x)}\frac{\delta g}{\delta \psi_a^\dag (x)}-\frac{\delta g}{\delta \psi_a(x)}\frac{\delta f}{\delta\psi_a^\dag (x)}\right],
\end{equation}
\normalsize
where it should be noted that these PBs are \emph{not} defined at a given time as usual. Instead, they encompass all the spacetime variables in contrast to the standard Poisson brackets (denoted as pb) 
\fontsize{9.5}{11.5}\selectfont
$$\{f,g\}_{\text{pb}}=\sum_a\!\int \!d^{3}x \left[\frac{\delta f}{\delta \psi_a(\textbf{x})}\frac{\delta g}{\delta \psi_a^\dag (\textbf{x})}-\frac{\delta g}{\delta \psi_a(\textbf{x})}\frac{\delta f}{\delta\psi_a^\dag (\textbf{x})}\right]\,.$$
\normalsize
In particular, for the spacetime approach we have
\begin{equation}
    \{\psi_a(x), \psi^\dag_b(y)\}_{\text{PB}}=\delta_{ab}\delta^{(4)}(x-y)
\end{equation}
which is the classical version of Eq.\ \eqref{eq:stalgcont} (here the brackets indicate PBs and not anticommutators). Instead, the conventional equal time algebra is given by $  \{\psi_a(t,\textbf{x}), \psi^\dag_b(t,\textbf{y})\}_{\text{pb}}=\delta_{ab}\delta^{(3)}(\textbf{x}-\textbf{y})$.

Now consider the \emph{classical} Dirac action 
\begin{equation}
    S_\tau=\tau\int d^4x\, \bar{\psi}(x)(\gamma^\mu i\partial_\mu -m)\psi(x)\,.
\end{equation}
One obtains by direct calculation
$
  \{\psi(x),S_\tau\}_{\text{PB}}= \tau \gamma^0 (i\gamma^\mu \partial_\mu-m) \psi(x) 
$
which, notably,  leads to the Dirac equation when equaled to zero. Let us discuss this claim: the extended approach to classical mechanics dictates that, in principle, all fields at different spacetime points are independent. Clearly, this, by itself, does not describe physical evolution. Instead, we \emph{impose} that the fields satisfy the constraints $\{\psi(x),S_\tau\}_{\text{PB}}\approx 0$. Within this subspace we get 
\begin{equation}
  \{\psi(x),S_\tau\}_{\text{PB}}=\tau \gamma^0 (i\gamma^\mu \partial_\mu-m) \psi(x) \approx 0\,,
\end{equation}
where the constraints are to be imposed for all $x$ and after all PBs have been evaluated (this is indicated with the ``weak equality'' notation ``$\approx$'' \cite{pam.50}). 
In this sense, classical evolution corresponds to a subset of all possible fields configurations in spacetime. Notably, these can be determined straightforwardly by using the classical action in extended phase-space variables. This can be understood from the following considerations:
notice that $\mathcal{S}_\tau=\tau \mathcal{P}_0-\tau \int dt\, H_D(t)$ with the Dirac action corresponding to the Dirac Hamiltonian $H_D$. On the other hand, we have
\begin{equation}\label{eqapp:P0}
    \{\psi(x),\mathcal{P}_0\}_{PB}=i\dot{\psi}(t,\textbf{x})
\end{equation}
an infinitesimal (classical) time translation, while 
\begin{equation}\label{eqapp:hampb}
\begin{split}
     \{\psi(x),\int dt\, H_D(t)\}_{PB}&=\int dt\,  \{\psi(t,\textbf{x}), H_D(t)\}_{PB}\\&=\{\psi(t,\textbf{x}), H_D\}_{pb}
\end{split}
\end{equation}
By combining Eqs.\ \eqref{eqapp:P0} and \eqref{eqapp:hampb} we obtain
\begin{equation}
   \{\psi(x),\mathcal{S}_\tau\}_{\text{PB}}=\tau\left[\,i\dot{\psi}(t,\textbf{x})-\{\psi(t,\textbf{x}), H_D\}_{pb}\,\right]
\end{equation}
which becomes one of the \emph{Hamilton} equations when set to zero ($i\dot{\psi}\equiv \pi\equiv i\psi^\dag$). The other Hamilton equation here simply gives the  
conjugated Dirac equation.

Let us make a final remark. One can verify that the constrained obtained are of the second kind (in clear contrast e.g. with the gauge constraint arising from parameterizing the time variable \cite{qg}, a scheme closely related to quantum time formalisms \cite{dia.19, diaz.21}). In that sense, a direct quantization following Dirac approach to constrained systems \cite{pam.50} is not feasible. Yet, within ``expectation values'' of the form $\langle \dots \rangle \propto{\rm Tr}[Pe^{i\mathcal{S}_\tau}\dots]$ the constraints are satisfied: 
\begin{equation}
    \langle \{\psi(x),\mathcal{S}_\tau\}\rangle=0
\end{equation}
 as it follows from the cyclicity of the trace and by noting that $\langle \psi(x)\mathcal{S}_\tau\rangle=\langle P\psi(x)P\mathcal{S}_\tau\rangle=-\langle \psi(x)\mathcal{S}_\tau\rangle$ where we used $P^2=\mathbbm{1}$, $[P,\mathcal{S}_\tau]=0$, and $P\psi(x)P=-\psi(x)$ (see also Appendix \ref{app:PIs}). In this sense, the constraints are automatically satisfied in the quantum case, as long as we evaluate physical quantities within brackets.

\section{Diagonalization of the Dirac quantum action operator}\label{app:actiondiag}
Here we explain how to derive the notion of particle that emerges from the diagonalization of the Dirac action described in section \ref{sec:paw} . 

The proof of Eq.\ (\ref{eq:abalg}), and the vanishing of the other anticommutators, can be obtained straightforwardly by employing $\{\psi_\sigma(p),\psi^\dag_{\sigma'}(p')\}=(2\pi)^4\delta^{(4)}(p-p')$ and conventional properties of the Dirac spinors. 
One can write the transformations (\ref{eq:uvmodes}) in compact form and understand them as a Bogoliubov transformation:
\begin{equation}
    \begin{pmatrix}
\textbf{a}(p)\\
\textbf{b}^\dag(\shortminus p)
\end{pmatrix}=W(p)\psi(p)= \frac{1}{\sqrt{2E_{\textbf{p}}}}    \begin{pmatrix}
\textbf{u}^\dag_\textbf{p}\\
\textbf{v}^\dag_{\shortminus \textbf{p}}
\end{pmatrix}   
\Psi(p)
\end{equation}
where we have defined $\textbf{a}(p)=(a^1(p),a^2(p))^t$, $\textbf{b}(p)=(b^1(p),b^2(p))^t$ and $\textbf{u}_\textbf{p}=(u^1_\textbf{p},u^2_\textbf{p})$, $\textbf{v}_\textbf{p}=(v^1_\textbf{p},v^2_\textbf{p})$. Note that $\textbf{u}_\textbf{p}$ and $\textbf{v}_\textbf{p}$ are $4\times 2$ matrices. The standard unitarity condition of the transformation 
requires $W^\dag(p)W(p)=W(p)W^\dag(p)=\mathbbm{1}_4$. In this case we obtain
\begin{align}
    W^\dag(p)W(p)&=\frac{1}{2E_{\textbf{p}}}(\textbf{u}_{\textbf{p}}\textbf{u}^\dag_{\textbf{p}}+\textbf{v}_{\shortminus\textbf{p}}\textbf{v}^\dag_{\shortminus\textbf{p}})\label{eq:firstuuvv}\\
    W(p)W^\dag(p)&=\frac{1}{2E_{\textbf{p}}}\begin{pmatrix}
\textbf{u}^\dag_\textbf{p}\textbf{u}_\textbf{p}&&\textbf{u}^\dag_\textbf{p}\textbf{v}_{\shortminus\textbf{p}}\\
\textbf{v}^\dag_{\shortminus \textbf{p}}\textbf{u}_\textbf{p}&&\textbf{v}^\dag_{\shortminus \textbf{p}}\textbf{v}_{\shortminus \textbf{p}}\label{eq:seconduuvv}
\end{pmatrix}
\end{align}
which in both cases yield $\mathbbm{1}_4$: in \eqref{eq:firstuuvv} one can use the completeness relation $$\textbf{u}_{\textbf{p}}\textbf{u}^\dag_{\textbf{p}}+\textbf{v}_{\shortminus\textbf{p}}\textbf{v}^\dag_{\shortminus\textbf{p}}=\sum_s( u^s_{\textbf{p}}u^{s\dag}_\textbf{p}+v^s_{\shortminus\textbf{p}}v^{s\dag}_{\shortminus\textbf{p}})=2E_{\textbf{p}}\mathbbm{1}_4\,,$$ 
while in \eqref{eq:seconduuvv} the orthogonality relations $u^{s\dag}_{\textbf{p}}u^r_{\textbf{p}}=v^{s\dag}_{\textbf{p}}v^r_{\textbf{p}}=2E_{\textbf{p}}\delta_{sr}$ and $u^{s\dag}_\textbf{p}v^r_{\shortminus\textbf{p}}=v^{s\dag}_{\textbf{p}}u^r_{\shortminus \textbf{p}}=0$ yield the desired result. We see that various fundamental properties of the Dirac spinors can be compactly expressed as the unitarity condition of $W_m(p)$, required for it to represent a Bogoliubov transformation. This same property leads to \eqref{eq:abalg}. 

It is now straightforward to obtain the inverse relation 
\begin{equation}\label{eq:boginverse}
\begin{split}
        \psi(p)&=W^\dag(p)\begin{pmatrix}
\textbf{a}(p)\\
\textbf{b}^\dag(\shortminus p)
\end{pmatrix}\\&=\frac{1}{\sqrt{2E_\textbf{p}}}\begin{pmatrix}
\textbf{u}_\textbf{p}\;\;\textbf{v}_{\shortminus \textbf{p}}
\end{pmatrix}\begin{pmatrix}
\textbf{a}(p)\\
\textbf{b}^\dag(\shortminus p)
\end{pmatrix}\,.
\end{split}
\end{equation}
This can be used to show that the action takes the diagonal form. In fact, one can 
use that, by definition, the spinors $\textbf{u}_\textbf{p},\textbf{v}_{\textbf{p}}$ diagonalize the Dirac Hamiltonian at fixed momentum to write $[p_0-(\boldsymbol{\alpha} \cdot\textbf{p}+\beta m)]\begin{pmatrix}
\textbf{u}_\textbf{p}\;\;\textbf{v}_{\shortminus \textbf{p}}
\end{pmatrix}=p_0 \begin{pmatrix}
\textbf{u}_\textbf{p}\;\;\textbf{v}_{\shortminus \textbf{p}}
\end{pmatrix}-E_\textbf{p} \begin{pmatrix}
\textbf{u}_\textbf{p}\;\;-\textbf{v}_{\shortminus \textbf{p}}
\end{pmatrix}$. Then, using \eqref{eq:boginverse} one obtains 
 \fontsize{9.5}{11.5}
\begin{equation}\label{eq:diagactionmatrices}
\begin{split}
     &\psi^\dag(p)[p_0-(\boldsymbol{\alpha} \cdot\textbf{p}+\beta m)]\psi(p)=\\&=\begin{pmatrix}
\textbf{a}^\dag(p) 
\textbf{b}(\shortminus p)
\end{pmatrix} \begin{pmatrix}
[p_0-E_\textbf{p}]\mathbbm{1}_2& 0\\
0& [p_0+E_\textbf{p}]\mathbbm{1}_2
\end{pmatrix}\begin{pmatrix}
\textbf{a}(p)\\
\textbf{b}^\dag(\shortminus p)
\end{pmatrix}
\\&=\sum_s \,\big[(p_0-E_\textbf{p}) a^{s \dag}(p) a^s(p)+ (p_0+E_\textbf{p}) b^s(\shortminus p)b^{s \dag}( \shortminus p)\big]
\end{split}
\end{equation}
\normalsize
where we used that $W(p)W^\dag(p)= \mathbbm{1}$ and similar orthogonality relations. Now, we notice that $E_\textbf{p} [a^{s \dag}(p) a^s(p)-b^s(\shortminus p)b^{s \dag}( \shortminus p)]=E_\textbf{p} [a^{s \dag}(p) a^s(p)+b^{s \dag} (\shortminus p)b^{s}( \shortminus p))-(2\pi)^4\delta^{(4)}(0)]$ with $(2\pi)^4\delta^{(4)}(0)=(2\pi)^4\int \frac{d^4x}{(2\pi)^4}$ the spacetime volume. We see that only positive energies are added by creating particles. On the other hand, when one integrates in $p$, one can interchange $p\to -p$ which affects the sign of $p_0$
 in \eqref{eq:diagactionmatrices} thus compensating the normal ordering sign. In summary, 
 \begin{equation}
    \mathcal{S}_\tau= \!\int \!\frac{d^4p}{(2\pi)^4} \,\sum_s \tau(p_0-E_\textbf{p}) \big(a^{s \dag}(p) a^s(p)+b^{s \dag} (p)b^{s}(p)\big)
 \end{equation}
up to the usual negative vacuum energy (integrated over time). No additional constant arises from normal ordering.

\section{Differences and connections to the consistent histories approach and related proposals}\label{app:consistenth}
Given the similarities in motivation between our work and the results presented in the context of the consistent history approach to QM \cite{griffiths1984consistent, omnes1988logical,gell2010quantum, ish.93} we dedicate this appendix to briefly introduce the concept and discuss its differences and potential connections to our work. The main object of these proposal is the so called decoherence functional which for standard QM takes the form
\begin{equation}
    d(\boldsymbol{\alpha},\boldsymbol{\beta}):={\rm tr}[C^\dag_\alpha\rho C_\beta]
\end{equation}
for $C_\alpha=\alpha_{t_0}(t_0)\alpha_{t_1}(t_1)\dots$ where the different operators are evolving in the Heisenberg picture. Each operator $\alpha_{t_i}$ is a projector corresponding to ``propositions'' about the system. As a matter of fact, one can show that $d(\boldsymbol{\alpha},\boldsymbol{\alpha})$ corresponds to a joint probability of sequential measurements at different times (a quantity that also appears in the context of pseudo density matrices \cite{fit.15,fullwood2024operator}). Let us remark that the decoherence functional does not contain just a time ordered composition of operators but both a forward and backward in time evolution.

The essential point of the coherent history approach is to postulate that the decoherence functional is a fundamental entity from which other physical quantities may be derived. Then, a series of criteria (such as hermiticity and positivity for diagonal entries) for what constitutes a valid physical decoherence functional can be introduced.
In this context, a useful characterization of possible decoherence functional can be accomplished by means of tensor products in time. This is one of the main motivations of \cite{ish.93} which makes use of a Hilbert space $\mathcal{H}=\otimes_t h$. In particular, it was shown in \cite{isham1994classification} that a valid decoherence functional may always be written as
\begin{equation}\label{eq:decoherenceX}
   d(\boldsymbol{\alpha},\boldsymbol{\beta})={\rm Tr}_{\mathcal{H}\otimes \mathcal{H}}\left[X(\otimes_t \alpha_t)\otimes (\otimes_t \beta_t)\right]\,, 
\end{equation}
for $X$ properly defined. This is stated in \cite{isham1994classification} as a generalization of Gleason Theorem of standard QM \cite{gleason1975measures}.

Let us now remark that two copies of the spacetime Hilbert space $\mathcal{H}$ are required to represent the decoherence functional, in contrast with the schemes we presented in the main text: We focused on time ordered correlation functions, which thus contains ``half'' the operators. In particular, the concept of spacetime states we studied thus differ from the object $X$ introduced in \cite{isham1994classification}. This is what separates our work from previous proposals, and in particular, allowed us to connect it with the PI formulation instead. Moreover, throughout the main text Theorems we showed that employing $\mathcal{H}$ is sufficient to recover all the standard pictures of evolution in QM.

\begin{figure}[t!]
    \centering
    \includegraphics[width=\linewidth]{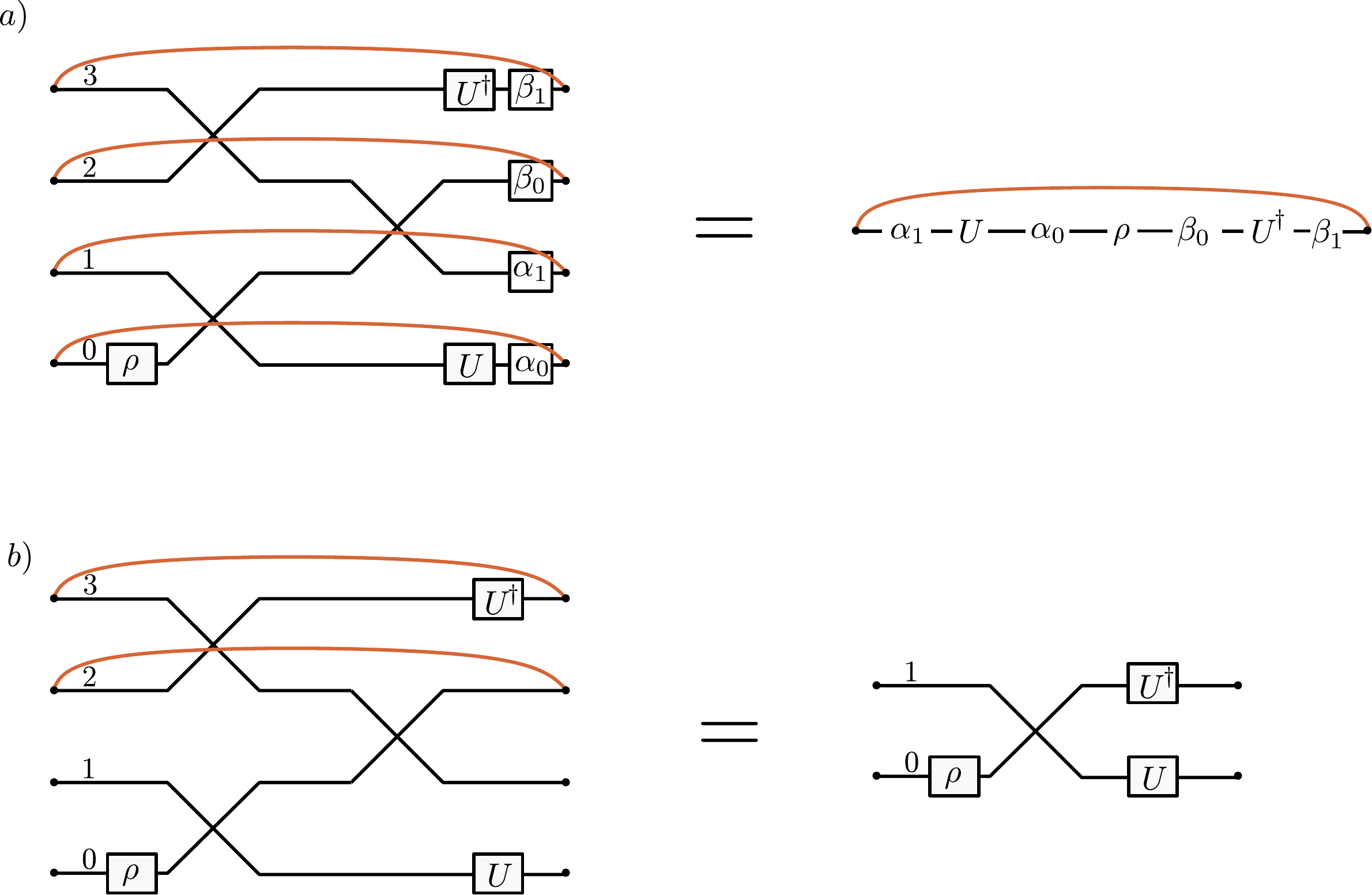}
    \caption{Graphical proofs of relations involving $X$ for two time slices. The orange lines correspond to traces. On panel a) we show that Eq.\ \eqref{eq:Xexpl} satisfies Eq.\ \eqref{eq:decoherenceX}. One recognized on the r.h.s. a decoherence functional for two times. On panel b) we prove Eq.\ \eqref{eq:ishampuri}. Notice that the ensuing operator on the r.h.s. is precisely the spacetime state: $(\rho \otimes \mathbbm{1}) \,\text{SWAP}\, U\otimes U^\dag=(\rho \,(U^\dag)^2 \otimes \mathbbm{1}) \,\text{SWAP}\,U\otimes U=\rho_0 e^{i\tilde{\mathcal{S}}}$. }
    \label{fig:tnX}
\end{figure}

 Having said this, it is interesting to take $\alpha_{t_i}\equiv \mathbbm{1}$ in Eq.\ \eqref{eq:decoherenceX} to study the relation between $X$ and our proposal.
Let us then denote the two fold copy of the spacetime Hilbert space $\mathcal{H}_T \equiv \mathcal{H}_1\otimes \mathcal{H}_2$. Lemma \ref{th:lemmabosonmap} in combination with Eq.\ \eqref{eq:decoherenceX} suggests the following relation
\begin{equation}\label{eq:ishampuri}
  \rho_0 e^{i\tilde{\mathcal{S}}}={\rm Tr}_{\mathcal{H}_2}[X]\,.
\end{equation}
We find that $X$ must be a ``purification'' of our spacetime states. So far this discussion has been abstract. We can actually employ the proof in \cite{isham1994classification} (which did not include a compact final form for $X$) to represent $X$ explicitly and verify Eq.\ \eqref{eq:ishampuri}. Let us discuss this briefly in the particular case of $N=2$. A general discussion will be presented elsewhere \cite{unifying}. Following the tensor network notation introduced in Figure \ref{fig:tn} we find that $X$ can be represented as 
\begin{equation}\label{eq:Xexpl}
X=\rho_0\,\text{SWAP}_{01}\text{SWAP}_{23}\text{SWAP}_{12}(U\otimes \mathbbm{1}\otimes \mathbbm{1}\otimes U^\dag)\,,
\end{equation}
 where we have labeled the Hilbert spaces as $\mathcal{H}_T=h_0\otimes h_1\otimes h_2\otimes h_3$ with $\mathcal{H}_1=h_0\otimes h_1$, $\mathcal{H}_2=h_2\otimes h_3$. Namely, this representation of $X$ satisfies Eq. \eqref{eq:decoherenceX} for all operators $\boldsymbol{\alpha}$, $\boldsymbol{\beta}$. This is proven graphically in Figure \ref{fig:tnX}. Therein it is also proven that $X$ leads to $\rho_0 e^{i\tilde{\mathcal{S}}}$ through partial trace.

Interestingly, we find that a small extension of our scheme allows one to describe decoherence functionals, thus relating our work to previous proposals. As a matter of fact, the decoherence functional might be regarded as a particular subset of possible out-of-order time correlations. A fully explicit version of $X$ can actually be connected to the simplest non-trivial Schwinger–Keldys contour (see also the discussion in section \ref{sec:entanglementintime}), so, indeed, $X$ is still connected to PIs once these types of contour are allowed, a result to be discussed in future work \cite{unifying}. This discussion shows that working in $\mathcal{H}$ is the first natural step from which more exotic possibilities might be considered. 
Let us also mention that in \cite{savvidou1999action} the concept of quantum action operator was also introduced in the consistent history context for continuous time, where a spacetime algebra was also discussed. However, to the authors knowledge, the relation with time ordered correlation functions and the PI formulation was not discussed previously. Moreover, the action introduced  in \cite{savvidou1999action}  acted on $\mathcal{H}$, so its relation to $X$ was not discussed either.

Let us finally briefly discuss possible relations  between $X$, our scheme, and superdensity operators $\varrho$ \cite{cot.18}. The latter are density matrices acting as $\mathcal{L}(\mathcal{H})\to \mathcal{L}(\mathcal{H})$, namely they are super operators in the spacetime Hilbert space. In this sense, $\varrho$ and our definition of spacetime state are not straightforwardly related. However, $X$ and 
$\varrho$ do indeed have the same dimensions, and, considering that superoperators can be related with the timelike correlators of the form of Eq.\ \eqref{eq:decoherenceX} (for general operators, not necessarily projectors), through Eq.\ \eqref{eq:ishampuri} a relation may be unveiled. Nonetheless, the relation between $X$ and 
$\varrho$ is also subtle: One can show that for the operationally  $\varrho$ introduced in \cite{cot.18} one might recover $\varrho$  from $X$ by a ``realignment'' of indices \cite{unifying}. This is related with the isomorphism of operator space $\mathcal{L}(\mathcal{H})\simeq \mathcal{H}\otimes \mathcal{H}^\ast$ and the way these two operators are interpreted as linear maps. A realignment does not preserve positivity (see e.g. the computable cross-norm or realignment criterium in standard quantum information \cite{chen2002matrix}), explaining why while $\varrho\geq 0$, even in our simple example of two times, $X$ can have negative (and complex) eigenvalues.

Interestingly, when introducing  superdensity operators  the authors also introduced the concept of anticommuting fermions at different times, another common aspect with our proposal (while the consistent history approach has not been developed for fermions as it is based on tensor products in time). This suggests to develop a general theory on $X$, applicable  to both bosons and fermions, to elucidate the connection to $\varrho$ and finally our concept of spacetime state.

The considerations of this section, together with the main text results relating our formalism to PIs, the PaW mechanism and the entanglement in time approach,
 suggest that the different proposals to generalize QM to spacetime might be unified. Namely, while different schemes provide a different perspective on the problem, after more research they might prove to be different aspects of a more basic structure. The common element in these schemes is the presence of a spacetime Hilbert space $\mathcal{H}$. As such, it is natural to view our framework as a useful step in this unifying direction, since the tools needed to recover standard QM and to relate these perspectives appear to be already present (perhaps implicitly) within a single spacetime Hilbert space 
$\mathcal{H}$. In particular, approaches formulated on several copies of 
$\mathcal{H}$ or on operator space may ultimately benefit from the kind of structure we presented throughout the manuscript.

\clearpage

%\bibliography{bibliography}

\begin{thebibliography}{80}%
\makeatletter
\providecommand \@ifxundefined [1]{%
 \@ifx{#1\undefined}
}%
\providecommand \@ifnum [1]{%
 \ifnum #1\expandafter \@firstoftwo
 \else \expandafter \@secondoftwo
 \fi
}%
\providecommand \@ifx [1]{%
 \ifx #1\expandafter \@firstoftwo
 \else \expandafter \@secondoftwo
 \fi
}%
\providecommand \natexlab [1]{#1}%
\providecommand \enquote  [1]{``#1''}%
\providecommand \bibnamefont  [1]{#1}%
\providecommand \bibfnamefont [1]{#1}%
\providecommand \citenamefont [1]{#1}%
\providecommand \href@noop [0]{\@secondoftwo}%
\providecommand \href [0]{\begingroup \@sanitize@url \@href}%
\providecommand \@href[1]{\@@startlink{#1}\@@href}%
\providecommand \@@href[1]{\endgroup#1\@@endlink}%
\providecommand \@sanitize@url [0]{\catcode `\\12\catcode `\$12\catcode `\&12\catcode `\#12\catcode `\^12\catcode `\_12\catcode `\%12\relax}%
\providecommand \@@startlink[1]{}%
\providecommand \@@endlink[0]{}%
\providecommand \url  [0]{\begingroup\@sanitize@url \@url }%
\providecommand \@url [1]{\endgroup\@href {#1}{\urlprefix }}%
\providecommand \urlprefix  [0]{URL }%
\providecommand \Eprint [0]{\href }%
\providecommand \doibase [0]{https://doi.org/}%
\providecommand \selectlanguage [0]{\@gobble}%
\providecommand \bibinfo  [0]{\@secondoftwo}%
\providecommand \bibfield  [0]{\@secondoftwo}%
\providecommand \translation [1]{[#1]}%
\providecommand \BibitemOpen [0]{}%
\providecommand \bibitemStop [0]{}%
\providecommand \bibitemNoStop [0]{.\EOS\space}%
\providecommand \EOS [0]{\spacefactor3000\relax}%
\providecommand \BibitemShut  [1]{\csname bibitem#1\endcsname}%
\let\auto@bib@innerbib\@empty
%</preamble>
\bibitem [{\citenamefont {Bell}(1964)}]{bell1964einstein}%
  \BibitemOpen
  \bibfield  {author} {\bibinfo {author} {\bibfnamefont {J.~S.}\ \bibnamefont {Bell}},\ }\bibfield  {title} {\bibinfo {title} {On the {E}instein {P}odolsky {R}osen paradox},\ }\href@noop {} {\bibfield  {journal} {\bibinfo  {journal} {Physics Physique Fizika}\ }\textbf {\bibinfo {volume} {1}},\ \bibinfo {pages} {195} (\bibinfo {year} {1964})}\BibitemShut {NoStop}%
\bibitem [{\citenamefont {Einstein}(1905)}]{E.05}%
  \BibitemOpen
  \bibfield  {author} {\bibinfo {author} {\bibfnamefont {A.}~\bibnamefont {Einstein}},\ }\bibfield  {title} {\bibinfo {title} {On the electrodynamics of moving bodies},\ }\href@noop {} {\bibfield  {journal} {\bibinfo  {journal} {Ann. Phys. (Berl.)}\ }\textbf {\bibinfo {volume} {17}},\ \bibinfo {pages} {50} (\bibinfo {year} {1905})}\BibitemShut {NoStop}%
\bibitem [{\citenamefont {Diaz}\ \emph {et~al.}(2024)\citenamefont {Diaz}, \citenamefont {Matera},\ and\ \citenamefont {Rossignoli}}]{diaz2024spacetime}%
  \BibitemOpen
  \bibfield  {author} {\bibinfo {author} {\bibfnamefont {N.~L.}\ \bibnamefont {Diaz}}, \bibinfo {author} {\bibfnamefont {J.~M.}\ \bibnamefont {Matera}},\ and\ \bibinfo {author} {\bibfnamefont {R.}~\bibnamefont {Rossignoli}},\ }\bibfield  {title} {\bibinfo {title} {Spacetime quantum and classical mechanics with dynamical foliation},\ }\href {https://doi.org/10.1103/PhysRevD.109.105008} {\bibfield  {journal} {\bibinfo  {journal} {Phys. Rev. D}\ }\textbf {\bibinfo {volume} {109}},\ \bibinfo {pages} {105008} (\bibinfo {year} {2024})}\BibitemShut {NoStop}%
\bibitem [{\citenamefont {Diaz}\ \emph {et~al.}(2025{\natexlab{a}})\citenamefont {Diaz}, \citenamefont {Matera},\ and\ \citenamefont {Rossignoli}}]{DIAZ2025170052}%
  \BibitemOpen
  \bibfield  {author} {\bibinfo {author} {\bibfnamefont {N.}~\bibnamefont {Diaz}}, \bibinfo {author} {\bibfnamefont {J.}~\bibnamefont {Matera}},\ and\ \bibinfo {author} {\bibfnamefont {R.}~\bibnamefont {Rossignoli}},\ }\bibfield  {title} {\bibinfo {title} {Path integrals from spacetime quantum actions},\ }\href {https://doi.org/https://doi.org/10.1016/j.aop.2025.170052} {\bibfield  {journal} {\bibinfo  {journal} {Ann. Phys. (N.Y.)}\ }\textbf {\bibinfo {volume} {479}},\ \bibinfo {pages} {170052} (\bibinfo {year} {2025}{\natexlab{a}})}\BibitemShut {NoStop}%
\bibitem [{\citenamefont {Ryu}\ and\ \citenamefont {Takayanagi}(2006)}]{ryu.06}%
  \BibitemOpen
  \bibfield  {author} {\bibinfo {author} {\bibfnamefont {S.}~\bibnamefont {Ryu}}\ and\ \bibinfo {author} {\bibfnamefont {T.}~\bibnamefont {Takayanagi}},\ }\bibfield  {title} {\bibinfo {title} {Holographic derivation of entanglement entropy from the anti--de {S}itter space/conformal field theory correspondence},\ }\href {https://doi.org/10.1103/PhysRevLett.96.181602} {\bibfield  {journal} {\bibinfo  {journal} {Phys.\ Rev.\ Lett.}\ }\textbf {\bibinfo {volume} {96}},\ \bibinfo {pages} {181602} (\bibinfo {year} {2006})}\BibitemShut {NoStop}%
\bibitem [{\citenamefont {Van~Raamsdonk}(2010)}]{van2010building}%
  \BibitemOpen
  \bibfield  {author} {\bibinfo {author} {\bibfnamefont {M.}~\bibnamefont {Van~Raamsdonk}},\ }\bibfield  {title} {\bibinfo {title} {Building up space--time with quantum entanglement},\ }\href@noop {} {\bibfield  {journal} {\bibinfo  {journal} {Int. J. Mod. Phys. D}\ }\textbf {\bibinfo {volume} {19}},\ \bibinfo {pages} {2429} (\bibinfo {year} {2010})}\BibitemShut {NoStop}%
\bibitem [{\citenamefont {Evenbly}\ and\ \citenamefont {Vidal}(2011)}]{evenbly2011tensor}%
  \BibitemOpen
  \bibfield  {author} {\bibinfo {author} {\bibfnamefont {G.}~\bibnamefont {Evenbly}}\ and\ \bibinfo {author} {\bibfnamefont {G.}~\bibnamefont {Vidal}},\ }\bibfield  {title} {\bibinfo {title} {Tensor network states and geometry},\ }\href@noop {} {\bibfield  {journal} {\bibinfo  {journal} {J. Stat. Phys.}\ }\textbf {\bibinfo {volume} {145}},\ \bibinfo {pages} {891} (\bibinfo {year} {2011})}\BibitemShut {NoStop}%
\bibitem [{\citenamefont {Cao}\ \emph {et~al.}(2017)\citenamefont {Cao}, \citenamefont {Carroll},\ and\ \citenamefont {Michalakis}}]{Ca.17}%
  \BibitemOpen
  \bibfield  {author} {\bibinfo {author} {\bibfnamefont {C.}~\bibnamefont {Cao}}, \bibinfo {author} {\bibfnamefont {S.~M.}\ \bibnamefont {Carroll}},\ and\ \bibinfo {author} {\bibfnamefont {S.}~\bibnamefont {Michalakis}},\ }\bibfield  {title} {\bibinfo {title} {Space from {H}ilbert space: Recovering geometry from bulk entanglement},\ }\href {https://journals.aps.org/prd/abstract/10.1103/PhysRevD.95.024031} {\bibfield  {journal} {\bibinfo  {journal} {Phys. Rev. D}\ }\textbf {\bibinfo {volume} {95}},\ \bibinfo {pages} {024031} (\bibinfo {year} {2017})}\BibitemShut {NoStop}%
\bibitem [{\citenamefont {Feynman}(1948)}]{Feynm.1948a}%
  \BibitemOpen
  \bibfield  {author} {\bibinfo {author} {\bibfnamefont {R.~P.}\ \bibnamefont {Feynman}},\ }\bibfield  {title} {\bibinfo {title} {Space-time approach to non-relativistic quantum mechanics},\ }\href {https://doi.org/10.1103/RevModPhys.20.367} {\bibfield  {journal} {\bibinfo  {journal} {Rev. Mod. Phys.}\ }\textbf {\bibinfo {volume} {20}},\ \bibinfo {pages} {367} (\bibinfo {year} {1948})}\BibitemShut {NoStop}%
\bibitem [{\citenamefont {Diaz}\ \emph {et~al.}(2021)\citenamefont {Diaz}, \citenamefont {Matera},\ and\ \citenamefont {Rossignoli}}]{diaz.21}%
  \BibitemOpen
  \bibfield  {author} {\bibinfo {author} {\bibfnamefont {N.~L.}\ \bibnamefont {Diaz}}, \bibinfo {author} {\bibfnamefont {J.~M.}\ \bibnamefont {Matera}},\ and\ \bibinfo {author} {\bibfnamefont {R.}~\bibnamefont {Rossignoli}},\ }\bibfield  {title} {\bibinfo {title} {Spacetime quantum actions},\ }\href {https://journals.aps.org/prd/abstract/10.1103/PhysRevD.103.065011} {\bibfield  {journal} {\bibinfo  {journal} {Phys. Rev. D}\ }\textbf {\bibinfo {volume} {103}},\ \bibinfo {pages} {065011} (\bibinfo {year} {2021})}\BibitemShut {NoStop}%
\bibitem [{\citenamefont {Page}\ and\ \citenamefont {Wootters}(1983)}]{PaW.83}%
  \BibitemOpen
  \bibfield  {author} {\bibinfo {author} {\bibfnamefont {D.~N.}\ \bibnamefont {Page}}\ and\ \bibinfo {author} {\bibfnamefont {W.~K.}\ \bibnamefont {Wootters}},\ }\bibfield  {title} {\bibinfo {title} {Evolution without evolution: Dynamics described by stationary observables},\ }\href {https://journals.aps.org/prd/abstract/10.1103/PhysRevD.27.2885} {\bibfield  {journal} {\bibinfo  {journal} {Phys. Rev. D}\ }\textbf {\bibinfo {volume} {27}},\ \bibinfo {pages} {2885} (\bibinfo {year} {1983})}\BibitemShut {NoStop}%
\bibitem [{\citenamefont {Giovannetti}\ \emph {et~al.}(2015)\citenamefont {Giovannetti}, \citenamefont {Lloyd},\ and\ \citenamefont {Maccone}}]{QT.15}%
  \BibitemOpen
  \bibfield  {author} {\bibinfo {author} {\bibfnamefont {V.}~\bibnamefont {Giovannetti}}, \bibinfo {author} {\bibfnamefont {S.}~\bibnamefont {Lloyd}},\ and\ \bibinfo {author} {\bibfnamefont {L.}~\bibnamefont {Maccone}},\ }\bibfield  {title} {\bibinfo {title} {Quantum time},\ }\href {https://journals.aps.org/prd/abstract/10.1103/PhysRevD.92.045033} {\bibfield  {journal} {\bibinfo  {journal} {Phys. Rev. D}\ }\textbf {\bibinfo {volume} {92}},\ \bibinfo {pages} {045033} (\bibinfo {year} {2015})}\BibitemShut {NoStop}%
\bibitem [{\citenamefont {Boette}\ \emph {et~al.}(2016)\citenamefont {Boette}, \citenamefont {Rossignoli}, \citenamefont {Gigena},\ and\ \citenamefont {Cerezo}}]{b.16}%
  \BibitemOpen
  \bibfield  {author} {\bibinfo {author} {\bibfnamefont {A.}~\bibnamefont {Boette}}, \bibinfo {author} {\bibfnamefont {R.}~\bibnamefont {Rossignoli}}, \bibinfo {author} {\bibfnamefont {N.}~\bibnamefont {Gigena}},\ and\ \bibinfo {author} {\bibfnamefont {M.}~\bibnamefont {Cerezo}},\ }\bibfield  {title} {\bibinfo {title} {System-time entanglement in a discrete-time model},\ }\href {https://journals.aps.org/pra/abstract/10.1103/PhysRevA.93.062127} {\bibfield  {journal} {\bibinfo  {journal} {Phys. Rev. A}\ }\textbf {\bibinfo {volume} {93}},\ \bibinfo {pages} {062127} (\bibinfo {year} {2016})}\BibitemShut {NoStop}%
\bibitem [{\citenamefont {Diaz}\ and\ \citenamefont {Rossignoli}(2019)}]{di.19}%
  \BibitemOpen
  \bibfield  {author} {\bibinfo {author} {\bibfnamefont {N.~L.}\ \bibnamefont {Diaz}}\ and\ \bibinfo {author} {\bibfnamefont {R.}~\bibnamefont {Rossignoli}},\ }\bibfield  {title} {\bibinfo {title} {History state formalism for {D}irac’s theory},\ }\href {https://journals.aps.org/prd/abstract/10.1103/PhysRevD.99.045008} {\bibfield  {journal} {\bibinfo  {journal} {Phys. Rev. D}\ }\textbf {\bibinfo {volume} {99}},\ \bibinfo {pages} {045008} (\bibinfo {year} {2019})}\BibitemShut {NoStop}%
\bibitem [{\citenamefont {Diaz}\ \emph {et~al.}(2019)\citenamefont {Diaz}, \citenamefont {Matera},\ and\ \citenamefont {Rossignoli}}]{dia.19}%
  \BibitemOpen
  \bibfield  {author} {\bibinfo {author} {\bibfnamefont {N.~L.}\ \bibnamefont {Diaz}}, \bibinfo {author} {\bibfnamefont {J.~M.}\ \bibnamefont {Matera}},\ and\ \bibinfo {author} {\bibfnamefont {R.}~\bibnamefont {Rossignoli}},\ }\bibfield  {title} {\bibinfo {title} {History state formalism for scalar particles},\ }\href {https://journals.aps.org/prd/abstract/10.1103/PhysRevD.100.125020} {\bibfield  {journal} {\bibinfo  {journal} {Phys. Rev. D}\ }\textbf {\bibinfo {volume} {100}},\ \bibinfo {pages} {125020} (\bibinfo {year} {2019})}\BibitemShut {NoStop}%
\bibitem [{\citenamefont {H\"ohn}\ \emph {et~al.}(2021)\citenamefont {H\"ohn}, \citenamefont {Smith},\ and\ \citenamefont {Lock}}]{hoh.21}%
  \BibitemOpen
  \bibfield  {author} {\bibinfo {author} {\bibfnamefont {P.~A.}\ \bibnamefont {H\"ohn}}, \bibinfo {author} {\bibfnamefont {A.~R.~H.}\ \bibnamefont {Smith}},\ and\ \bibinfo {author} {\bibfnamefont {M.~P.~E.}\ \bibnamefont {Lock}},\ }\bibfield  {title} {\bibinfo {title} {Trinity of relational quantum dynamics},\ }\href {https://doi.org/10.1103/PhysRevD.104.066001} {\bibfield  {journal} {\bibinfo  {journal} {Phys. Rev. D}\ }\textbf {\bibinfo {volume} {104}},\ \bibinfo {pages} {066001} (\bibinfo {year} {2021})}\BibitemShut {NoStop}%
\bibitem [{\citenamefont {Favalli}\ and\ \citenamefont {Smerzi}(2022)}]{fav.22}%
  \BibitemOpen
  \bibfield  {author} {\bibinfo {author} {\bibfnamefont {T.}~\bibnamefont {Favalli}}\ and\ \bibinfo {author} {\bibfnamefont {A.}~\bibnamefont {Smerzi}},\ }\bibfield  {title} {\bibinfo {title} {A model of quantum spacetime},\ }\href@noop {} {\bibfield  {journal} {\bibinfo  {journal} {AVS Quantum Science}\ }\textbf {\bibinfo {volume} {4}},\ \bibinfo {pages} {044403} (\bibinfo {year} {2022})}\BibitemShut {NoStop}%
\bibitem [{\citenamefont {Diaz}\ \emph {et~al.}(2025{\natexlab{b}})\citenamefont {Diaz}, \citenamefont {Braccia}, \citenamefont {Larocca}, \citenamefont {Matera}, \citenamefont {Rossignoli},\ and\ \citenamefont {Cerezo}}]{diaz2023parallel}%
  \BibitemOpen
  \bibfield  {author} {\bibinfo {author} {\bibfnamefont {N.~L.}\ \bibnamefont {Diaz}}, \bibinfo {author} {\bibfnamefont {P.}~\bibnamefont {Braccia}}, \bibinfo {author} {\bibfnamefont {M.}~\bibnamefont {Larocca}}, \bibinfo {author} {\bibfnamefont {J.}~\bibnamefont {Matera}}, \bibinfo {author} {\bibfnamefont {R.}~\bibnamefont {Rossignoli}},\ and\ \bibinfo {author} {\bibfnamefont {M.}~\bibnamefont {Cerezo}},\ }\bibfield  {title} {\bibinfo {title} {Parallel-in-time quantum simulation via {P}age and {W}ootters quantum time},\ }\href {https://doi.org/10.1103/wpnf-4nnn} {\bibfield  {journal} {\bibinfo  {journal} {Phys. Rev. Research}\ } (\bibinfo {year} {2025}{\natexlab{b}})}\BibitemShut {NoStop}%
\bibitem [{\citenamefont {Giovannetti}\ \emph {et~al.}(2023)\citenamefont {Giovannetti}, \citenamefont {Lloyd},\ and\ \citenamefont {Maccone}}]{giovannetti2023geometric}%
  \BibitemOpen
  \bibfield  {author} {\bibinfo {author} {\bibfnamefont {V.}~\bibnamefont {Giovannetti}}, \bibinfo {author} {\bibfnamefont {S.}~\bibnamefont {Lloyd}},\ and\ \bibinfo {author} {\bibfnamefont {L.}~\bibnamefont {Maccone}},\ }\bibfield  {title} {\bibinfo {title} {Geometric event-based quantum mechanics},\ }\href {https://iopscience.iop.org/article/10.1088/1367-2630/acb793} {\bibfield  {journal} {\bibinfo  {journal} {New J.\ Phys.}\ }\textbf {\bibinfo {volume} {25}},\ \bibinfo {pages} {023027} (\bibinfo {year} {2023})}\BibitemShut {NoStop}%
\bibitem [{\citenamefont {Suleymanov}\ \emph {et~al.}(2024)\citenamefont {Suleymanov}, \citenamefont {Paiva},\ and\ \citenamefont {Cohen}}]{suleymanov2024nonrelativistic}%
  \BibitemOpen
  \bibfield  {author} {\bibinfo {author} {\bibfnamefont {M.}~\bibnamefont {Suleymanov}}, \bibinfo {author} {\bibfnamefont {I.~L.}\ \bibnamefont {Paiva}},\ and\ \bibinfo {author} {\bibfnamefont {E.}~\bibnamefont {Cohen}},\ }\bibfield  {title} {\bibinfo {title} {Nonrelativistic spatiotemporal quantum reference frames},\ }\href@noop {} {\bibfield  {journal} {\bibinfo  {journal} {Phys. Rev. A}\ }\textbf {\bibinfo {volume} {109}},\ \bibinfo {pages} {032205} (\bibinfo {year} {2024})}\BibitemShut {NoStop}%
\bibitem [{\citenamefont {Cafasso}\ \emph {et~al.}(2024)\citenamefont {Cafasso}, \citenamefont {Pranzini}, \citenamefont {Malo}, \citenamefont {Giovannetti},\ and\ \citenamefont {Chiofalo}}]{cafasso2024quantum}%
  \BibitemOpen
  \bibfield  {author} {\bibinfo {author} {\bibfnamefont {D.}~\bibnamefont {Cafasso}}, \bibinfo {author} {\bibfnamefont {N.}~\bibnamefont {Pranzini}}, \bibinfo {author} {\bibfnamefont {J.~Y.}\ \bibnamefont {Malo}}, \bibinfo {author} {\bibfnamefont {V.}~\bibnamefont {Giovannetti}},\ and\ \bibinfo {author} {\bibfnamefont {M.}~\bibnamefont {Chiofalo}},\ }\bibfield  {title} {\bibinfo {title} {Quantum time and the time-dilation induced interaction transfer mechanism},\ }\href {https://doi.org/10.1103/PhysRevD.110.106014} {\bibfield  {journal} {\bibinfo  {journal} {Phys. Rev. D}\ }\textbf {\bibinfo {volume} {110}},\ \bibinfo {pages} {106014} (\bibinfo {year} {2024})}\BibitemShut {NoStop}%
\bibitem [{\citenamefont {Milekhin}\ \emph {et~al.}(2025)\citenamefont {Milekhin}, \citenamefont {Adamska},\ and\ \citenamefont {Preskill}}]{MAP.25}%
  \BibitemOpen
  \bibfield  {author} {\bibinfo {author} {\bibfnamefont {A.}~\bibnamefont {Milekhin}}, \bibinfo {author} {\bibfnamefont {Z.}~\bibnamefont {Adamska}},\ and\ \bibinfo {author} {\bibfnamefont {J.}~\bibnamefont {Preskill}},\ }\href@noop {} {\bibinfo {title} {Observable and computable entanglement in time}} (\bibinfo {year} {2025}),\ \Eprint {https://arxiv.org/abs/2502.12240} {arXiv:2502.12240 [quant-ph]} \BibitemShut {NoStop}%
\bibitem [{\citenamefont {{Wu-zhong Guo}}(2025)}]{Guo.25}%
  \BibitemOpen
  \bibfield  {author} {\bibinfo {author} {\bibnamefont {{Wu-zhong Guo}}},\ }\href@noop {} {\bibinfo {title} {Spacetime density matrix: Formalism and properties}} (\bibinfo {year} {2025}),\ \Eprint {https://doi.org/10.1007/JHEP01(2026)128} {JHEP 01(2026)128} \BibitemShut {NoStop}%
\bibitem [{\citenamefont {Doi}\ \emph {et~al.}(2023{\natexlab{a}})\citenamefont {Doi}, \citenamefont {Harper}, \citenamefont {Mollabashi}, \citenamefont {Takayanagi},\ and\ \citenamefont {Taki}}]{tak.23}%
  \BibitemOpen
  \bibfield  {author} {\bibinfo {author} {\bibfnamefont {K.}~\bibnamefont {Doi}}, \bibinfo {author} {\bibfnamefont {J.}~\bibnamefont {Harper}}, \bibinfo {author} {\bibfnamefont {A.}~\bibnamefont {Mollabashi}}, \bibinfo {author} {\bibfnamefont {T.}~\bibnamefont {Takayanagi}},\ and\ \bibinfo {author} {\bibfnamefont {Y.}~\bibnamefont {Taki}},\ }\bibfield  {title} {\bibinfo {title} {Pseudoentropy in d{S/CFT} and timelike entanglement entropy},\ }\href {https://doi.org/10.1103/PhysRevLett.130.031601} {\bibfield  {journal} {\bibinfo  {journal} {Phys.\ Rev.\ Lett.}\ }\textbf {\bibinfo {volume} {130}},\ \bibinfo {pages} {031601} (\bibinfo {year} {2023}{\natexlab{a}})}\BibitemShut {NoStop}%
\bibitem [{\citenamefont {Doi}\ \emph {et~al.}(2023{\natexlab{b}})\citenamefont {Doi}, \citenamefont {Harper}, \citenamefont {Mollabashi}, \citenamefont {Takayanagi},\ and\ \citenamefont {Taki}}]{har.23}%
  \BibitemOpen
  \bibfield  {author} {\bibinfo {author} {\bibfnamefont {K.}~\bibnamefont {Doi}}, \bibinfo {author} {\bibfnamefont {J.}~\bibnamefont {Harper}}, \bibinfo {author} {\bibfnamefont {A.}~\bibnamefont {Mollabashi}}, \bibinfo {author} {\bibfnamefont {T.}~\bibnamefont {Takayanagi}},\ and\ \bibinfo {author} {\bibfnamefont {Y.}~\bibnamefont {Taki}},\ }\bibfield  {title} {\bibinfo {title} {Timelike entanglement entropy},\ }\href {https://link.springer.com/article/10.1007/JHEP05(2023)052} {\bibfield  {journal} {\bibinfo  {journal} {J.\ High Energy Phys.}\ }\textbf {\bibinfo {volume} {2023}},\ \bibinfo {pages} {52} (\bibinfo {year} {2023}{\natexlab{b}})}\BibitemShut {NoStop}%
\bibitem [{\citenamefont {Narayan}(2023)}]{nar.22}%
  \BibitemOpen
  \bibfield  {author} {\bibinfo {author} {\bibfnamefont {K.}~\bibnamefont {Narayan}},\ }\bibfield  {title} {\bibinfo {title} {de {S}itter space, extremal surfaces, and time entanglement},\ }\href {https://doi.org/10.1103/PhysRevD.107.126004} {\bibfield  {journal} {\bibinfo  {journal} {Phys.\ Rev.\ D}\ }\textbf {\bibinfo {volume} {107}},\ \bibinfo {pages} {126004} (\bibinfo {year} {2023})}\BibitemShut {NoStop}%
\bibitem [{\citenamefont {Chu}\ and\ \citenamefont {Parihar}(2023)}]{chu2023time}%
  \BibitemOpen
  \bibfield  {author} {\bibinfo {author} {\bibfnamefont {C.-S.}\ \bibnamefont {Chu}}\ and\ \bibinfo {author} {\bibfnamefont {H.}~\bibnamefont {Parihar}},\ }\bibfield  {title} {\bibinfo {title} {Time-like entanglement entropy in ads/bcft},\ }\href {https://doi.org/10.1007/JHEP06%282023%29173} {\bibfield  {journal} {\bibinfo  {journal} {arXiv:2304.10907}\ } (\bibinfo {year} {2023})}\BibitemShut {NoStop}%
\bibitem [{\citenamefont {Fitzsimons}\ \emph {et~al.}(2015)\citenamefont {Fitzsimons}, \citenamefont {Jones},\ and\ \citenamefont {Vedral}}]{fit.15}%
  \BibitemOpen
  \bibfield  {author} {\bibinfo {author} {\bibfnamefont {J.~F.}\ \bibnamefont {Fitzsimons}}, \bibinfo {author} {\bibfnamefont {J.~A.}\ \bibnamefont {Jones}},\ and\ \bibinfo {author} {\bibfnamefont {V.}~\bibnamefont {Vedral}},\ }\bibfield  {title} {\bibinfo {title} {Quantum correlations which imply causation},\ }\href {https://www.nature.com/articles/srep18281} {\bibfield  {journal} {\bibinfo  {journal} {Sci. Rep.}\ }\textbf {\bibinfo {volume} {5}},\ \bibinfo {pages} {18281} (\bibinfo {year} {2015})}\BibitemShut {NoStop}%
\bibitem [{\citenamefont {Horsman}\ \emph {et~al.}(2017)\citenamefont {Horsman}, \citenamefont {Heunen}, \citenamefont {Pusey}, \citenamefont {Barrett},\ and\ \citenamefont {Spekkens}}]{ho.17}%
  \BibitemOpen
  \bibfield  {author} {\bibinfo {author} {\bibfnamefont {D.}~\bibnamefont {Horsman}}, \bibinfo {author} {\bibfnamefont {C.}~\bibnamefont {Heunen}}, \bibinfo {author} {\bibfnamefont {M.~F.}\ \bibnamefont {Pusey}}, \bibinfo {author} {\bibfnamefont {J.}~\bibnamefont {Barrett}},\ and\ \bibinfo {author} {\bibfnamefont {R.~W.}\ \bibnamefont {Spekkens}},\ }\bibfield  {title} {\bibinfo {title} {Can a quantum state over time resemble a quantum state at a single time?},\ }\href {https://royalsocietypublishing.org/doi/10.1098/rspa.2017.0395} {\bibfield  {journal} {\bibinfo  {journal} {Proc. R. Soc. A}\ }\textbf {\bibinfo {volume} {473}},\ \bibinfo {pages} {20170395} (\bibinfo {year} {2017})}\BibitemShut {NoStop}%
\bibitem [{\citenamefont {Cotler}\ \emph {et~al.}(2018)\citenamefont {Cotler}, \citenamefont {Jian}, \citenamefont {Qi},\ and\ \citenamefont {Wilczek}}]{cot.18}%
  \BibitemOpen
  \bibfield  {author} {\bibinfo {author} {\bibfnamefont {J.}~\bibnamefont {Cotler}}, \bibinfo {author} {\bibfnamefont {C.~M.}\ \bibnamefont {Jian}}, \bibinfo {author} {\bibfnamefont {X.}~\bibnamefont {Qi}},\ and\ \bibinfo {author} {\bibfnamefont {F.}~\bibnamefont {Wilczek}},\ }\bibfield  {title} {\bibinfo {title} {Superdensity operators for spacetime quantum mechanics},\ }\href {https://link.springer.com/article/10.1007/JHEP09(2018)093} {\bibfield  {journal} {\bibinfo  {journal} {J. High Energy Phys.}\ }\textbf {\bibinfo {volume} {2018}},\ \bibinfo {pages} {93}}\BibitemShut {NoStop}%
\bibitem [{\citenamefont {Nielsen}\ and\ \citenamefont {Chuang}(2001)}]{nie.01}%
  \BibitemOpen
  \bibfield  {author} {\bibinfo {author} {\bibfnamefont {M.~A.}\ \bibnamefont {Nielsen}}\ and\ \bibinfo {author} {\bibfnamefont {I.~L.}\ \bibnamefont {Chuang}},\ }\bibfield  {title} {\bibinfo {title} {Quantum computation and quantum information},\ }\href@noop {} {\bibfield  {journal} {\bibinfo  {journal} {Cambridge University Press}\ } (\bibinfo {year} {2010})}\BibitemShut {NoStop}%
\bibitem [{\citenamefont {Buhrman}\ \emph {et~al.}(2001)\citenamefont {Buhrman}, \citenamefont {Cleve}, \citenamefont {Watrous},\ and\ \citenamefont {De~Wolf}}]{buhrman2001quantum}%
  \BibitemOpen
  \bibfield  {author} {\bibinfo {author} {\bibfnamefont {H.}~\bibnamefont {Buhrman}}, \bibinfo {author} {\bibfnamefont {R.}~\bibnamefont {Cleve}}, \bibinfo {author} {\bibfnamefont {J.}~\bibnamefont {Watrous}},\ and\ \bibinfo {author} {\bibfnamefont {R.}~\bibnamefont {De~Wolf}},\ }\bibfield  {title} {\bibinfo {title} {Quantum fingerprinting},\ }\href {https://doi.org/10.1103/PhysRevLett.87.167902} {\bibfield  {journal} {\bibinfo  {journal} {Phys.\ Rev.\ Lett.}\ }\textbf {\bibinfo {volume} {87}},\ \bibinfo {pages} {167902} (\bibinfo {year} {2001})}\BibitemShut {NoStop}%
\bibitem [{\citenamefont {Bultrini}\ \emph {et~al.}(2023)\citenamefont {Bultrini}, \citenamefont {Gordon}, \citenamefont {Czarnik}, \citenamefont {Arrasmith}, \citenamefont {Cerezo}, \citenamefont {Coles},\ and\ \citenamefont {Cincio}}]{bultrini2023unifying}%
  \BibitemOpen
  \bibfield  {author} {\bibinfo {author} {\bibfnamefont {D.}~\bibnamefont {Bultrini}}, \bibinfo {author} {\bibfnamefont {M.~H.}\ \bibnamefont {Gordon}}, \bibinfo {author} {\bibfnamefont {P.}~\bibnamefont {Czarnik}}, \bibinfo {author} {\bibfnamefont {A.}~\bibnamefont {Arrasmith}}, \bibinfo {author} {\bibfnamefont {M.}~\bibnamefont {Cerezo}}, \bibinfo {author} {\bibfnamefont {P.~J.}\ \bibnamefont {Coles}},\ and\ \bibinfo {author} {\bibfnamefont {L.}~\bibnamefont {Cincio}},\ }\bibfield  {title} {\bibinfo {title} {Unifying and benchmarking state-of-the-art quantum error mitigation techniques},\ }\href@noop {} {\bibfield  {journal} {\bibinfo  {journal} {Quantum}\ }\textbf {\bibinfo {volume} {7}},\ \bibinfo {pages} {1034} (\bibinfo {year} {2023})}\BibitemShut {NoStop}%
\bibitem [{\citenamefont {Holmes}\ \emph {et~al.}(2023)\citenamefont {Holmes}, \citenamefont {Coble}, \citenamefont {Sornborger},\ and\ \citenamefont {Suba{\c{s}}{\i}}}]{holmes2023nonlinear}%
  \BibitemOpen
  \bibfield  {author} {\bibinfo {author} {\bibfnamefont {Z.}~\bibnamefont {Holmes}}, \bibinfo {author} {\bibfnamefont {N.~J.}\ \bibnamefont {Coble}}, \bibinfo {author} {\bibfnamefont {A.~T.}\ \bibnamefont {Sornborger}},\ and\ \bibinfo {author} {\bibfnamefont {Y.}~\bibnamefont {Suba{\c{s}}{\i}}},\ }\bibfield  {title} {\bibinfo {title} {Nonlinear transformations in quantum computation},\ }\href@noop {} {\bibfield  {journal} {\bibinfo  {journal} {Phys. Rev. Res.}\ }\textbf {\bibinfo {volume} {5}},\ \bibinfo {pages} {013105} (\bibinfo {year} {2023})}\BibitemShut {NoStop}%
\bibitem [{\citenamefont {Diaz}\ \emph {et~al.}(2023)\citenamefont {Diaz}, \citenamefont {Garc{\'\i}a-Mart{\'\i}n}, \citenamefont {Kazi}, \citenamefont {Larocca},\ and\ \citenamefont {Cerezo}}]{diaz2023showcasing}%
  \BibitemOpen
  \bibfield  {author} {\bibinfo {author} {\bibfnamefont {N.}~\bibnamefont {Diaz}}, \bibinfo {author} {\bibfnamefont {D.}~\bibnamefont {Garc{\'\i}a-Mart{\'\i}n}}, \bibinfo {author} {\bibfnamefont {S.}~\bibnamefont {Kazi}}, \bibinfo {author} {\bibfnamefont {M.}~\bibnamefont {Larocca}},\ and\ \bibinfo {author} {\bibfnamefont {M.}~\bibnamefont {Cerezo}},\ }\bibfield  {title} {\bibinfo {title} {Showcasing a barren plateau theory beyond the dynamical lie algebra},\ }\href {https://doi.org/10.48550/arXiv.2310.11505} {\bibfield  {journal} {\bibinfo  {journal} {arXiv:2310.11505}\ } (\bibinfo {year} {2023})}\BibitemShut {NoStop}%
\bibitem [{\citenamefont {Jordan}\ and\ \citenamefont {Wigner}(1928)}]{JW0.28}%
  \BibitemOpen
  \bibfield  {author} {\bibinfo {author} {\bibfnamefont {P.}~\bibnamefont {Jordan}}\ and\ \bibinfo {author} {\bibfnamefont {E.}~\bibnamefont {Wigner}},\ }\bibfield  {title} {\bibinfo {title} {{\"U}ber das {P}aulische \"aquivalenzverbot},\ }\href@noop {} {\bibfield  {journal} {\bibinfo  {journal} {Zeit. Phys.}\ }\textbf {\bibinfo {volume} {47}},\ \bibinfo {pages} {631} (\bibinfo {year} {1928})}\BibitemShut {NoStop}%
\bibitem [{\citenamefont {Lieb}\ \emph {et~al.}(1961)\citenamefont {Lieb}, \citenamefont {Schultz},\ and\ \citenamefont {Mattis}}]{LSM.61}%
  \BibitemOpen
  \bibfield  {author} {\bibinfo {author} {\bibfnamefont {E.}~\bibnamefont {Lieb}}, \bibinfo {author} {\bibfnamefont {T.}~\bibnamefont {Schultz}},\ and\ \bibinfo {author} {\bibfnamefont {D.}~\bibnamefont {Mattis}},\ }\bibfield  {title} {\bibinfo {title} {Two soluble models of an antiferromagnetic chain},\ }\href@noop {} {\bibfield  {journal} {\bibinfo  {journal} {Ann. Phys. (N.Y.)}\ }\textbf {\bibinfo {volume} {16}},\ \bibinfo {pages} {407} (\bibinfo {year} {1961})}\BibitemShut {NoStop}%
\bibitem [{\citenamefont {Jozsa}\ and\ \citenamefont {Miyake}(2008)}]{jozsa2008matchgates}%
  \BibitemOpen
  \bibfield  {author} {\bibinfo {author} {\bibfnamefont {R.}~\bibnamefont {Jozsa}}\ and\ \bibinfo {author} {\bibfnamefont {A.}~\bibnamefont {Miyake}},\ }\bibfield  {title} {\bibinfo {title} {Matchgates and classical simulation of quantum circuits},\ }\href@noop {} {\bibfield  {journal} {\bibinfo  {journal} {Proc. R. Soc. A}\ }\textbf {\bibinfo {volume} {464}},\ \bibinfo {pages} {3089} (\bibinfo {year} {2008})}\BibitemShut {NoStop}%
\bibitem [{Note1()}]{Note1}%
  \BibitemOpen
  \bibinfo {note} {It is clear that if we think of $e^{i\epsilon \protect \mathcal {P}}$ as a two-qubit operation, it is a Matchgate with the convention of \cite {jozsa2008matchgates,hebenstreit2019all, diaz2023showcasing}: $e^{i\epsilon \protect \mathcal {P}}=\exp \{i\protect \frac {\pi }{4}(\sigma _2 \otimes \sigma _1-\sigma _1\otimes \sigma _2)\}$ with $\sigma _2 \otimes \sigma _1, \sigma _1\otimes \sigma _2$ both a product of two Majoranas \cite {diaz2023showcasing}. In this sense, $e^{i\epsilon \protect \mathcal {P}}$ is a variation of the fSWAP gate \cite {hebenstreit2019all}.}\BibitemShut {Stop}%
\bibitem [{\citenamefont {{Gaudin}}(1960)}]{1960NucPh}%
  \BibitemOpen
  \bibfield  {author} {\bibinfo {author} {\bibfnamefont {M.}~\bibnamefont {{Gaudin}}},\ }\bibfield  {title} {\bibinfo {title} {{Une d{\'e}monstration simplifi{\'e}e du th{\'e}or{\`e}me de {W}ick en m{\'e}canique statistique}},\ }\href {https://doi.org/10.1016/0029-5582(60)90285-6} {\bibfield  {journal} {\bibinfo  {journal} {Nucl. Phys.}\ }\textbf {\bibinfo {volume} {15}},\ \bibinfo {pages} {89} (\bibinfo {year} {1960})}\BibitemShut {NoStop}%
\bibitem [{\citenamefont {Evans}\ and\ \citenamefont {Steer}(1996)}]{evans1996wick}%
  \BibitemOpen
  \bibfield  {author} {\bibinfo {author} {\bibfnamefont {T.}~\bibnamefont {Evans}}\ and\ \bibinfo {author} {\bibfnamefont {D.~A.}\ \bibnamefont {Steer}},\ }\bibfield  {title} {\bibinfo {title} {Wick's theorem at finite temperature},\ }\href@noop {} {\bibfield  {journal} {\bibinfo  {journal} {Nucl. Phys. B}\ }\textbf {\bibinfo {volume} {474}},\ \bibinfo {pages} {481} (\bibinfo {year} {1996})}\BibitemShut {NoStop}%
\bibitem [{\citenamefont {Altland}\ and\ \citenamefont {Simons}(2010)}]{altland.2010}%
  \BibitemOpen
  \bibfield  {author} {\bibinfo {author} {\bibfnamefont {A.}~\bibnamefont {Altland}}\ and\ \bibinfo {author} {\bibfnamefont {B.}~\bibnamefont {Simons}},\ }\href@noop {} {\emph {\bibinfo {title} {Condensed Matter Field Theory}}}\ (\bibinfo  {publisher} {Cambridge University Press},\ \bibinfo {year} {2010})\BibitemShut {NoStop}%
\bibitem [{Note2()}]{Note2}%
  \BibitemOpen
  \bibinfo {note} {One can show that while the quantum action is related to Feynman PIs, the operator $\protect \tilde {\protect \mathcal {S}}$ is instead related to Schwinger–Keldysh PIs \cite {schwinger1961brownian}. We employ the second to discuss generalized states since here we are interested in expectation values at a single initial time. See also section \ref {sec:entanglementintime}.}\BibitemShut {Stop}%
\bibitem [{\citenamefont {Aharonov}\ \emph {et~al.}(1988)\citenamefont {Aharonov}, \citenamefont {Albert},\ and\ \citenamefont {Vaidman}}]{yak.88}%
  \BibitemOpen
  \bibfield  {author} {\bibinfo {author} {\bibfnamefont {Y.}~\bibnamefont {Aharonov}}, \bibinfo {author} {\bibfnamefont {D.~Z.}\ \bibnamefont {Albert}},\ and\ \bibinfo {author} {\bibfnamefont {L.}~\bibnamefont {Vaidman}},\ }\bibfield  {title} {\bibinfo {title} {How the result of a measurement of a component of the spin of a spin-1/2 particle can turn out to be 100},\ }\href {https://doi.org/10.1103/PhysRevLett.60.1351} {\bibfield  {journal} {\bibinfo  {journal} {Phys. Rev. Lett.}\ }\textbf {\bibinfo {volume} {60}},\ \bibinfo {pages} {1351} (\bibinfo {year} {1988})}\BibitemShut {NoStop}%
\bibitem [{\citenamefont {Dressel}\ \emph {et~al.}(2014)\citenamefont {Dressel}, \citenamefont {Malik}, \citenamefont {Miatto}, \citenamefont {Jordan},\ and\ \citenamefont {Boyd}}]{dres.14}%
  \BibitemOpen
  \bibfield  {author} {\bibinfo {author} {\bibfnamefont {J.}~\bibnamefont {Dressel}}, \bibinfo {author} {\bibfnamefont {M.}~\bibnamefont {Malik}}, \bibinfo {author} {\bibfnamefont {F.~M.}\ \bibnamefont {Miatto}}, \bibinfo {author} {\bibfnamefont {A.~N.}\ \bibnamefont {Jordan}},\ and\ \bibinfo {author} {\bibfnamefont {R.~W.}\ \bibnamefont {Boyd}},\ }\bibfield  {title} {\bibinfo {title} {Colloquium: Understanding quantum weak values: Basics and applications},\ }\href {https://doi.org/10.1103/RevModPhys.86.307} {\bibfield  {journal} {\bibinfo  {journal} {Rev.\ Mod.\ Phys.}\ }\textbf {\bibinfo {volume} {86}},\ \bibinfo {pages} {307} (\bibinfo {year} {2014})}\BibitemShut {NoStop}%
\bibitem [{\citenamefont {Ring}\ and\ \citenamefont {Schuck}(2004)}]{ring2004nuclear}%
  \BibitemOpen
  \bibfield  {author} {\bibinfo {author} {\bibfnamefont {P.}~\bibnamefont {Ring}}\ and\ \bibinfo {author} {\bibfnamefont {P.}~\bibnamefont {Schuck}},\ }\href@noop {} {\emph {\bibinfo {title} {The nuclear many-body problem}}}\ (\bibinfo  {publisher} {Springer Science \& Business Media},\ \bibinfo {year} {2004})\BibitemShut {NoStop}%
\bibitem [{Note3()}]{Note3}%
  \BibitemOpen
  \bibinfo {note} {This type of entropy was introduced in the context of dS-CFT correspondence \cite {tak.23} (for standard QM).}\BibitemShut {Stop}%
\bibitem [{\citenamefont {Balian}(2005)}]{balian2005information}%
  \BibitemOpen
  \bibfield  {author} {\bibinfo {author} {\bibfnamefont {R.}~\bibnamefont {Balian}},\ }\bibfield  {title} {\bibinfo {title} {Information in statistical physics},\ }\href@noop {} {\bibfield  {journal} {\bibinfo  {journal} {Studies in History and Philosophy of Science Part B: Studies in History and Philosophy of Modern Physics}\ }\textbf {\bibinfo {volume} {36}},\ \bibinfo {pages} {323} (\bibinfo {year} {2005})}\BibitemShut {NoStop}%
\bibitem [{\citenamefont {Balian}\ \emph {et~al.}(2006)\citenamefont {Balian}, \citenamefont {Haar},\ and\ \citenamefont {Gregg}}]{balian2006microphysics}%
  \BibitemOpen
  \bibfield  {author} {\bibinfo {author} {\bibfnamefont {R.}~\bibnamefont {Balian}}, \bibinfo {author} {\bibfnamefont {D.}~\bibnamefont {Haar}},\ and\ \bibinfo {author} {\bibfnamefont {J.}~\bibnamefont {Gregg}},\ }\href {https://books.google.com/books?id=xsQzLxiatyMC} {\emph {\bibinfo {title} {From Microphysics to Macrophysics: Methods and Applications of Statistical Physics}}},\ \bibinfo {series} {Theoretical and Mathematical Physics}\ No.\ \bibinfo {number} {v. 1}\ (\bibinfo  {publisher} {Springer Berlin Heidelberg},\ \bibinfo {year} {2006})\BibitemShut {NoStop}%
\bibitem [{\citenamefont {Nielsen}\ \emph {et~al.}(2003)\citenamefont {Nielsen}, \citenamefont {Dawson}, \citenamefont {Dodd}, \citenamefont {Gilchrist}, \citenamefont {Mortimer}, \citenamefont {Osborne}, \citenamefont {Bremner}, \citenamefont {Harrow},\ and\ \citenamefont {Hines}}]{nielsen2003quantum}%
  \BibitemOpen
  \bibfield  {author} {\bibinfo {author} {\bibfnamefont {M.~A.}\ \bibnamefont {Nielsen}}, \bibinfo {author} {\bibfnamefont {C.~M.}\ \bibnamefont {Dawson}}, \bibinfo {author} {\bibfnamefont {J.~L.}\ \bibnamefont {Dodd}}, \bibinfo {author} {\bibfnamefont {A.}~\bibnamefont {Gilchrist}}, \bibinfo {author} {\bibfnamefont {D.}~\bibnamefont {Mortimer}}, \bibinfo {author} {\bibfnamefont {T.~J.}\ \bibnamefont {Osborne}}, \bibinfo {author} {\bibfnamefont {M.~J.}\ \bibnamefont {Bremner}}, \bibinfo {author} {\bibfnamefont {A.~W.}\ \bibnamefont {Harrow}},\ and\ \bibinfo {author} {\bibfnamefont {A.}~\bibnamefont {Hines}},\ }\bibfield  {title} {\bibinfo {title} {Quantum dynamics as a physical resource},\ }\href@noop {} {\bibfield  {journal} {\bibinfo  {journal} {Phys. Rev. A}\ }\textbf {\bibinfo {volume} {67}},\ \bibinfo {pages} {052301} (\bibinfo {year} {2003})}\BibitemShut {NoStop}%
\bibitem [{\citenamefont {Gigena}\ and\ \citenamefont {Rossignoli}(2015)}]{gigena2015entanglement}%
  \BibitemOpen
  \bibfield  {author} {\bibinfo {author} {\bibfnamefont {N.}~\bibnamefont {Gigena}}\ and\ \bibinfo {author} {\bibfnamefont {R.}~\bibnamefont {Rossignoli}},\ }\bibfield  {title} {\bibinfo {title} {Entanglement in fermion systems},\ }\href@noop {} {\bibfield  {journal} {\bibinfo  {journal} {Phys. Rev. A}\ }\textbf {\bibinfo {volume} {92}},\ \bibinfo {pages} {042326} (\bibinfo {year} {2015})}\BibitemShut {NoStop}%
\bibitem [{\citenamefont {Kleinert}(2006)}]{kleinert2006path}%
  \BibitemOpen
  \bibfield  {author} {\bibinfo {author} {\bibfnamefont {H.}~\bibnamefont {Kleinert}},\ }\href@noop {} {\emph {\bibinfo {title} {Path integrals in quantum mechanics, statistics, polymer physics, and financial markets}}}\ (\bibinfo  {publisher} {World Scientific Publishing Company},\ \bibinfo {year} {2006})\BibitemShut {NoStop}%
\bibitem [{\citenamefont {Puddu}\ \emph {et~al.}(1991)\citenamefont {Puddu}, \citenamefont {Bortignon},\ and\ \citenamefont {Broglia}}]{puddu1991rpa}%
  \BibitemOpen
  \bibfield  {author} {\bibinfo {author} {\bibfnamefont {G.}~\bibnamefont {Puddu}}, \bibinfo {author} {\bibfnamefont {P.}~\bibnamefont {Bortignon}},\ and\ \bibinfo {author} {\bibfnamefont {R.}~\bibnamefont {Broglia}},\ }\bibfield  {title} {\bibinfo {title} {The rpa-spa approximation to level densities},\ }\href@noop {} {\bibfield  {journal} {\bibinfo  {journal} {Ann. Phys. (N.Y.)}\ }\textbf {\bibinfo {volume} {206}},\ \bibinfo {pages} {409} (\bibinfo {year} {1991})}\BibitemShut {NoStop}%
\bibitem [{\citenamefont {Rossignoli}\ \emph {et~al.}(1998)\citenamefont {Rossignoli}, \citenamefont {Canosa},\ and\ \citenamefont {Ring}}]{rossignoli1998thermal}%
  \BibitemOpen
  \bibfield  {author} {\bibinfo {author} {\bibfnamefont {R.}~\bibnamefont {Rossignoli}}, \bibinfo {author} {\bibfnamefont {N.}~\bibnamefont {Canosa}},\ and\ \bibinfo {author} {\bibfnamefont {P.}~\bibnamefont {Ring}},\ }\bibfield  {title} {\bibinfo {title} {Thermal and quantal fluctuations for fixed particle number in finite superfluid systems},\ }\href@noop {} {\bibfield  {journal} {\bibinfo  {journal} {Phys. Rev. Lett.}\ }\textbf {\bibinfo {volume} {80}},\ \bibinfo {pages} {1853} (\bibinfo {year} {1998})}\BibitemShut {NoStop}%
\bibitem [{\citenamefont {Nakata}\ \emph {et~al.}(2021)\citenamefont {Nakata}, \citenamefont {Takayanagi}, \citenamefont {Taki}, \citenamefont {Tamaoka},\ and\ \citenamefont {Wei}}]{nak.21}%
  \BibitemOpen
  \bibfield  {author} {\bibinfo {author} {\bibfnamefont {Y.}~\bibnamefont {Nakata}}, \bibinfo {author} {\bibfnamefont {T.}~\bibnamefont {Takayanagi}}, \bibinfo {author} {\bibfnamefont {Y.}~\bibnamefont {Taki}}, \bibinfo {author} {\bibfnamefont {K.}~\bibnamefont {Tamaoka}},\ and\ \bibinfo {author} {\bibfnamefont {Z.}~\bibnamefont {Wei}},\ }\bibfield  {title} {\bibinfo {title} {New holographic generalization of entanglement entropy},\ }\href {https://doi.org/10.1103/PhysRevD.103.026005} {\bibfield  {journal} {\bibinfo  {journal} {Phys. Rev. D}\ }\textbf {\bibinfo {volume} {103}},\ \bibinfo {pages} {026005} (\bibinfo {year} {2021})}\BibitemShut {NoStop}%
\bibitem [{\citenamefont {Peskin}(2018)}]{P.18}%
  \BibitemOpen
  \bibfield  {author} {\bibinfo {author} {\bibfnamefont {M.~E.}\ \bibnamefont {Peskin}},\ }\href@noop {} {\emph {\bibinfo {title} {An introduction to quantum field theory}}}\ (\bibinfo  {publisher} {CRC Press},\ \bibinfo {year} {2018})\BibitemShut {NoStop}%
\bibitem [{\citenamefont {Schwabl}(2008)}]{Sc.97}%
  \BibitemOpen
  \bibfield  {author} {\bibinfo {author} {\bibfnamefont {F.}~\bibnamefont {Schwabl}},\ }\href@noop {} {\emph {\bibinfo {title} {Advanced Quantum Mechanics}}}\ (\bibinfo  {publisher} {Springer},\ \bibinfo {year} {2008})\BibitemShut {NoStop}%
\bibitem [{\citenamefont {Thaller}(1992)}]{T.92}%
  \BibitemOpen
  \bibfield  {author} {\bibinfo {author} {\bibfnamefont {B.}~\bibnamefont {Thaller}},\ }\href@noop {} {\emph {\bibinfo {title} {The Dirac Equation}}}\ (\bibinfo  {publisher} {Springer-Verlag},\ \bibinfo {address} {Berlin Heidelberg},\ \bibinfo {year} {1992})\BibitemShut {NoStop}%
\bibitem [{\citenamefont {Feynman}(2018)}]{feynman2018theory}%
  \BibitemOpen
  \bibfield  {author} {\bibinfo {author} {\bibfnamefont {R.~P.}\ \bibnamefont {Feynman}},\ }\bibfield  {title} {\bibinfo {title} {The theory of positrons},\ }in\ \href@noop {} {\emph {\bibinfo {booktitle} {Quantum Electrodynamics}}}\ (\bibinfo  {publisher} {CRC Press},\ \bibinfo {year} {2018})\ pp.\ \bibinfo {pages} {167--177}\BibitemShut {NoStop}%
\bibitem [{\citenamefont {C{\'e}leri}\ \emph {et~al.}(2016)\citenamefont {C{\'e}leri}, \citenamefont {Kiosses},\ and\ \citenamefont {Terno}}]{cel.16}%
  \BibitemOpen
  \bibfield  {author} {\bibinfo {author} {\bibfnamefont {L.~C.}\ \bibnamefont {C{\'e}leri}}, \bibinfo {author} {\bibfnamefont {V.}~\bibnamefont {Kiosses}},\ and\ \bibinfo {author} {\bibfnamefont {D.~R.}\ \bibnamefont {Terno}},\ }\bibfield  {title} {\bibinfo {title} {Spin and localization of relativistic fermions and uncertainty relations},\ }\href@noop {} {\bibfield  {journal} {\bibinfo  {journal} {Phys. Rev. A}\ }\textbf {\bibinfo {volume} {94}},\ \bibinfo {pages} {062115} (\bibinfo {year} {2016})}\BibitemShut {NoStop}%
\bibitem [{\citenamefont {Schwinger}(1961)}]{schwinger1961brownian}%
  \BibitemOpen
  \bibfield  {author} {\bibinfo {author} {\bibfnamefont {J.}~\bibnamefont {Schwinger}},\ }\bibfield  {title} {\bibinfo {title} {Brownian motion of a quantum oscillator},\ }\href@noop {} {\bibfield  {journal} {\bibinfo  {journal} {J. Math. Phys.}\ }\textbf {\bibinfo {volume} {2}},\ \bibinfo {pages} {407} (\bibinfo {year} {1961})}\BibitemShut {NoStop}%
\bibitem [{\citenamefont {Parzygnat}\ \emph {et~al.}(2023)\citenamefont {Parzygnat}, \citenamefont {Takayanagi}, \citenamefont {Taki},\ and\ \citenamefont {Wei}}]{parzygnat2023svd}%
  \BibitemOpen
  \bibfield  {author} {\bibinfo {author} {\bibfnamefont {A.~J.}\ \bibnamefont {Parzygnat}}, \bibinfo {author} {\bibfnamefont {T.}~\bibnamefont {Takayanagi}}, \bibinfo {author} {\bibfnamefont {Y.}~\bibnamefont {Taki}},\ and\ \bibinfo {author} {\bibfnamefont {Z.}~\bibnamefont {Wei}},\ }\bibfield  {title} {\bibinfo {title} {{SVD} entanglement entropy},\ }\href@noop {} {\bibfield  {journal} {\bibinfo  {journal} {J. High Energy Phys.}\ }\textbf {\bibinfo {volume} {2023}}\bibinfo  {number} { (12)},\ \bibinfo {pages} {123}}\BibitemShut {NoStop}%
\bibitem [{\citenamefont {Fullwood}\ and\ \citenamefont {Parzygnat}(2024)}]{fullwood2024operator}%
  \BibitemOpen
\bibfield  {number} {  }\bibfield  {author} {\bibinfo {author} {\bibfnamefont {J.}~\bibnamefont {Fullwood}}\ and\ \bibinfo {author} {\bibfnamefont {A.~J.}\ \bibnamefont {Parzygnat}},\ }\bibfield  {title} {\bibinfo {title} {Operator representation of spatiotemporal quantum correlations},\ }\href@noop {} {\bibfield  {journal} {\bibinfo  {journal} {arXiv:2405.17555}\ } (\bibinfo {year} {2024})}\BibitemShut {NoStop}%
\bibitem [{\citenamefont {Zurek}(2003)}]{zurek2003environment}%
  \BibitemOpen
  \bibfield  {author} {\bibinfo {author} {\bibfnamefont {W.~H.}\ \bibnamefont {Zurek}},\ }\bibfield  {title} {\bibinfo {title} {Environment-assisted invariance, entanglement, and probabilities in quantum physics},\ }\href@noop {} {\bibfield  {journal} {\bibinfo  {journal} {Phys. Rev. Lett.}\ }\textbf {\bibinfo {volume} {90}},\ \bibinfo {pages} {120404} (\bibinfo {year} {2003})}\BibitemShut {NoStop}%
\bibitem [{\citenamefont {Zurek}(2013)}]{zurek2013wave}%
  \BibitemOpen
  \bibfield  {author} {\bibinfo {author} {\bibfnamefont {W.~H.}\ \bibnamefont {Zurek}},\ }\bibfield  {title} {\bibinfo {title} {Wave-packet collapse and the core quantum postulates: Discreteness of quantum jumps from unitarity, repeatability, and actionable information},\ }\href@noop {} {\bibfield  {journal} {\bibinfo  {journal} {Phys, Rev. A}\ }\textbf {\bibinfo {volume} {87}},\ \bibinfo {pages} {052111} (\bibinfo {year} {2013})}\BibitemShut {NoStop}%
\bibitem [{\citenamefont {Gigena}\ and\ \citenamefont {Rossignoli}(2016)}]{gigena2016one}%
  \BibitemOpen
  \bibfield  {author} {\bibinfo {author} {\bibfnamefont {N.}~\bibnamefont {Gigena}}\ and\ \bibinfo {author} {\bibfnamefont {R.}~\bibnamefont {Rossignoli}},\ }\bibfield  {title} {\bibinfo {title} {One-body information loss in fermion systems},\ }\href@noop {} {\bibfield  {journal} {\bibinfo  {journal} {Phys. Rev. A}\ }\textbf {\bibinfo {volume} {94}},\ \bibinfo {pages} {042315} (\bibinfo {year} {2016})}\BibitemShut {NoStop}%
\bibitem [{\citenamefont {Shankar}(2012)}]{shankar2012principles}%
  \BibitemOpen
  \bibfield  {author} {\bibinfo {author} {\bibfnamefont {R.}~\bibnamefont {Shankar}},\ }\href@noop {} {\emph {\bibinfo {title} {Principles of quantum mechanics}}}\ (\bibinfo  {publisher} {Springer Science \& Business Media},\ \bibinfo {year} {2012})\BibitemShut {NoStop}%
\bibitem [{\citenamefont {Dirac}(1950)}]{pam.50}%
  \BibitemOpen
  \bibfield  {author} {\bibinfo {author} {\bibfnamefont {P.~A.~M.}\ \bibnamefont {Dirac}},\ }\bibfield  {title} {\bibinfo {title} {Generalized hamiltonian dynamics},\ }\href@noop {} {\bibfield  {journal} {\bibinfo  {journal} {Can. J. Math.}\ }\textbf {\bibinfo {volume} {2}},\ \bibinfo {pages} {129–148} (\bibinfo {year} {1950})}\BibitemShut {NoStop}%
\bibitem [{\citenamefont {Kiefer}(2004)}]{qg}%
  \BibitemOpen
  \bibfield  {author} {\bibinfo {author} {\bibfnamefont {C.}~\bibnamefont {Kiefer}},\ }\bibfield  {title} {\bibinfo {title} {Quantum gravity},\ }\href@noop {} {\bibfield  {journal} {\bibinfo  {journal} {Int. Ser. Monogr. Phys.}\ }\textbf {\bibinfo {volume} {155}},\ \bibinfo {pages} {1} (\bibinfo {year} {2004})}\BibitemShut {NoStop}%
\bibitem [{\citenamefont {Griffiths}(1984)}]{griffiths1984consistent}%
  \BibitemOpen
  \bibfield  {author} {\bibinfo {author} {\bibfnamefont {R.~B.}\ \bibnamefont {Griffiths}},\ }\bibfield  {title} {\bibinfo {title} {Consistent histories and the interpretation of quantum mechanics},\ }\href@noop {} {\bibfield  {journal} {\bibinfo  {journal} {Journal of Statistical Physics}\ }\textbf {\bibinfo {volume} {36}},\ \bibinfo {pages} {219} (\bibinfo {year} {1984})}\BibitemShut {NoStop}%
\bibitem [{\citenamefont {Omnes}(1988)}]{omnes1988logical}%
  \BibitemOpen
  \bibfield  {author} {\bibinfo {author} {\bibfnamefont {R.}~\bibnamefont {Omnes}},\ }\bibfield  {title} {\bibinfo {title} {Logical reformulation of quantum mechanics. i. foundations},\ }\href@noop {} {\bibfield  {journal} {\bibinfo  {journal} {Journal of Statistical Physics}\ }\textbf {\bibinfo {volume} {53}},\ \bibinfo {pages} {893} (\bibinfo {year} {1988})}\BibitemShut {NoStop}%
\bibitem [{\citenamefont {Gell-Mann}\ and\ \citenamefont {Hartle}(2010)}]{gell2010quantum}%
  \BibitemOpen
  \bibfield  {author} {\bibinfo {author} {\bibfnamefont {M.}~\bibnamefont {Gell-Mann}}\ and\ \bibinfo {author} {\bibfnamefont {J.~B.}\ \bibnamefont {Hartle}},\ }\bibfield  {title} {\bibinfo {title} {Quantum mechanics in the light of quantum cosmology},\ }in\ \href@noop {} {\emph {\bibinfo {booktitle} {Murray Gell-Mann: Selected Papers}}}\ (\bibinfo  {publisher} {World Scientific},\ \bibinfo {year} {2010})\ pp.\ \bibinfo {pages} {303--325}\BibitemShut {NoStop}%
\bibitem [{\citenamefont {Isham}(1994{\natexlab{b}})}]{ish.93}%
  \BibitemOpen
  \bibfield  {author} {\bibinfo {author} {\bibfnamefont {C.~J.}\ \bibnamefont {Isham}},\ }\bibfield  {title} {\bibinfo {title} {Quantum logic and the histories approach to quantum theory},\ }\href {https://doi.org/10.1063/1.530544} {\bibfield  {journal} {\bibinfo  {journal} {J.\ Math.\ Phys.}\ }\textbf {\bibinfo {volume} {35}},\ \bibinfo {pages} {2157} (\bibinfo {year} {1994}{\natexlab{b}})}\BibitemShut {NoStop}%
\bibitem [{\citenamefont {Isham}\ \emph {et~al.}(1994)\citenamefont {Isham}, \citenamefont {Linden},\ and\ \citenamefont {Schreckenberg}}]{isham1994classification}%
  \BibitemOpen
  \bibfield  {author} {\bibinfo {author} {\bibfnamefont {C.~J.}\ \bibnamefont {Isham}}, \bibinfo {author} {\bibfnamefont {N.}~\bibnamefont {Linden}},\ and\ \bibinfo {author} {\bibfnamefont {S.}~\bibnamefont {Schreckenberg}},\ }\bibfield  {title} {\bibinfo {title} {The classification of decoherence functionals: An analog of gleason’s theorem},\ }\href@noop {} {\bibfield  {journal} {\bibinfo  {journal} {Journal of Mathematical Physics}\ }\textbf {\bibinfo {volume} {35}},\ \bibinfo {pages} {6360} (\bibinfo {year} {1994})}\BibitemShut {NoStop}%
\bibitem [{\citenamefont {Gleason}(1975)}]{gleason1975measures}%
  \BibitemOpen
  \bibfield  {author} {\bibinfo {author} {\bibfnamefont {A.~M.}\ \bibnamefont {Gleason}},\ }\bibfield  {title} {\bibinfo {title} {Measures on the closed subspaces of a hilbert space},\ }in\ \href@noop {} {\emph {\bibinfo {booktitle} {The Logico-Algebraic Approach to Quantum Mechanics: Volume I: Historical Evolution}}}\ (\bibinfo  {publisher} {Springer},\ \bibinfo {year} {1975})\ pp.\ \bibinfo {pages} {123--133}\BibitemShut {NoStop}%
\bibitem [{\citenamefont {et~al.}()}]{unifying}%
  \BibitemOpen
  \bibfield  {author} {\bibinfo {author} {\bibfnamefont {N. L. Diaz}\ \bibnamefont {et~al}},\ }\bibfield  {title} {\bibinfo {title} {In preparation}}\href@noop {} {\ }\BibitemShut {NoStop}%
\bibitem [{\citenamefont {Savvidou}(1999)}]{savvidou1999action}%
  \BibitemOpen
  \bibfield  {author} {\bibinfo {author} {\bibfnamefont {K.}~\bibnamefont {Savvidou}},\ }\bibfield  {title} {\bibinfo {title} {The action operator for continuous-time histories},\ }\href@noop {} {\bibfield  {journal} {\bibinfo  {journal} {Journal of Mathematical Physics}\ }\textbf {\bibinfo {volume} {40}},\ \bibinfo {pages} {5657} (\bibinfo {year} {1999})}\BibitemShut {NoStop}%
\bibitem [{\citenamefont {Chen}\ and\ \citenamefont {Wu}(2003)}]{chen2002matrix}%
  \BibitemOpen
  \bibfield  {author} {\bibinfo {author} {\bibfnamefont {K.}~\bibnamefont {Chen}}\ and\ \bibinfo {author} {\bibfnamefont {L.-A.}\ \bibnamefont {Wu}},\ }\bibfield  {title} {\bibinfo {title} {A matrix realignment method for recognizing entanglement},\ }\href@noop {} {\bibfield  {journal} {\bibinfo  {journal} {Quantum Information and Computation}\ }\textbf {\bibinfo {volume} {3}},\ \bibinfo {pages} {193} (\bibinfo {year} {2003})}\BibitemShut {NoStop}%
\bibitem [{\citenamefont {Hebenstreit}\ \emph {et~al.}(2019)\citenamefont {Hebenstreit}, \citenamefont {Jozsa}, \citenamefont {Kraus}, \citenamefont {Strelchuk},\ and\ \citenamefont {Yoganathan}}]{hebenstreit2019all}%
  \BibitemOpen
  \bibfield  {author} {\bibinfo {author} {\bibfnamefont {M.}~\bibnamefont {Hebenstreit}}, \bibinfo {author} {\bibfnamefont {R.}~\bibnamefont {Jozsa}}, \bibinfo {author} {\bibfnamefont {B.}~\bibnamefont {Kraus}}, \bibinfo {author} {\bibfnamefont {S.}~\bibnamefont {Strelchuk}},\ and\ \bibinfo {author} {\bibfnamefont {M.}~\bibnamefont {Yoganathan}},\ }\bibfield  {title} {\bibinfo {title} {All pure fermionic non-gaussian states are magic states for matchgate computations},\ }\href@noop {} {\bibfield  {journal} {\bibinfo  {journal} {Phys. Rev. Lett.}\ }\textbf {\bibinfo {volume} {123}},\ \bibinfo {pages} {080503} (\bibinfo {year} {2019})}\BibitemShut {NoStop}%
\end{thebibliography}

%apsrev4-2.bst 2019-01-14 (MD) hand-edited version of apsrev4-1.bst
%Control: key (0)
%Control: author (8) initials jnrlst
%Control: editor formatted (1) identically to author
%Control: production of article title (0) allowed
%Control: page (0) single
%Control: year (1) truncated
%Control: production of eprint (0) enabled
%

\end{document}